\documentclass[useAMS,fleqn,usenatbib]{mnras}
 
\usepackage{graphicx}
\usepackage{psfig}
\usepackage{makeidx,amsmath,mathrsfs}
\usepackage{verbatim}
\usepackage{newtxtext,newtxmath}      
\usepackage[T1]{fontenc}
\usepackage{ae,aecompl}
\usepackage[dvipsnames]{xcolor}	
\usepackage{comment}

\title[eBOSS Galaxy Mock Challenge]{The Completed SDSS-IV Extended Baryon Oscillation Spectroscopic Survey: $N$-body Mock Challenge for Galaxy Clustering Measurements}

\author[G. Rossi et al. (2020)]{\parbox{\textwidth}{
Graziano Rossi$^1$\thanks{Corresponding Author: graziano@sejong.ac.kr}, 
Peter D. Choi$^1$, 
Jeongin Moon$^1$,  
Julian E. Bautista$^2$, 
Hector Gil-Mar\'{i}n$^{3, 4}$,
Romain Paviot$^{5,6}$,
Mariana Vargas-Maga\~na$^{7}$,
Sylvain de la Torre$^{5}$,
Sebastien Fromenteau$^{8}$,
Ashley J. Ross$^{9}$,
Santiago {\'A}vila$^{10,11}$,
Etienne Burtin$^{12}$,
Kyle~S. Dawson$^{13}$,
St{\'e}phanie Escoffier$^{6}$, 
Salman Habib$^{14,15}$,
Katrin Heitmann$^{14}$,
Jiamin Hou$^{16}$,
Eva-Maria Mueller$^{17}$,
Will J. Percival$^{18, 19, 20}$,
Alex Smith$^{12}$,
Cheng Zhao$^{21}$,
Gong-Bo Zhao$^{22, 23, 2}$\\
} \vspace*{4pt} \\ 
$^{1}${
Department of Physics and Astronomy, Sejong University, Seoul, 143-747, Korea
}\\
$^{2}${
Institute of Cosmology \& Gravitation, 
Dennis Sciama Building, University of Portsmouth, 
Portsmouth, PO1 3FX, UK
}\\
$^{3}${
Institut de Ci\`encies del Cosmos,  
Universitat  de  Barcelona,  ICCUB,  
Mart\'i  i  Franqu\`es  1,  E08028  Barcelona,  Spain
}\\
$^{4}${
Institut  d'Estudis  Espacials  de  Catalunya  (IEEC),  
E08034  Barcelona,  Spain
}\\
$^{5}${
Aix Marseille Univ., CNRS, CNES, LAM, Marseille, France
}\\
$^{6}${
Aix Marseille Univ., CNRS/IN2P3, CPPM, Marseille, France
}\\
$^{7}${
Instituto de F\'isica, 
Universidad Nacional Aut\'onoma de M\'exico, 
Apdo. Postal 20-364, Ciudad de M\'exico, M\'exico
}\\
$^{8}${ 
Instituto de Ciencias F\'isicas, 
Universidad Nacional Aut\'onoma de M\'exico, 
Av. Universidad s/n, 62210 Cuernavaca, Mor., Mexico
}\\
$^{9}${
Center for Cosmology and Astro-Particle Physics,
Ohio State University, Columbus, OH 43210
}\\
$^{10}${
Universidad Aut\'onoma de Madrid, 28049, Madrid, Spain
}\\
$^{11}${
Instituto de Fisica Teorica UAM/CSIC, Universidad Autonoma de Madrid, 28049 Madrid, Spain
}\\
$^{12}${
CEA, Centre de Saclay, Irfu/SPP,  F-91191 Gif-sur-Yvette, France
}\\
$^{13}${
Department of Physics and Astronomy, 
University of Utah, Salt Lake City, UT 84112, USA
}\\
$^{14}${
High Energy Physics Division, 
Argonne National Laboratory, Lemont, IL 60439, USA 
}\\
$^{15}${
Computational Science Division, 
Argonne National Laboratory, Lemont, IL 60439, USA 
}\\
$^{16}${ 
Max-Planck-Institut f\"ur Extraterrestrische Physik, Postfach 1312, 
Giessenbachstrasse 1, 85748 Garching bei M\"unchen, Germany
}\\
$^{17}${ 
Sub-department of Astrophysics, Department of Physics,
University of Oxford, Denys Wilkinson Building, Keble Road, Oxford OX1 3RH
}\\
$^{18}${
Waterloo Centre for Astrophysics, 
University of Waterloo, Waterloo, ON N2L 3G1, Canada
}\\
$^{19}${ 
Department of Physics and Astronomy, University of Waterloo, 
Waterloo, ON N2L 3G1, Canada
}\\
$^{20}${
Perimeter Institute for Theoretical Physics, 
31 Caroline St. North, Waterloo, ON N2L 2Y5, Canada
}\\
$^{21}${
Laboratory of Astrophysics, \'Ecole Polytechnique F\'ed\'erale de Lausanne (EPFL), 
Observatoire de Sauverny, 1290 Versoix, Switzerland
}\\
$^{22}${
National Astronomy Observatories, 
Chinese Academy of Science, 
Beijing, 100012, P. R. China
}\\
$^{23}${
College of Astronomy and Space Sciences, 
University of Chinese Academy of Sciences, Beijing 100049,
China
}\\
}

\date{Accepted 2020 December 17. Received 2020 November 30; in original form 2020 July 18.}
\pagerange{\pageref{firstpage}--\pageref{lastpage}}
\pubyear{2021}

\begin{document}
\maketitle
\label{firstpage}



\clearpage
\begin{abstract}
We develop a series of $N$-body data challenges, functional to the final analysis of the  
extended Baryon Oscillation Spectroscopic Survey (eBOSS) Data Release 16 (DR16) galaxy sample. 
The challenges are primarily based  
on high-fidelity catalogs constructed from the Outer Rim simulation --  a large box size realization ($3 h^{-1} {\rm Gpc}$) 
characterized by an unprecedented combination of volume and mass resolution, down to $1.85 \cdot 10^9 h^{-1} M_{\odot}$.  
We generate synthetic galaxy mocks by populating Outer Rim halos with a variety of halo occupation distribution 
(HOD) schemes of increasing complexity, spanning 
different redshift intervals. We then assess the performance of three complementary 
redshift space distortion (RSD) models in configuration
and Fourier space, adopted for the analysis of the complete DR16 eBOSS sample of Luminous Red Galaxies (LRGs). 
We find all the methods mutually consistent, with comparable systematic errors on the Alcock-Paczynski parameters and the growth of structure, and robust to 
different HOD prescriptions 
-- thus validating the robustness
of the models and the pipelines used for the baryon acoustic oscillation (BAO) and full shape clustering analysis. 
In particular, all the techniques are able to recover $\alpha_{\parallel}$ and $\alpha_{\perp}$ to within $0.9\%$, and 
$f \sigma_8$ to within $1.5\%$.
As a by-product of our work, we are also able to gain interesting insights  
on the galaxy-halo connection.
Our study is relevant for the final eBOSS DR16 `consensus cosmology',
as the  systematic error budget 
is informed by testing the results of analyses against these high-resolution mocks.  
In addition, it is also useful for future large-volume surveys, since similar mock-making techniques and 
systematic corrections can be readily extended to model for instance the Dark Energy Spectroscopic Instrument (DESI) galaxy sample.  
\end{abstract}






\section{Introduction}


The Sloan Digital Sky Survey \citep[SDSS;][]{York2000}, currently in its fourth generation
(SDSS-IV; see \citealt{Blanton2017} for a review), has established a remarkable legacy in astronomy 
and set new standards for precision cosmology.  
A key component of the SDSS-IV, the Extended Baryon Oscillation Spectroscopic Survey \citep[eBOSS;][]{Dawson2016}
is now releasing the final cosmological catalogs 
\citep{Lyke2020,Raichoor2020,Ross2020} with the Data Release 16 (DR16), 
summarizing the efforts of more than 10 years of operations. 
eBOSS spectroscopically targets four distinct astrophysical populations: luminous red galaxies (LRGs, the primary focus of this work), 
emission line galaxies (ELGs), clustering quasars (QSOs), and the Lyman-$\alpha$ (Ly$\alpha$) forest of quasars at high redshift. 
In a novel and yet uncharted redshift interval, 
eBOSS has built the most complete, unprecedented  
large volume map of the universe usable for large-scale structure (LSS) to date.

Exquisite high-quality data from the SDSS have been pivotal in 
firmly establishing the standard minimal six-parameter concordance cosmological scenario dominated by cold 
dark matter (CDM) and a dark energy (DE) 
component in the form of a cosmological constant $\Lambda$, known as the $\Lambda$CDM model. 
Traditionally, this has been achieved by using the 
baryon acoustic oscillation (BAO) feature as measured in galaxy and quasar clustering,
to estimate the angular diameter distance $D_{\rm M}$ and the Hubble parameter $H$,  as well as their 
product  from the Alcock-Paczynski effect \citep[AP;][]{Alcock1979}, and the growth of structure 
quantified by $f \sigma_8(z)$ from redshift-space distortions (RSD) -- with $f(z)$ the 
logarithmic growth rate of the linear fluctuation amplitude with respect to the expansion factor, and 
$\sigma_8 (z)$ the normalization of the linear theory matter power spectrum at redshift $z$ via the rms fluctuation in $8h^{-1}{\rm Mpc}$ spheres.
Since the very first BAO detections \citep{Colless2003,Cole2005,Eisenstein2005}, 
measurements of the BAO peak have been sharpening and expanding in redshift range, 
allowing for multiple accurate cosmological constraints and solid confirmations of the $\Lambda$CDM framework.  
Noticeably, the eBOSS team has recently presented the first measurement of the BAO signal in a novel uncharted redshift range 
($0.8<z<2.2$) using the clustering properties of $147,000$ new quasars \citep{Ata2018}, and reported
a BAO detection with a significance $> 2.8 \sigma$ along with detailed high-$z$ distance measurements (within 3.8\%), a remarkable result that confirms and 
extends the validity of the standard $\Lambda$CDM cosmological model to an unprecedented large-volume.

To this end, multiple techniques involving 
RSD methods and clustering estimators along with BAO reconstructions in configuration or Fourier 
space are generally adopted for the analysis 
of the various LSS tracers, to extract cosmological information.
The most up-to-date SDSS results involving LRGs can be found in \citet{Beutler2017a,Beutler2017b}, \citet{Gil-Marin2017}, \citet{Bautista2018}, 
\citet{Mueller2018}, \citet{Vargas-Magana2018}, \citet{WangY2018}, \citet{Zheng2019}, and \citet{Icaza-Lizaola2020}.
Regarding ELGs, one of the novelties in eBOSS, recent studies have been carried out by 
\citet{Comparat2016}, \citet{Raichoor2017}, \citet{Guo2019}.
For the QSO population, see e.g.  \citet{Gil-Marin2018,Ruggeri2019}, \citet{Hou2018}, \citet{WangD2018}, \citet{Zarrouk2018},  
\citet{Zhu2018}, \citet{Zhao2019}; and for Ly$\alpha$ QSOs see \citet{Blomqvist2019} and \citet{deSainteAgathe2019}.

Traditionally, all the main results from different SDSS tracers are eventually combined in a `consensus' publication
\citep[e.g][]{Aubourg2015, Alam2017,Ata2018}, 
and confronted with measurements from other state-of-the-art surveys -- such as Planck (2018). 
This consensus is then of utmost importance, as it represents a legacy for the entire science community.
We are now releasing the final eBOSS DR16 consensus analysis 
that summarizes the full impact of the SDSS spectroscopic surveys on the cosmological model
\citep{eBOSS_Cosmology2020}, 
which encapsulates all the supporting clustering measurements presented in 
\citet{LRG_corr2020} and \citet{Gil-Marin2020} for LRGs,
\citet{Hou2020} and \citet{Neveux2020} for QSOs, \citet{DeMattia2020} and \citet{Tamone2020} for ELGs, 
as well as \citet{duMasdesBourboux2020} for the Ly$\alpha$ forest.\footnote{A description of eBOSS and a link to its associated 
publications can be found at this URL: \url{https://www.sdss.org/surveys/eboss/}.}  


In this respect, quantifying the 
systematic error budget in RSD methods and
BAO clustering estimators for all of the eBOSS tracers as well as characterizing the robustness of the analysis pipelines are essential tasks, in order to 
obtain unbiased cosmological parameters, accurate $f\sigma_8$ constraints, and reliable consensus likelihoods. 
This is indeed the central aim of our work: here  
we focus on galaxies,
and assess the 
performance and robustness of the BAO fitting methods and of three complementary RSD 
full shape (FS)  models in configuration and redshift space, adopted in \citet{LRG_corr2020} and \citet{Gil-Marin2020}
for the analysis of the complete DR16 eBOSS LRG sample -- briefly described in Section \ref{sec_analysis_methods}. 
See  also \citet{Smith2020} for an analogous effort on the QSO sample, and
\citet{Alam2020}, \citet{Avila2020}, and \citet{Lin2020} for ELGs. 


With this primary goal in mind, we have devised a targeted galaxy mock challenge.
In embryonic form, a similar mini-challenge was already present in the consensus eBOSS Data Release 12 (DR12) 
LRG analysis \citep[][see their Section 7]{Alam2017}.
Here we expand on that, and carry out a more systematic investigation. 
Specifically, in our challenge  (detailed in Section \ref{sec_unblind_challenge})
we test the performance of BAO/RSD LRG fitting techniques against
different galaxy population schemes and bias models having analogous
clustering properties, with the main  objective of
validation and calibration of such methods and the quantification of theoretical systematics. 


Assessing the robustness and accuracy of   
RSD models is
only possible via high-fidelity ($N$-body-based) synthetic realizations.
In this work, we construct new heterogeneous sets of 
galaxy mocks from the Outer Rim
\citep[][see Section \ref{sec_tools_methodology}]{Heitmann2019}
--  a large box size run ($3 h^{-1} {\rm Gpc}$) 
characterized by a high mass resolution, down to $1.85 \cdot 10^9 h^{-1} M_{\odot}$.  
We base our methodology on Halo Occupation Distribution (HOD) techniques, in an increasing level of complexity 
(as thoroughly explained in Section \ref{sec_theory}): in particular, moving from the most conventional HOD framework,
we explore more sophisticated scenarios able to distinguish between quiescent and star-forming galaxies  
and with the inclusion of assembly bias, that generalize further the standard HOD framework.
Since our primary goal is to test and validate LRG analysis pipelines under a common set of high-resolution mocks sharing similar clustering properties, rather than improve the HOD modeling,
for this study we select a few representative galaxy-halo connection schemes from the plethora of HODs available in the literature:
the inclusion of  further models that go beyond the more conventional HOD formulation
is left to future studies. 
We also exploit a small homogeneous set of cut-sky mocks (the \textsc{Nseries}) -- which has been previously used in the SDSS DR12 galaxy 
clustering analysis \citep{Alam2017} -- to address the impact of cosmic variance and related theoretical systematics, and
make use of a new series of DR16 \textsc{EZmocks} \citep{Zhao2020}
for determining the rescaled covariance matrices functional to all the
analyses. The mock-making procedure is explained in detail in Section \ref{sec_tools_methodology}.


By confronting the
different BAO and RSD LRG fitting techniques on a common ground against a subset of those 
high-fidelity mocks having different HOD prescriptions,
we are thus able to assess their performance,
quantify the systematic errors on the AP  
parameters and the growth of structure, 
and eventually confirm the effectiveness of the LRG analysis pipelines. 
In particular, we anticipate that we find all the methods mutually consistent, and 
robust to different HOD prescriptions, validating  the models used for the LRG clustering analysis.


Furthermore, the mock challenge developed here is suitable to a number of applications. 
Beside being  directly useful  for the final eBOSS DR16 `consensus cosmology' 
\citep{eBOSS_Cosmology2020},
as the systematic error budget for the ultimate $f\sigma_8$ constraint are informed by testing the results of analyses against these 
high-resolution mocks, our work may be relevant  for future large-volume surveys. For example, similar 
 mock-making techniques and 
systematic corrections can be readily extended to model the Dark Energy Spectroscopic 
Instrument \citep[DESI;][]{DESICollaboration2016a}
and the Large Synoptic Survey Telescope 
\citep[LSST;][]{Ivezic2019}
galaxy samples.   
 

The layout of the paper is organized as follows. 
Section \ref{sec_data}  briefly describes the eBOSS DR16 data release, and the final LRG sample.
Section \ref{sec_theory} provides the theoretical foundation for modeling the galaxy-halo connection,
and explains the different HOD schemes adopted in the mock-making procedure -- along with the rationale behind our choices. 
Section \ref{sec_tools_methodology} describes the tools and methodology used to construct 
high-fidelity mocks; the  expert reader may wish to jump directly to the next sections, while
a reader less familiar with HOD modeling could benefit from these parts, without the need of consulting extensive  literature works.
Section \ref{sec_analysis_methods} briefly presents the different 
RSD models, that are described in depth in companion papers. 
Section \ref{sec_unblind_challenge} shows selected results from the mock challenge, and compares the various LRG BAO and RSD models in configuration and Fourier space. 
Section \ref{sec_global_sys} presents the global error budget for the completed eBOSS DR16 LRG sample, with a primary  focus on theoretical systematics. 
Finally, we conclude in Section  \ref{sec_conclusions}, where we summarize the main findings and
indicate future avenues.  
We leave in
Appendix \ref{sec_appendix_mock_products}
an extensive description of all of the mock products developed and publicly released with this study, 
and report in Appendix \ref{sec_appendix_tables} some useful tables.


Throughout the paper and if not specified otherwise, all numerical values of length and mass are understood to be in $h=1$ units.



\section{SDSS-IV eBOSS and DR16 LRG Sample} \label{sec_data}
 

\subsection{SDSS Legacy and eBOSS}

SDSS observations, carried out 
on the 2.5-meter Sloan Foundation telescope at Apache 
Point Observatory \citep{Gunn2006}, first begun in July 2014 \citep[SDSS-I and SDSS-II;][]{York2000}. 
Since then, thanks to the
remarkable efforts of more than 10 years of operations, 
the survey has evolved till its current fourth generation (SDSS-IV), collecting an
increasing number of high-quality data for high-precision cosmology -- outperforming on the targets that drove the initial survey design. 
eBOSS, a key component of the SDSS-IV and ranked in the highest tier in the 2018 DOE-HEP Portfolio Review, 
is a continuation of the Baryon Oscillation Spectroscopic Survey (BOSS) -- part of the SDSS-III \citep{Eisenstein2011} -- 
and a pre-cursor for DESI \citep{DESICollaboration2016a}.
eBOSS lies at the leading edge of cosmological experimentation: by spectroscopically targeting four distinct astrophysical populations 
in a unique redshift interval, eBOSS has built the largest volume and most complete map of the Universe  to date of any redshift survey. 
The primary innovation in eBOSS is extending BAO measurements with ELGs and a much larger number of quasars, enabling
a percent-level measurement in the critical epoch of transition from deceleration to acceleration (i.e., $0.8 < z < 2.2$). 
This is why  the eBOSS data set allows exploration of DE in epochs where no precision cosmological 
measurements currently exist (improving the DE Figure of Merit by a factor of 3), 
by addressing three 
Particle Physics Project Prioritization Panel (P5)
science drivers
and pursuing four key goals: BAO measurements of the Hubble parameter and distance as a function of $z$, RSD measurements of the gravitational growth 
of structure, constraints on the neutrino mass sum, and constraints on inflation through measurements of primordial non-Gaussianity.  
In particular, the exquisite BAO and RSD measurements that eBOSS provide \citep[see e.g.][]{eBOSS_Cosmology2020}
are key for DE and gravity studies. 
Moreover, eBOSS has the spectroscopic capabilities to complement and enhance other current and future cosmological probes, 
representing a strategic asset and a pathfinder for upcoming experiments. 


\subsection{The eBOSS DR16 LRG Sample} 

The LRG spectroscopic sample allowed the first SDSS detection of the BAO peak in 
the galaxy large-scale correlation function \citep{Eisenstein2005}.
Since its original version, comprised by 46,748 LRGs over $3816$ deg$^2$ at $0.16<z<0.47$,
the SDSS LRG catalog has considerably grown both in size and redshift depth, thanks to over about a decade of observations. 
In particular, BOSS was designed to measure BAOs with LRGs over the redshift range $0.2 < z <0.75$, while 
eBOSS increases the redshift coverage up to $z=1$. 
With the final eBOSS DR16, completed on March 1, 2019, 
the LRG eBOSS-only released sample contains 174,816 galaxies with good redshifts  
in the interval range $0.6 < z <1.0$, with an effective redshift $\bar{z}= 0.698$, 
spanning a total area
of  4104 deg$^2$ and an effective volume of  1.241 Gpc$^3$. 
LRG targets were selected {\bf via} optical and infrared imaging
over 7500 deg$^2$ angular footprint, using photometry with updated calibration
\citep{Dawson2016}:  full details of LRG selections are provided in \citet{Prakash2016}.
To this end,  \citet{Bautista2018} recently demonstrated that the sample is
well-suited for LSS studies.
 The final DR16 eBOSS-only LRG sample is combined with the
 high redshift tail of the BOSS galaxy sample (denoted as CMASS), in order to
 provide one catalog of luminous galaxies with $z>0.6$.
 Overall, BOSS CMASS galaxies
 make up slightly more than half of the total sample, and the area they occupy
 is more than twice that of eBOSS LRGs -- over an effective volume of  1.445 Gpc$^3$, hence the total effective volume of the
 combined DR16 LRG sample is 2.654 Gpc$^3$.
 Most of the BOSS CMASS footprint was re-observed by the eBOSS LRG program, which covered $37\%$  
 and $65\%$ of the original Northern Galactic Cup (NGC)   and Southern Galactic Cup (SGC)  CMASS areas, respectively. 
 The projected number density of galaxies with $0.6 < z < 1.0$
 is more than twice as high for the eBOSS LRGs 
 (44 deg$^{-2}$ compared to 21 deg$^{-2}$).

Regarding redshift assignment, a different
philosophy both for redshift estimates and spectral classification has been 
designed specifically
for the eBOSS clustering catalogs, motivated by new challenges
due to low signal-to-noise eBOSS galaxy spectra. In fact, 
previous routines used for BOSS were not optimized
for the fainter and higher redshift LRG galaxies that comprise the eBOSS LRG sample,
and therefore a new approach and software development to provide accurate 
redshift estimation (indicated as \textsc{Redrock}\footnote{See \url{github.com/desihub/redrock}}) was necessary.
As a result, the redshift completeness approaches 98\% for the eBOSS LRG sample with a 
rate of `catastrophic failures' estimated to be less than 1\% -- hence such redshift failures are not a concern for the LRG sample. 
 
About sector completeness (a sector being an area covered by a unique set of plates), for the eBOSS LRG sample
the 100\% completeness requirement 
was relaxed, to increase the fiber efficiency and total survey area.
To this end, the completeness of the eBOSS
LRG sample exceeds 95\% in every relevant chunk (i.e. an area tiled in a single software run)
of the survey,  where completeness statistics are determined on a per-sector basis.
 
A technical description of the LRG observational strategy, 
and on how spectra are turned into redshift estimates, can be found in \citet{Ross2020}.
Extensive details on the LRG catalog creation, observing strategy, matching targets and spectroscopic observations, veto masks, etc., as well as
observational effects such as varying completeness, collision priority, close pairs, redshift failures, systematics related to imaging
and their correction are also presented in \citet{Ross2020}.



\section{Modeling the Galaxy-Halo Connection: Theoretical Background} \label{sec_theory}
 
 
In this section, we provide a concise overview of the theoretical formalism 
underlying our mock-making procedure, in an increasing 
level of complexity. Starting from the most conventional HOD approach,  
we  then consider more sophisticated scenarios 
able to distinguish between quiescent or star-forming galaxies, and
with the inclusion of assembly bias -- that generalize further the standard HOD framework. 

 
\subsection{HOD Modeling: Basics}


\begin{figure*}
\centering                                                              
\includegraphics[angle=0,width=0.85\textwidth]{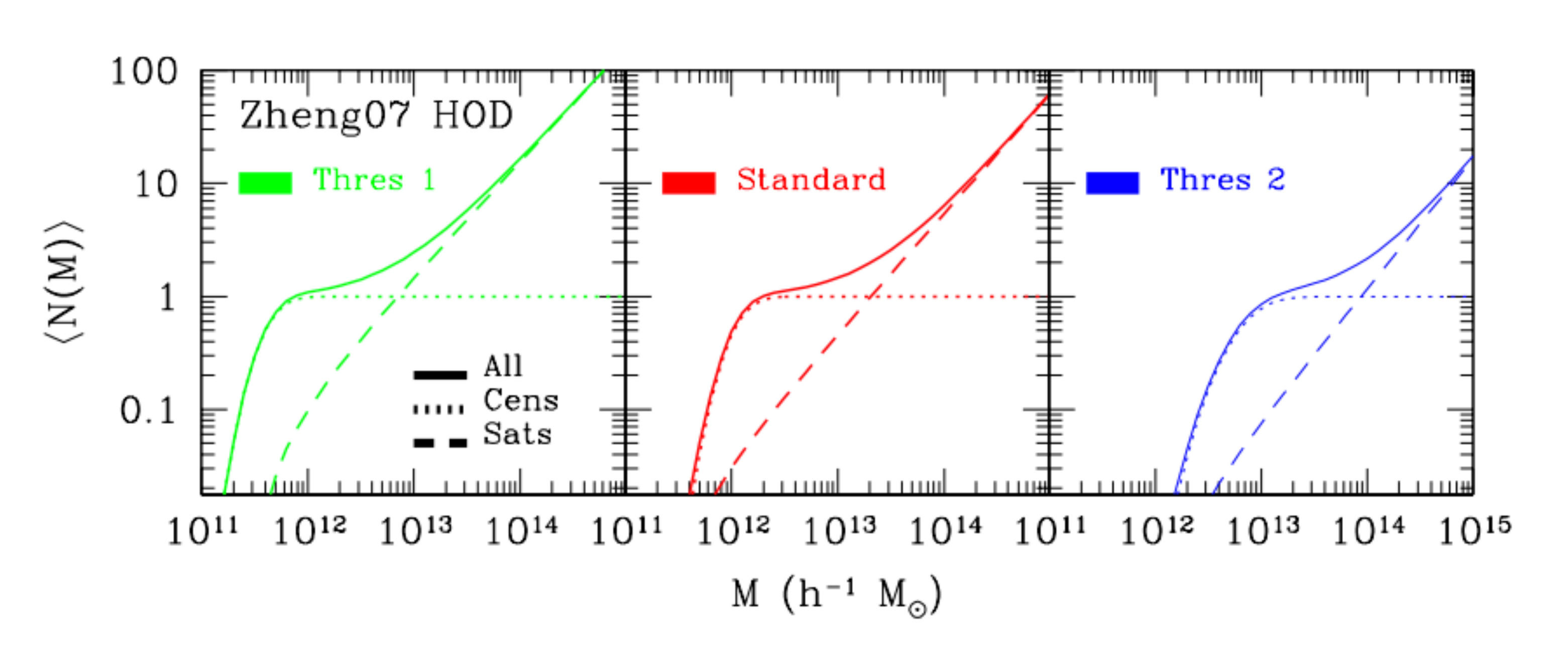}
\caption{HOD shapes in the \citet{Zheng2007} model, used for the production of galaxy mocks. In the various panels, 
dotted lines describe the central occupation statistics (Equation \ref{eq_zheng_hod_cens}),
dashed lines are used for the satellite occupation statistics (Equation \ref{eq_zheng_hod_sats}), and solid lines represent the
composite HODs. Three luminosity thresholds are considered, corresponding to
different HOD parameter choices, as reported in Table \ref{table_zheng07}. See the main text for more details.}
\label{fig_zheng07_hod}
\end{figure*}


The Halo Occupation Distribution 
(HOD;  see e.g., \citealp{Peacock2000,Seljak2000,Scoccimarro2001,Berlind2002,Kravtsov2004,Zheng2005} 
for pioneering works, and e.g., \citealp{Guo2016,Yuan2018,Tinker2019,Alpaslan2020}
for more recent implementations and extensions)
is a popular framework able to establish
a statistical connection between galaxies and dark matter halos bypassing the complex galaxy formation physics, useful 
to inform models of galaxy formation, interpret  LSS measurements, and eventually 
constrain cosmological parameters. 
The core assumption underlying any HOD modeling is that 
all galaxies reside in dark matter halos, and that halos
are biased tracers of the dark matter density field. In this regards,
knowledge of how galaxies populate, and are distributed within, dark matter halos
enables a complete description of all the statistics of the observed galaxy distribution.  

In the most conventional HOD formulation, 
the central quantity is  the probability distribution function (PDF) of galaxies within halos $P(N_{\rm g} | M_{\rm h})$, namely the
probability that a halo of mass $M_{\rm h}$ hosts -- on average -- $N_{\rm g}$ galaxies
in a pre-defined sample. Galaxies are further split into centrals and satellites, and 
conventionally the occupation statistics of central galaxies $\langle N_{\rm cen} \rangle$ are modeled separately 
from satellites $\langle N_{\rm sat} \rangle$, so that:
\begin{equation}
\langle N_{\rm g} | M_{\rm h} \rangle = \langle N_{\rm cen} | M_{\rm h} \rangle + \langle N_{\rm sat} | M_{\rm h} \rangle. 
\end{equation}
In the standard mass-only `ansatz' (i.e., halo bias $b_{\rm h}$ is only a function of halo mass),
central galaxies are commonly assumed to reside at the center of their host halos, inheriting the corresponding halo velocity and concentration values, while 
satellite galaxies are typically designed to follow a radial number density distribution   
that traces the NFW density distribution
\citep{Navarro1997} of the underlying dark matter halo.  
Hence, the starting point for any HOD-style model is choosing an analytic form 
for $\langle N_{\rm cen} \rangle$ and  $\langle N_{\rm sat} \rangle$.
In the vast majority of HOD studies it is assumed that centrals and satellite HODs are completely uncorrelated, so that 
 $\langle N_{\rm cen} N_{\rm sat} | M_{\rm h} \rangle =  \langle N_{\rm cen}|M_{\rm h} \rangle    \langle N_{\rm sat} |M_{\rm h} \rangle$.
This means that satellites have no knowledge of the central galaxy occupation of their
host halo. Moreover, motivated by the occupation statistics of subhalos in high-resolution $N$-body simulations, the PDF of
satellite occupation is commonly assumed to be Poissonian, so that 
$\langle N_{\rm sat} (N_{\rm sat} -1)|M_{\rm h} \rangle = \langle N_{\rm sat}|M_{\rm h} \rangle^2$. 
 
The widespread success of the standard HOD framework in  
interpreting galaxy clustering statistics, the galaxy-halo connection,  
and testing cosmology at small scales is not free from several drawbacks. 
There are in fact a number of simplifying assumptions that enter in the conventional HOD modeling, and
may represent a limitation of its predicative power -- see for example the recent interesting study by \citet{Hadzhiyska2020}.

To start with, while certainly halo mass is the dominant parameter governing the environmental demographics of galaxies, 
in reality semi-analytical models and hydrodynamical simulations of galaxy formation and evolution
predict significant correlations between galaxy properties and halo properties other then mass (i.e., halo formation time, concentration, halo spin, merger history, star formation rate, etc.).
Such dependence of the spatial distribution of dark matter halos upon properties besides mass in generically referred to as {\it halo assembly bias}.
To date, a clear detection of assembly bias remains still controversial, as some earlier claims 
of detection in a sample of SDSS galaxy clusters \citep{Miyatake2016} have been called into significant question -- see e.g. \citet{Zu2017}.
However, some level of assembly bias may be present in the eBOSS LRG sample, as recent studies seem to indicate
\citep{Zentner2019,Obuljen2020,Yuan2020}. Moreover, it has been shown that ignoring  assembly bias in HOD modeling yields constraints on the
galaxy-DM connection that may be plagued by significant systematic errors 
\citep{Yang2006,Blanton2007,Zentner2014}.  
In addition, while in standard HOD studies centrals and satellite HODs are completely uncorrelated, likely, a degree of central-satellite correlation is always present:   
such correlations are
induced by interesting astrophysics, rather than being simply  a nuisance systematics. 
In fact, the correlation encodes the extent to which the properties of satellite galaxies 
(stellar mass, color, etc.) may be correlated with the properties of its central galaxy at fixed halo mass (i.e., galactic cannibalism or conformity).
Moreover, central galaxies may not be located at the central (minimum) of the halo potential well, 
the occupation statistics of subhalos in host halos 
of fixed mass has been shown to deviate from a Poisson distribution especially in the limit where the first occupation
moment become large, the satellite distribution may not track the NFW spatial profile of the dark matter halo, 
and much more -- see for example \citet{Yuan2018} and the most recent study by \citet{Duan2019} for an extensive discussion on such challenges. 
 
Within this complex framework, the main goal of our study is to produce a series of synthetic galaxy catalogs spanning a variety of HODs 
exhibiting similar clustering properties, in order to
assess the robustness of different fitting methodologies relevant for LRG clustering. 
In this respect, the primary focus is not to 
improve the HOD modeling and the galaxy-halo connection. However, we have devised the mock challenge in an increasing order of HOD complexity by exploring various methodologies,
so that our study may be helpful for ameliorating the galaxy-halo connection in future works. 
Motivated by these reasons, we start from the simplest and most conventional HOD framework, and gradually increase the complexity
till considering models with assembly bias, particularly useful in exploring intermediate correlations between
central-satellites, as well as more generalized HOD approaches. 
As a byproduct of our work, we are thus able to draw some interesting conclusions regarding the galaxy-halo connection, 
based on our high-fidelity mocks.

In this work, unless specified otherwise, 
we always consider two galaxy populations (referred as centrals and satellites); moreover,  as commonly adopted in the most conventional HOD implementations, we generally assume that 
the central phase space model requires  central galaxies to be located at the exact center of the host halo with the same halo velocity, and that
the satellite phase space model follows an unbiased NFW profile with a phase space distribution of mass and/or galaxies 
in isotropic Jeans equilibrium, where the concentration of galaxies is identical to the one of the parent halo. 

 
\subsection{Traditional HOD: Zheng Model}


\begin{table}
\centering
\caption{HOD parameters adopted for the \citet{Zheng2007} model, corresponding to different `luminosity threshold' values.}
\doublerulesep2.0pt
\renewcommand\arraystretch{1.5}
\begin{tabular}{cccccc} 
\hline \hline
{\bf ZHENG MODEL} &&&&& \\
\hline \hline
Threshold & $\log M_{\rm min}$ & $\sigma_{ \log \rm M}$ & $\alpha$ & $M_0$ & $M_1$ \\
\hline 
Th1 ($M_{\rm r} =-19$) & 11.60 & 0.26 & 1.02 & 11.49 & 12.83 \\
Std ($M_{\rm r}=-20$) & 12.02 & 0.26 & 1.06 & 11.38 & 13.31 \\
Th2 ($M_{\rm r}=-21$) & 12.79 & 0.39 & 1.15 & 11.92 & 13.94 \\
\hline \hline
\label{table_zheng07}
\end{tabular}
\end{table}


The most traditional composite HOD model is the one first proposed by \citet{Zheng2007}: 
it represents the backbone for any other HOD framework, and the starting point of this work.
The central occupation statistic $\langle N_{\rm cen}  \rangle$ is described by a nearest integer distribution with first moment given by an error function introduced by \citet{Zheng2005}, 
namely\footnote{Throughout the paper, we implicitly assume base 10 for all the logarithmic notations indicated with \textit{log}, namely $\log \equiv \log_{10}$, and drop the understood
subscript for clarity of notation.}:
\begin{equation}
\langle N_{\rm cen} (M_{\rm h}) \rangle = {1 \over 2} \Big [  1 + {\rm erf} \Big \{   { \log (M_{\rm h}) -  \log (M_{\rm min}) \over {  \sigma_{\log_{\rm M}  } }}  \Big \}   \Big ] 
\label{eq_zheng_hod_cens}
\end{equation}
where $M_{\rm h}$ is the halo mass, 
$M_{\rm min}$ is the characteristic minimal mass for a halo to host a central galaxy above a luminosity threshold\footnote{As pointed 
out by \citet{Zheng2007},  $M_{\rm min}$ can also be interpreted as the mass of such halos for which half of them host galaxies above the given luminosity threshold, i.e., $\langle N_{\rm cen}(M_{\rm min}) \rangle = 0.5$.},
and $\sigma_{\log {\rm M}}$ is the rate of transition from $\langle N_{\rm cen} \rangle =0$ to  $\langle N_{\rm cen} \rangle =1$, representing the width of the cutoff profile.  
Hence, central galaxies are characterized just by two HOD parameters. 
The satellite occupation statistic $\langle N_{\rm sat}  \rangle$ is represented by a Poisson distribution with 
first moment given by a power law that has been truncated at the low-mass end \citep{Kravtsov2005}, 
and described by three parameters:
\begin{equation}
\langle N_{\rm sat} (M_{\rm h}) \rangle = \Big ( {M_{\rm h} - {{M_0}}   \over  M_1} \Big )^{ \alpha} 
\label{eq_zheng_hod_sats}
\end{equation}
where $\alpha$ is the power law slope of the relation between halo mass and $\langle N_{\rm sat} \rangle$,
$M_0$  a low-mass cutoff in  $\langle N_{\rm sat} \rangle$, and 
$M_1$ is the mass where approximately there is an average of one satellite galaxy per halo, namely
 $\langle N_{\rm sat}(M_{\rm h}=M_1) \rangle \sim 1$ -- or more specifically 
$\langle N_{\rm sat}(M_{\rm h}=M'_1) \rangle = 1$ 
with $M'_1 = M_1+M_0$.
Note that the previous distribution can be optionally modulated by the central distribution $\langle N_{\rm cen} \rangle$. 	
Redshift has no impact on this model. 

Following \textsc{Halotools} conventions (\citealt{Hearin2017}; see Section \ref{sec_tools_methodology_lrgs}), 
the setting of these parameters in our mock-making procedure is controlled by a luminosity threshold,  intended as 
the $r-$band absolute magnitude of the luminosity of the galaxy sample.  
The  HOD parameters used in our modeling are those of Table 2 in \citet{Zheng2007}, and conveniently 
reported in Table \ref{table_zheng07} as a function of threshold: specifically, we consider three thresholds in this work, 
referred globally as `threshold 1'  (Th1; $M_{\rm r}=-19$), `standard' (Std; $M_{\rm r}=-20$), and `threshold 2' (Th2; $M_{\rm r}=-21$); the latter one is
closer to the characteristics of the eBOSS DR16 LRG sample. 
Since the meaning of the parameter `threshold' in the Zheng model, according to \textsc{Halotools} conventions,
is effectively different from that of the other HOD models considered
(based on stellar mass rather than luminosity, thus not direct correspondent because we choose to maintain HOD literature parameters -- 
see again Section \ref{sec_tools_methodology_lrgs} and Table \ref{table_number_density_unblind_set}), we opted to 
keep the Zheng framework separate from the other models in the following presentation, in order to avoid confusion; we also notice that in general stellar mass is a more faithful
tracer of the halo mass than galaxy luminosity -- see, e.g., \citet{Leauthaud2011,Leauthaud2012,Tinker2013}.
However, we reiterate that the 5-parameter Zheng HOD framework is the starting point of our work, 
and results involving the Zheng model can be found in Appendix \ref{sec_appendix_tables} (Table \ref{table_lrg_team_zheng}),
as well as in  our companion paper 
\citet{Gil-Marin2020} -- see in particular their Figure 11. Simply, 
in the main analysis presented in Section \ref{sec_unblind_challenge} we have chosen to 
display only results involving the
HOD models discussed next, as their selected  HOD parameters for Th2 provide mocks
closer in terms of number density to the eBOSS LRG sample (i.e. Table \ref{table_number_density_unblind_set}), and thus more suitable for our main science targets.

The shapes of the Zheng HODs used in this work are shown in Figure \ref{fig_zheng07_hod}, 
for the three conventional choices of HOD parameters corresponding to the previously mentioned threshold values (see Table  \ref{table_zheng07}). 
Note also that in our HOD modeling we do not modulate the satellite distribution by the central one, unless specified otherwise. 


\subsection{Adding the SHMR Complexity: Leauthaud Model}


\begin{figure}
\includegraphics[angle=0,width=0.41\textwidth]{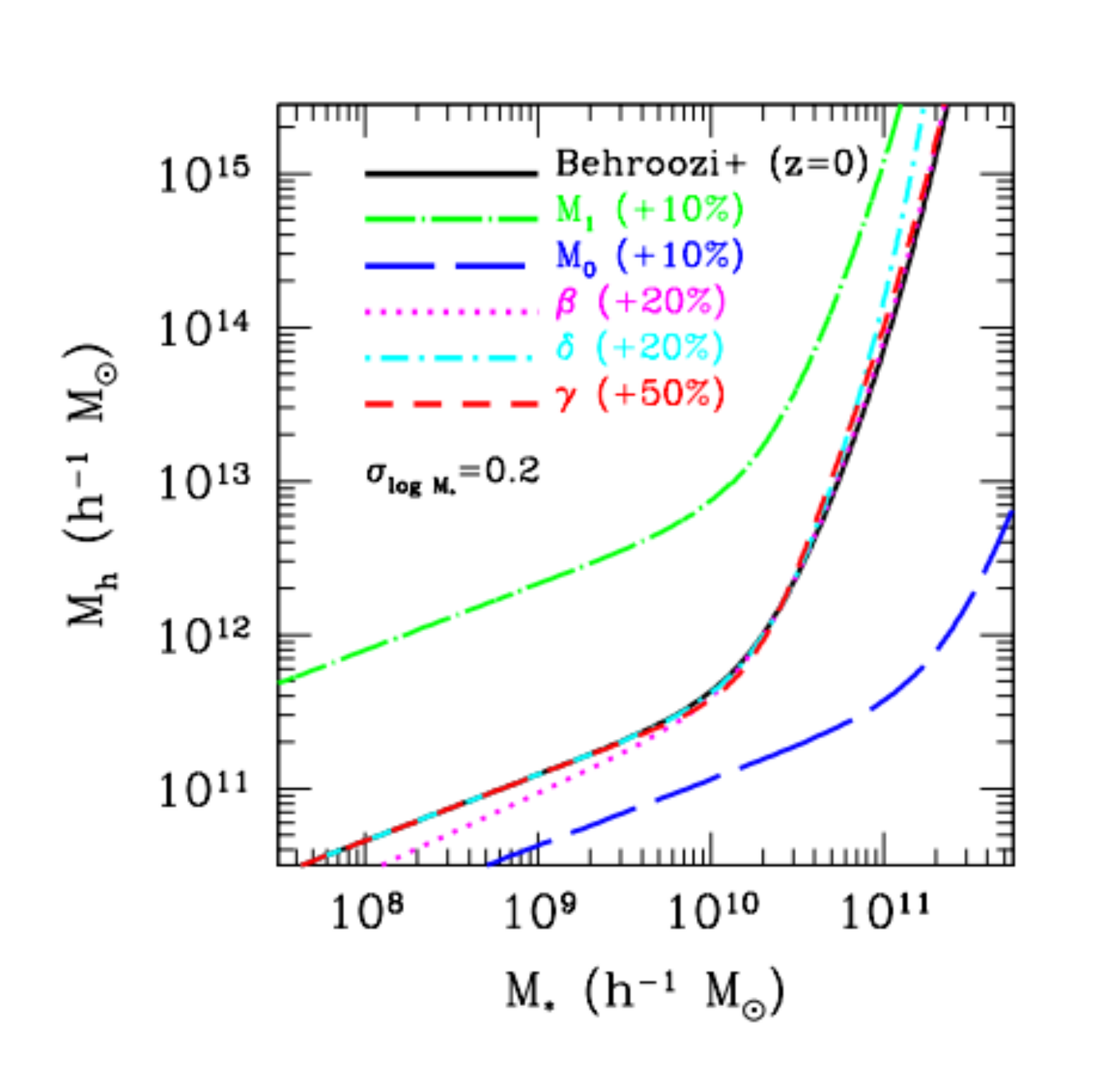}
\includegraphics[angle=0,width=0.42\textwidth]{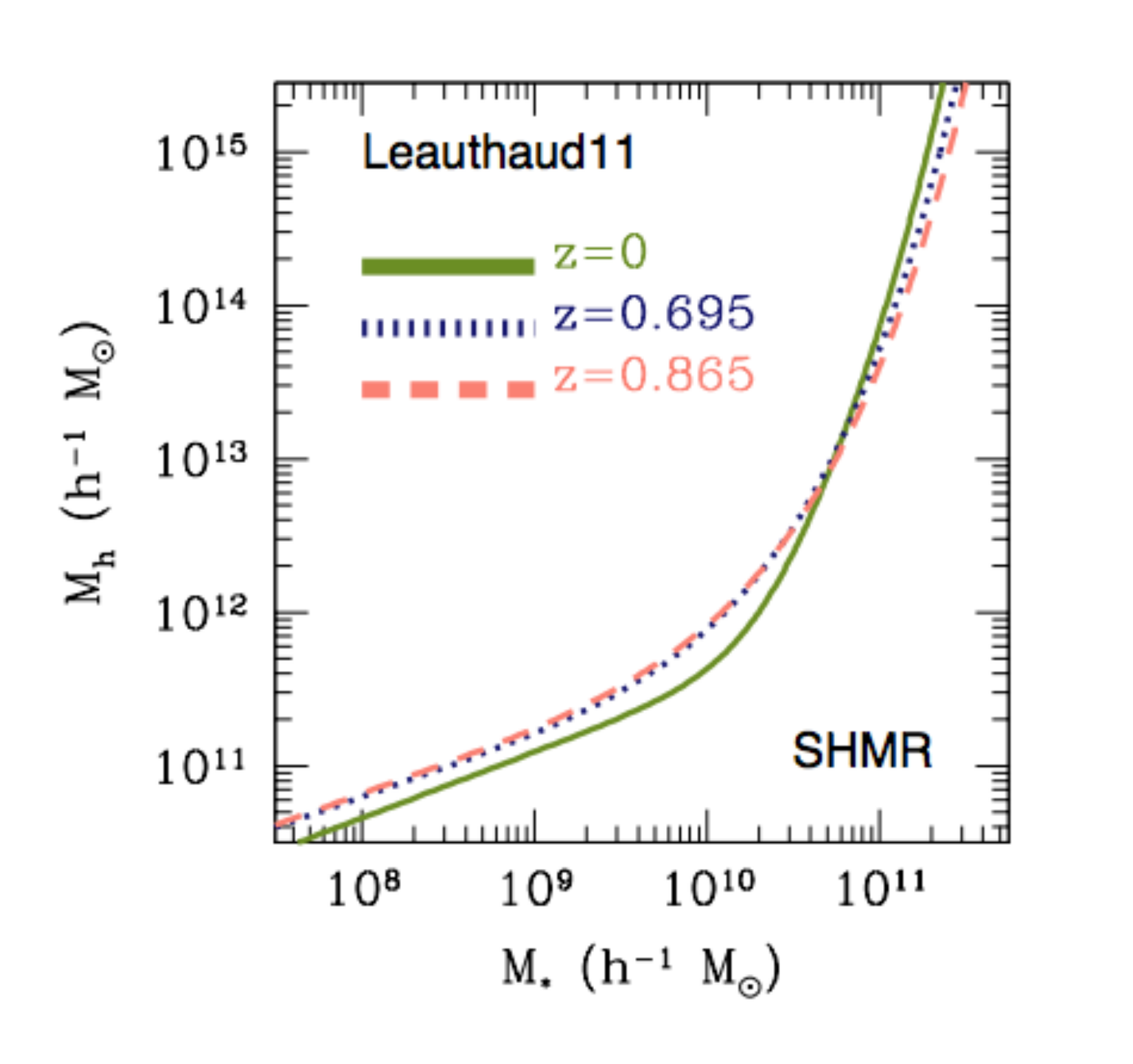}
\caption{[Top] Effects of varying the main parameters controlling the SHMF for central galaxies (Equation \ref{eq_leauthaud_fshmr}),  modeled as a mean-log relation in
the Leauthaud model and heavily based on \citet{Behroozi2010}.  The key shape parameters 
are altered in turn by $10\%$ ($M_{0}$, $M_{1}$), $20\%$ ($\beta$, $\delta$), and $50\%$ ($\gamma$), respectively,
from the baseline \citet{Behroozi2010} model at $z=0$ (solid black line and upper part in Table \ref{table_leauthaud11_hod}), when $\sigma_{\log \rm M_{*}}=0.20$. 
Different line styles and colors refer to such variations, as clearly indicated in the panel.  
[Bottom] SHMF underlying the Leauthaud model at $z=0$, and its
redshift evolution  at  $z=0.695$ and $z=0.865$: we use these SHMRs in our mock-making procedure.}
\label{fig_leauthaud11_shmr}
\end{figure}


The second model we consider is the Leauthaud prescription \citep{Leauthaud2011,Leauthaud2012}, 
a composite HOD framework that extends the standard Zheng formalism by 
including a parameterization of an underlying stellar-to-halo mass relation (SHMR),
which specifies the mean mass of a galaxy as a function of halo mass. 
The main assumption in this picture is that the SHMR is valid only for central galaxies, as
satellites and centrals experience distinct stellar growth rates 
and so it is necessary to model them separately. 
To this end, the conditional stellar mass function (CSMF) -- denoted as $\Phi(M_{*}|M_{\rm h})$, with $M_{*}$ the mass in stars and $M_{\rm h}$ the mass of the parent halo -- is 
divided into centrals and satellites, namely $\Phi(M_{*}|M_{\rm h}) = \Phi_{\rm cen}(M_{*}|M_{\rm h}) + \Phi_{\rm sat}(M_{*}|M_{\rm h})$.
Moreover, $\Phi_{\rm cen}(M_{*}|M_{\rm h})$ is modeled stochastically as 
a log-normal PDF with a log-normal scatter $\sigma_{\log \rm M_{*}}$, and normalized to unity.  
The exact form is \citep{Leauthaud2012}:
\begin{eqnarray}
\Phi_{\rm cen} (M_{*}|M_{\rm h}) &=& {1 \over \ln (10) \sigma_{\log \rm M_{*}}  \sqrt{2 \pi}}  \cdot \nonumber \\
 && \exp \Big [ -{ \{  \log{M_{*} - \log[f_{\rm SHMR} (M_{\rm h})  ]}        \}^2  \over 2 \sigma^2_{\log \rm M_{*}}  } \Big ]
\label{eq_leauthaud_phi_cen}
\end{eqnarray}
where $f_{\rm SHMR}$ is the logarithmic mean of the stellar mass given the halo mass for the $\Phi_{\rm cen}$ distribution function.
Equation (\ref{eq_leauthaud_phi_cen}) incorporates the scatter associated with the determination of stellar masses,
as well as the intrinsic scatter in stellar mass at fixed halo mass due to astrophysical processes. 
The functional form for  $f_{\rm SHMR}$  is described by 5 shape parameters ($M_{0}$, $M_{1}$, $\beta$, $\delta$, $\gamma$), and 
defined  via its inverse function  
 as \citep{Behroozi2010}:
\begin{eqnarray}
 \log [f^{-1}_{\rm SHMR} (M_{*}) ] &=& \log (M_1) + f_{\rm low}(M_{*}/M_{0}) +  f_{\rm high}(M_{*}/M_{0}) \nonumber \\
  &=& \log (M_1)  + \beta \log \Big ( {M_{*} \over M_{0} } \Big ) + \nonumber \\
    && { ( M_{*} /M_{0} )^{\delta} \over 1 + ( M_{*} /M_{0} )^{- \gamma} }- {1\over 2}, 
\label{eq_leauthaud_fshmr}
\end{eqnarray}
with $f_{\rm low}$ and $f_{\rm high}$ the low and high mass parts of the SHMF, respectively. 
In the previous expression, 
$M_{0}$ and $M_{1}$ are characteristic stellar and halo masses -- respectively -- in the $\langle M_{*} \rangle (M_{\rm h})$ map,  
$\beta$ is a low-mass slope   of the $\langle M_{*} \rangle (M_{\rm h})$ map,
$\delta$ is the high-mass slope   of the  same map, and
$\gamma$ represents the transition between the low- and high-mass behavior of the $\langle M_{*} \rangle (M_{\rm h})$  mapping. 
The redshift evolution of the SHMR is modeled 
by allowing the parameters that define $f_{\rm SHMR}$ in Equation (\ref{eq_leauthaud_fshmr}) to vary linearly with the scale factor $a$ as:
\begin{eqnarray}
\log [M_1(a)] &=&  \log(M_{1,0}) + \log (M_{1a}) (a-1) \\
\log [M_{0}(a)] &=&  \log(M_{0,0}) + \log (M_{0a}) (a-1) \\
\beta(a) &=&  \beta_{0} + \beta_{a} (a-1) \\
\delta(a) &=&  \delta_{0} + \delta_{a} (a-1) \\
\gamma(a) &=&  \gamma_{0} + \gamma_{a} (a-1).
\end{eqnarray}
Therefore, the model for the SHMR effectively requires 10 parameters.
For our modeling, we adopt the same literature values as in the second 
column of Table 2 in \citet{Behroozi2010} at $z=0$, and also reported in the upper part of 
Table \ref{table_leauthaud11_hod} for convenience. 
Furthermore, Equation (\ref{eq_leauthaud_phi_cen})  requires a functional form for 
the lognormal scatter $\sigma_{\log \rm M_{*}}$ in the SHMR: 
motivated by  \citet{Leauthaud2012}, who
found that a halo mass-varying scatter produced no better fit than a model with constant scatter, 
we assume a constant scatter  in this work -- which can be thought as 
the sum in quadrature of an intrinsic component
plus a measurement error component.  
Specifically, we set $\sigma_{\log \rm M_{*}}=0.20$ in our modeling for all the redshifts considered.

The top panel of Figure  \ref{fig_leauthaud11_shmr}
provides an example of the effects of varying in turn the 5 main parameters that
control the SHMF (Equation \ref{eq_leauthaud_fshmr}) at $z=0$ by 10\% and up to 50\% -- as specified in the plot with different line styles and colors,
when the lognormal scatter is kept constant.  
The baseline parameters are fixed as in \citet{Behroozi2010} at $z=0$ (solid black line, and upper part in Table \ref{table_leauthaud11_hod}). 
The bottom panel of the same figure displays the
SHMF  underlying the Leauthaud model at $z=0$, along with its
redshift evolution  for the two main
redshift snapshots considered in our  mock-making procedure ($z=0.695$ and $z=0.865$, respectively -- see Section \ref{sec_tools_methodology}):
we use these SHMRs in our modeling.

The key difference with respect to the Zheng model
relies in the assumption that the stellar mass, rather than the galaxy luminosity, is 
used to implement the HOD as a more reliable tracer of the halo mass -- see e.g. \citet{Leauthaud2011,Leauthaud2012,Tinker2013}.
To this end, for a volume-limited sample of galaxies
such that $M_{*} > M^{\rm thr}_{*}$,  
with  $M^{\rm thr}_{*}$ a galaxy threshold mass,  
the central occupation function $\langle N_{\rm cen} (M_{\rm h} | M_{*}^{\rm thr}) \rangle$ 
is fully specified given $\Phi_{\rm cen} (M_{*} | M_{\rm h})$
according to \citep{Leauthaud2011}: 
\begin{equation}
\langle N_{\rm cen} (M_{\rm h} | M_{*}^{\rm thr}) \rangle = \int_{M^{\rm thr}_{*}}^{\infty} \Phi_{\rm cen} (M_{*} | M_{\rm h}) {\rm d}M_{*}.   
\label{eq_LH}
\end{equation}
Assuming that $\sigma_{\log \rm M_{*}}$ is constant, the previous expression can be readily integrated and becomes:
\begin{equation}
\langle N_{\rm cen} (M_{\rm h} | M_{*}^{\rm thr}) \rangle = {1 \over 2}  \Big [  1 - {\rm erf} \Big \{   {  \log(M^{\rm thr}_{*} ) - \log [ f_{\rm SHMR}(M_{\rm h})  ]   \over \sqrt{2} \sigma_{\log \rm M_{*}}    }  \Big \} \Big ]. 
\label{eqLH_constant_scatter}
\end{equation}
Equation (\ref{eqLH_constant_scatter}) represents a 
generalization of the Zheng HOD formula (Equation \ref{eq_zheng_hod_cens}), and
it is controlled by 5 parameters that enter in the SHMR -- plus 1, if we allow for a varying scatter in the SHMR, and plus
additional 5 parameters if we also consider redshift evolution in the SHMR. 
Interestingly, Equation (\ref{eq_zheng_hod_cens}) -- i.e. the Zheng model -- can be readily
recovered from (\ref{eq_LH}) as a limiting case 
by assuming  a constant scatter in the SHMR and by setting
$f_{\rm SHMR}$ to be a power law; however, this latter assumption is not realistic. 


\begin{table}
\centering
\caption{HOD parameters for central and satellite galaxies assumed for the calibration of the \citet{Leauthaud2011} model.}
\doublerulesep2.0pt
\renewcommand\arraystretch{1.5}
\begin{tabular}{cc} 
\hline \hline
\hspace{1cm} {\bf LEAUTHAUD MODEL} \\
\hline \hline
Centrals \\
\hline  
$\log [M_{1,0}]$ & 12.35 \\
$\log [M_{1 \rm a}]$ & 0.30 \\
$\log [M_{0,0}]$ & 10.72 \\
$\log [M_{0 \rm a}]$ &  0.59 \\
$\beta_0$ & 0.43 \\
$\beta_{\rm a}$ &  0.18 \\
$\delta_0$ & 0.56 \\
$\delta_{\rm a}$ &  0.18 \\\
$\gamma_0$ &1.54 \\
$\gamma_{\rm a}$ &  2.52 \\
$\sigma_{\log M_{*}}$ & 0.20 \\ 
\hline \hline
\hspace{1cm} Satellites \\ 
\hline 
$\alpha_{\rm sat}$ & 1.0 \\
$\beta_{\rm sat}$ & 0.859 \\ 
$B_{\rm sat}$ & 10.62 \\ 
$\beta_{\rm cut}$ & -0.13 \\
$B_{\rm cut}$ &  1.47 \\
\hline \hline
\label{table_leauthaud11_hod}
\end{tabular}
\end{table}

  
Regarding the satellite occupation function, in the Leauthaud model
it is parameterized as a power law of host mass with an exponential cutoff, and can be optionally scaled by $\langle N_{\rm cen} \rangle$ (this is not done in our case).
Specifically:
\begin{equation}
\langle N_{\rm sat} (M_{\rm h} | M_{*}^{\rm thr}) \rangle = \Big ( { M_{\rm h} \over M_{\rm sat}}    \Big )^{\alpha_{\rm sat}} \exp \Big ( - {M_{\rm cut} \over M_{\rm h}}  \Big )
\label{}
\end{equation} 
where  $M_{\rm sat}$ defines the amplitude of the power law and $M_{\rm cut}$ sets the scale of the exponential cutoff;
halos with masses $M_{\rm h} < M_{\rm cut}$ are extremely unlikely to host a satellite galaxy.
Instead of simply modeling $M_{\rm sat}$ and $M_{\rm cut}$ as constant factors of $f^{-1}_{\rm SHMR} (M^{\rm thr}_{*})$, 
flexibility is added by enabling  $M_{\rm sat}$ and $M_{\rm cut}$ to vary as power law functions of  $f^{-1}_{\rm SHMR} (M^{\rm thr}_{*})$:
\begin{equation}
{M_{\rm sat} \over \bar{M}_{12} } = B_{\rm sat} \Big [ {f^{-1}_{\rm SHMR} (M^{\rm thr}_{*}) \over \bar{M}_{12} }     \Big ]^{\beta_{\rm sat}}
\label{}
\end{equation}
\begin{equation}
{M_{\rm cut} \over \bar{M}_{12} } = B_{\rm cut} \Big [ {f^{-1}_{\rm SHMR} (M^{\rm thr}_{*}) \over \bar{M}_{12} }     \Big ]^{\beta_{\rm cut}}
\label{}
\end{equation}
where $\bar{M}_{12} = 10^{12} M_{\odot}$. 
Hence, satellite occupation statistics, independent of binning schemes and parameterized with threshold samples, are
modeled by 5 parameters: $\alpha_{\rm sat}, \beta_{\rm sat}, B_{\rm sat}, \beta_{\rm cut}, B_{\rm cut}$. 
In detail,  
$\alpha_{\rm sat}$ is  the power law slope of the relation between halo mass and the satellite mean occupation function $\langle N_{\rm sat} \rangle$,
$\beta_{\rm sat}$ and $B_{\rm sat}$ control the amplitude of the power law slope of $\langle N_{\rm sat} \rangle$, and
$\beta_{\rm cut}$ and $B_{\rm cut}$ control the low-mass cut off in $\langle N_{\rm sat} \rangle$.
These satellites parameters are fixed as in Table 5 of \citet{Leauthaud2011} for the first redshift bin,
and also reported in the bottom part of Table \ref{table_leauthaud11_hod} for convenience.

In summary, the Leauthaud framework is completely determined by
11 HOD parameters: 6 controlling the central galaxies and 5 for satellite galaxies, plus additional 5 if
we also take into account the redshift evolution of the SHMR.  
In this framework, `threshold' should be intended as the minimal stellar mass of the galaxy sample,  rather than galaxy luminosity. 

The shapes of the HODs in the Leauthaud model adopted in this work are shown in the top panels of 
Figure \ref{fig_lth_hod_A} for
3 thresholds in mass, indicated as `Threshold 1' (Th1; $M^{\rm thr}_{*}=10^{10} h^{-1} M_{\odot}$), 
`Standard' (Std; $M^{\rm thr}_{*}=10^{10.5} h^{-1} M_{\odot}$), and
`Threshold 2' (Th2; $M^{\rm thr}_{*}=10^{11} h^{-1} M_{\odot}$),  when $z=0.695$.
In particular, Th2 is the one closer to the eBOSS LRG sample, and 
we mainly focus on this mass interval in our analysis. 
Note finally that in the Leauthaud framework $\langle N_{\rm sat} (M_{\rm h} | M_{*}^{\rm thr}) \rangle$ depends on
$\langle N_{\rm cen} (M_{\rm h} | M_{*}^{\rm thr}) \rangle$, indicating that in this particular model
the occupation statistics of centrals and satellites are correlated:
this constitutes  an important difference with respect to the traditional
5-parameter HOD. In fact, as pointed out by \citet{Contreras2017} and  \citet{Duan2019}, if the goal is not to constrain HOD parameters and this 
fitting difficulty is not of concern (as in our specific case), there is no advantage or motivation for insisting on no correlations between centrals and satellites. 

 
\subsection{HOD with Color/SFR:  Tinker Model}

 
\begin{figure}
\includegraphics[angle=0,width=0.45\textwidth]{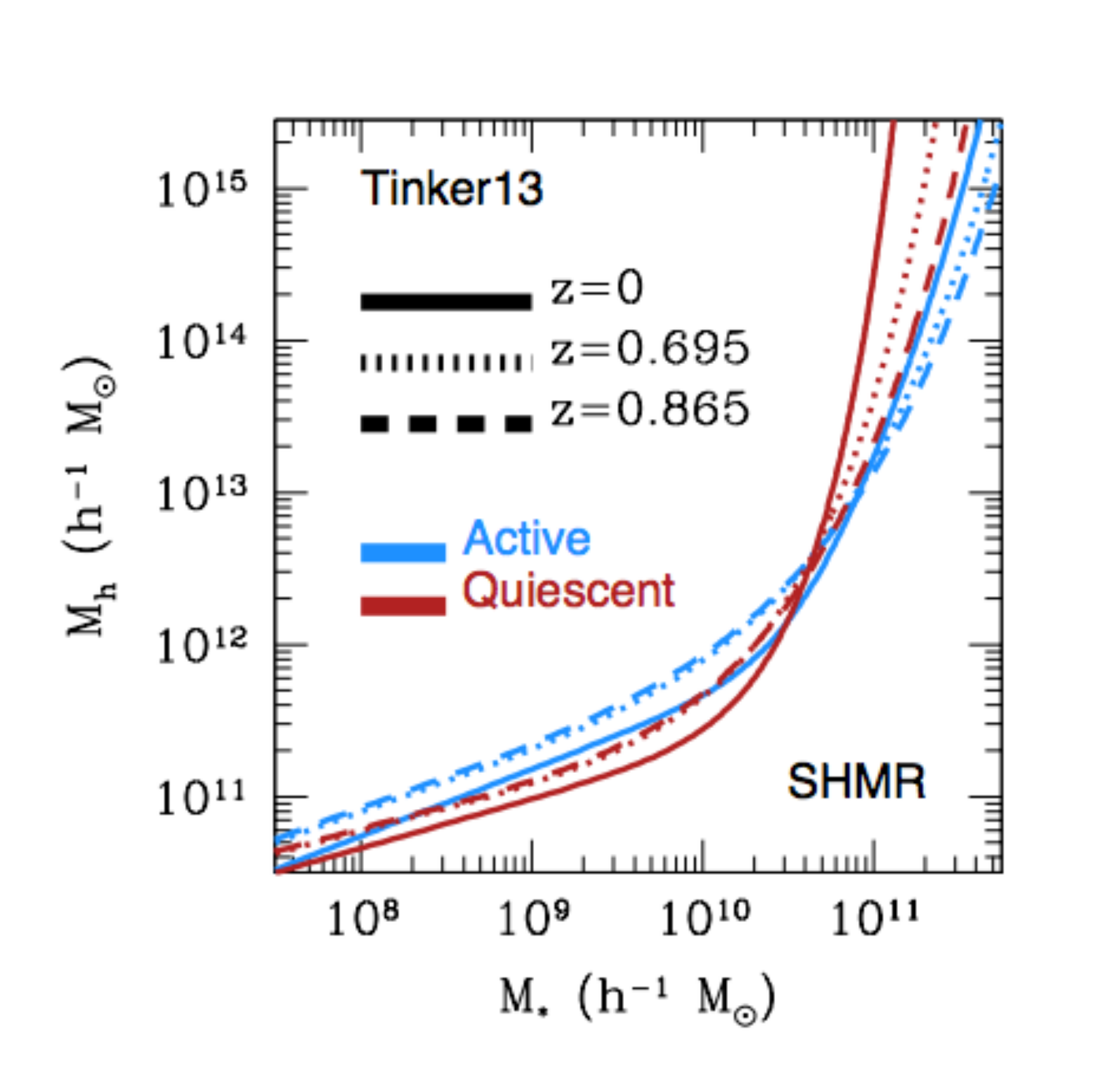}
\caption{SHMFs for central galaxies adopted in the
Tinker model and in our mock-making procedure at $z=0$ (solid lines),
$z=0.695$ (dotted lines), and $z=0.865$ (dashed lines). 
Active galaxies are displayed in blue, while quiescent galaxies are represented in brown.}
\label{fig_tinker_shmr}
\end{figure}


\begin{table}
\centering
\caption{HOD parameters for central and satellite galaxies assumed for the calibration of the \citet{Tinker2013} model.}
\doublerulesep2.0pt
\renewcommand\arraystretch{1.5}
\begin{tabular}{cc} 
\hline \hline
\hspace{1cm} {\bf TINKER MODEL} \\
\hline \hline
Centrals \\
\hline  
$\log [M_{1,0, \rm active}]$ & 12.56 \\
$\log [M_{1,0, \rm quiescent}]$ & 12.08 \\
$\log [M_{0,0, \rm active}]$ & 10.96 \\
$\log [M_{0, 0, \rm quiescent}]$ &  10.70 \\
$\beta_{0, \rm active}$ & 0.44 \\
$\beta_{0, \rm quiescent}$ &  0.32 \\
$\delta_{0, \rm active}$ & 0.52 \\
$\delta_{0, \rm quiescent}$ &  0.93 \\\
$\gamma_{0, \rm active}$ &1.48 \\
$\gamma_{0, \rm quiescent}$ &  0.81 \\
$\sigma_{\log \rm M_{*}, active}$ & 0.21 \\
$\sigma_{\log \rm M_{*}, quiescent}$ & 0.28 \\ 
\hline \hline
Satellites \\
\hline 
$\alpha_{\rm sat, active}$ & 0.99 \\
$\alpha_{\rm sat, quiescent}$ & 1.08 \\
$\beta_{\rm sat, active}$ & 1.05 \\ 
$\beta_{\rm sat, quiescent}$ & 0.62 \\ 
$B_{\rm sat, active}$ & 33.96 \\ 
$B_{\rm sat, quiescent}$ & 17.9 \\ 
$\beta_{\rm cut, active}$ & 0.77 \\
$\beta_{\rm cut, quiescent}$ & -0.12 \\
$B_{\rm cut, active}$ &  0.28 \\
$B_{\rm cut, quiescent}$ &  21.42 \\
\hline \hline
\label{table_tinker13_hod}
\end{tabular}
\end{table}


The next HOD framework we consider has been 
introduced by \citet{Tinker2013}: it
is an extension of the Leauthaud formalism previously described, to samples defined by both stellar mass and star formation (SF) activity.
In this context, galaxies can be roughly categorized into the star-forming sequence of blue, disky,
gas-rich galaxies, and the quiescent, ellipsoidal galaxies with old stellar populations and red colors:
the bimodality is firmly in place at $z=1$ \citep{Tinker2013, Tinker2019}. 
Therefore, in this model galaxies are divided into quiescent, to indicate
galaxies that have little to no star formation and are intrinsically located on the red sequence, 
and the set of star-forming galaxies.
Hence, the Tinker model represents a minimal modification of the Leauthaud prescription
to adapt it to passive and SF subsamples of galaxies.
The HOD behavior is in fact governed by an assumed underlying SHMR as first introduced in \citet{Behroozi2010}, but that is now
distinct for star-forming and quiescent populations: each subsample will then have a separate $f_{\rm SHMR}$, with
2 different sets of HOD parameters related to their specific SHMRs.
A constant scatter $\sigma_{\log \rm M_{*}}$ in the SHMR is adopted here, but
the scatter is different and independent for passive and SF central galaxies.
The main difference with respect to the Leauthaud formalism is the following requirement:
\begin{equation}
\int \Big \{ f_{\rm q}(M_{\rm h}) \times \Phi_{\rm cen}^{\rm q} (M_{*}|M_{\rm h})  + [1-f_{\rm q} (M_{\rm h})] \times \Phi_{\rm cen}^{\rm SF} (M_{*}| M_{\rm h}) \Big \} {\rm d}M_{*}=1
\end{equation}
where $f_{\rm q} (M_{\rm h})$ is a function specifying the fraction of times that a halo of mass
$M_{\rm h}$  contains a quenched central galaxy (independent of galaxy mass),
and $\Phi_{\rm cen}^{\rm x}$ is the conditional stellar mass function for central quiescent or star-forming galaxies, each normalized to unity.
The function $f_{\rm q} (M_{\rm h})$ does not have a parametric form, but
five halo mass points are
chosen at which to specify $f_{\rm q} (M_{\rm h})$ and smoothly interpolate between them, where the 5 masses are evenly spaced in $\log M_{\rm h}$.
Moreover, to avoid explicit dependencies of HOD parameters on bin sizes, all HODs are defined as threshold quantities, which
provides maximal flexibility.

Figure \ref{fig_tinker_shmr} shows the SHMFs adopted in the
Tinker model, as well as in our mock-making procedure, along with their redshift evolution: 
solid lines refer to $z=0$, dotted and dashed lines
are at $z=0.695$ and $z=0.865$, respectively. 
Active galaxies are represented in blue,
while quiescent galaxies are displayed in brown. 

As the Tinker model inherits almost all the features and methods of the
Leauthaud framework,  the HOD for centrals is the same as in Equation (\ref{eqLH_constant_scatter}), but the parameters of the $f_{\rm SHMR}$
are independent for each subsample: this is indeed an important aspect that clearly differentiates the two models. Moreover, 
for red central galaxies, the HOD is multiplied by $f_{\rm q} (M_{\rm h})$,
and by $1-f_{\rm q} (M_{\rm h})$ for SF central galaxies.

The occupation statistics of satellite galaxies as a function of halo mass are similar 
to those of \citet{Leauthaud2011}, although the satellite occupation of  passive and star-forming galaxies subsamples are treated independently.
Hence, a modification is introduced in order to produce a proper cutoff scale by including $f_{\rm SHMR}^{-1}$ to the numerator in the exponential cutoff, so that: 
\begin{equation}
\langle N_{\rm sat} (M_{\rm h} | M_{*}^{\rm thr}) \rangle = \Big ( { M_{\rm h} \over M_{\rm sat}}    \Big )^{\alpha_{\rm sat}} \exp \Big ( - {[M_{\rm cut} +  f_{\rm SHMR}^{-1} (M_{*})]\over M_{\rm h}}  \Big ).
\label{}
\end{equation} 
This guarantees that satellite occupation fully cuts off at the same halo mass scale as central galaxies of the same mass.
In addition, while in \citet{Leauthaud2011} $\alpha_{\rm sat}=1$, here the fraction of satellites
that are star forming depends on halo mass, so $\alpha_{\rm sat}=1$ is allowed to be free for both passive and star-forming subsamples.
 
In summary, the Tinker model is characterized by 27 free parameters: 
11 are needed  for the composite HOD of a given subsample (5 for the central SHMR, one additional for the SHMR scatter, plus 5 for the
satellite occupation statistics), and 5 pivot points are necessary to specify $f_{\rm q} (M_{\rm h})$. 
Each set of 27 parameters describes the galaxy-halo relation at a given redshift, and clearly additional quantities are required to 
characterize the redshift evolution. In this work, we adopt literature values 
from the lowest redshift bin in Table 2 of \citet{Tinker2013}, as reported in Table \ref{table_tinker13_hod};
we note that adopting these parameters makes the model differing from that of Leauthaud, even when the full HOD shape is
considered.

\begin{figure*}
\includegraphics[angle=0,width=0.87\textwidth]{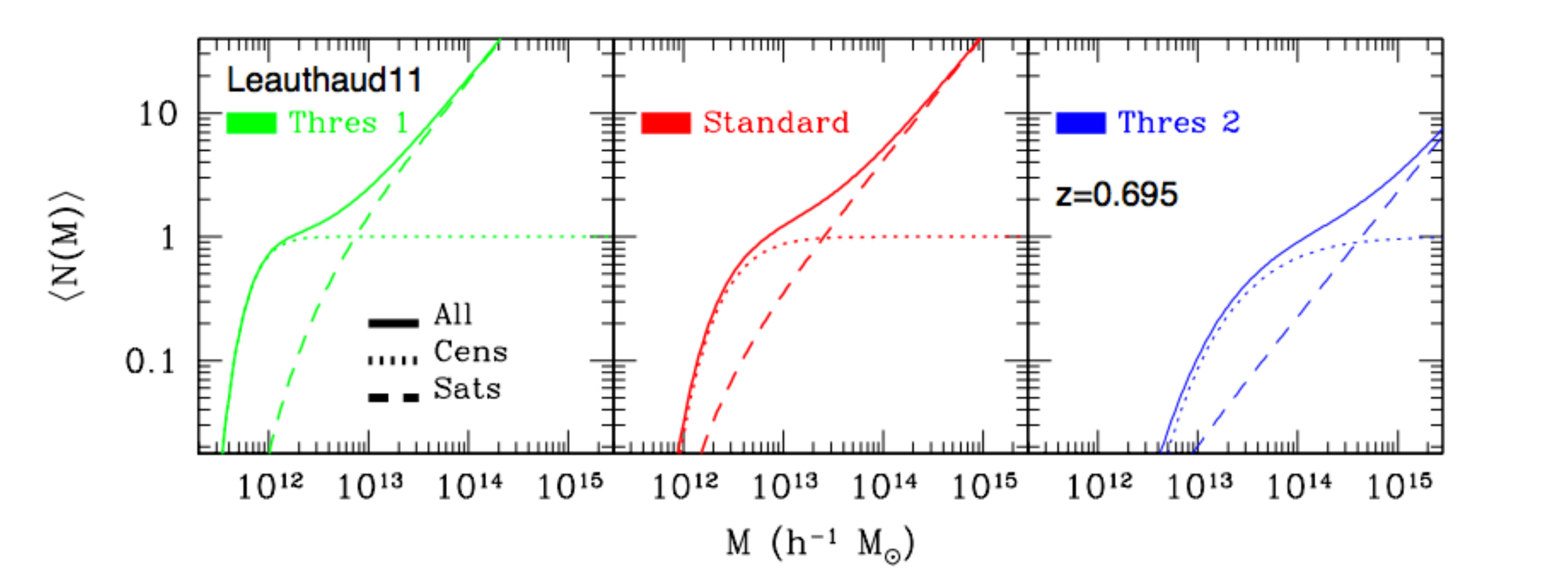}
\includegraphics[angle=0,width=0.87\textwidth]{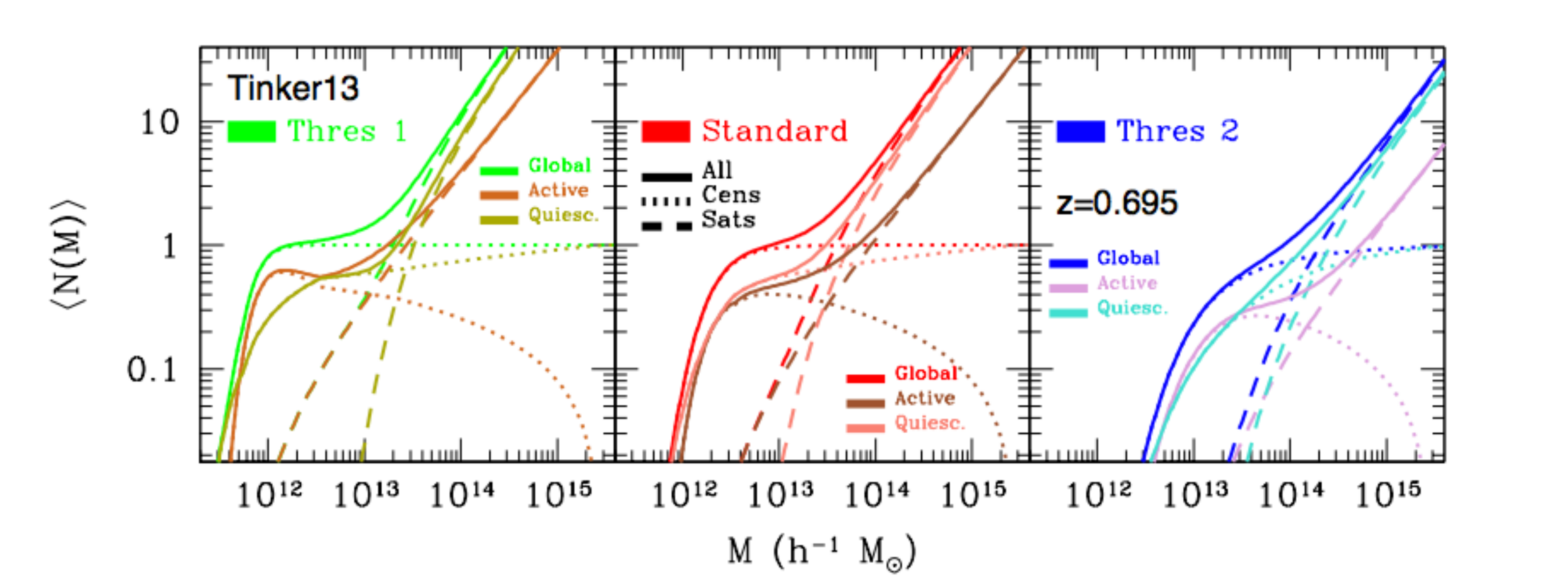}
\includegraphics[angle=0,width=0.87\textwidth]{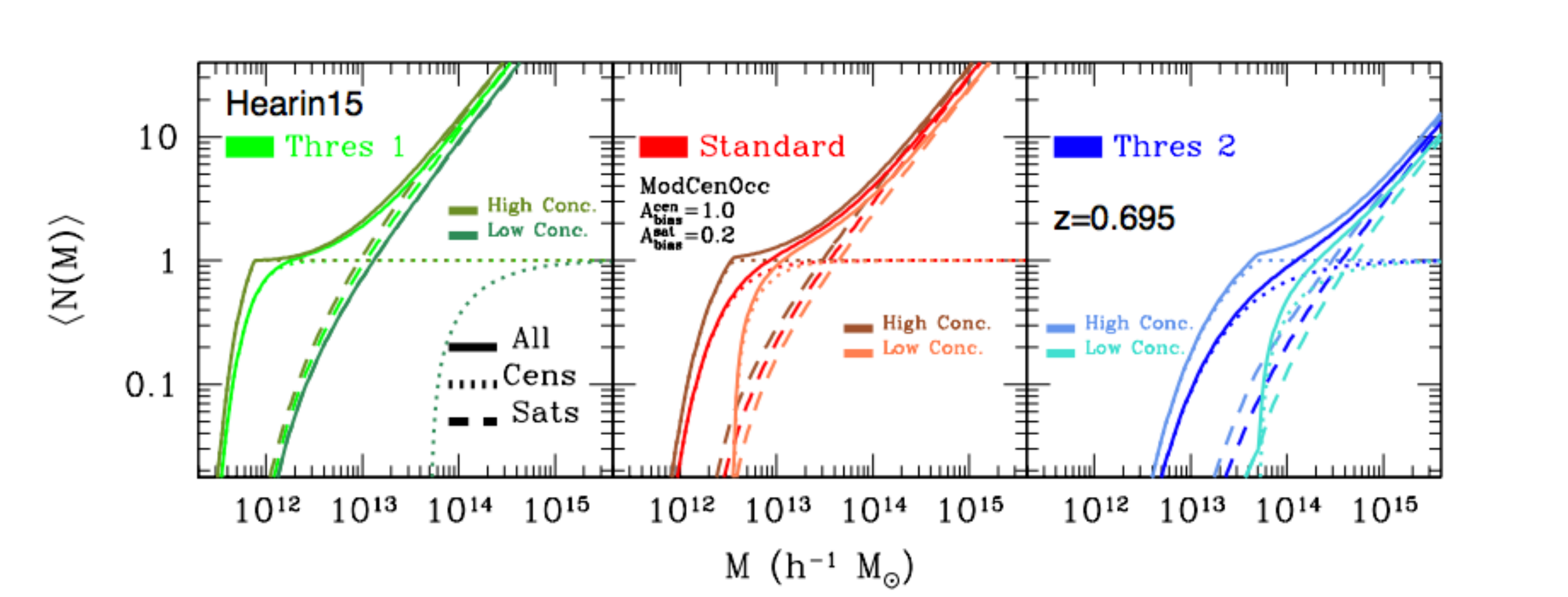}
\caption{HOD shapes adopted in our mock-making procedure, at $z=0.695$, for
3 thresholds in mass, denoted as `Thres 1' ($M^{\rm thr}_{*}=10^{10} h^{-1} M_{\odot}$), 
`Standard' ($M^{\rm thr}_{*}=10^{10.5} h^{-1} M_{\odot}$), and
`Thres 2' ($M^{\rm thr}_{*}=10^{11} h^{-1} M_{\odot}$). Top panels display the various HODs in the 
Leauthaud model. Central panels show the Tinker model, where active and quiescent galaxy HODs are represented by 
different colors, as indicated in the figure. 
Bottom panels  are for the Hearin HODs, where galaxies are split into 
upper- and lower-percentiles in terms of halo concentration, respectively, with different assembly bias strength for centrals and satellites
($A_{\rm bias}^{\rm cen} =1.0$ and $A_{\rm bias}^{\rm sat}=0.2$).
In all the plots,  
 the central occupation statistics is displayed with dotted lines,
the satellite occupation statistics with dashed lines, and the global HOD shapes with solid lines. }
\label{fig_lth_hod_A}
\end{figure*}

The central panels of  Figure \ref{fig_lth_hod_A}  
show the shapes of the HODs in the Tinker model adopted in this work at $z=0.695$,
for the same 3 thresholds in mass described before in relation to the Leauthaud framework, and also distinguishing between centrals and satellites. 
Active and quiescent galaxy HODs are represented by different colors, as indicated in the panels,
and the global HODs are also displayed. 
 As noted by \citet{Tinker2013} and also evident from our figures, the number of quiescent 
 satellites exhibits minimal redshift evolution; all evolution
in the red sequence is due to low-mass central galaxies being quenched of their star formation.
Moreover, the efficiency of quenching star formation for centrals increases with cosmic time, 
while the mechanisms that quench the star formation of satellite galaxies in groups and 
clusters is losing efficiency.


\subsection{Decorated HOD: Hearin Model}

The fourth galaxy-halo prescription we consider is the Hearin model \citep{Hearin2016}, a 
{\it decorated} HOD framework designed to account for 
{\it galaxy assembly bias}, that naturally extends the standard HOD approach, 
minimally expands the parameter space, and maximizes the independence between traditional and
novel HOD parameters. Galaxy assembly bias, namely the correlation between galaxy properties and 
halo properties at fixed halo mass,
is a  challenging and yet important ingredient in the galaxy-halo connection framework.
The model builds on early work by \citet{Wechsler2006}. 
The halo occupation statistics are described in 
terms of two halo properties rather than just one, and the extra degree of freedom
has relevant impact on galaxy clustering. 
The formalism of the model is  general and flexible, with parametric freedom, and 
it can be applied to any halo property in addition to halo mass; it is also
readily extendable to describe HODs that depend upon numerous additional halo properties. 
Interestingly, the decorated HOD formalism allows one to
characterize and quantify the 
degree of central-satellite correlation at fixed halo mass, which  
is an indication of compelling astrophysics -- such as galactic cannibalism or conformity.
We refer the reader to \citet{Hearin2016} for extensive modeling details, and report here only a
few key aspects relevant to our work --  see also, e.g. \citet{Xu+2020} and  \citet{XuZheng2020} for recent studies. 

In particular, the 
core idea is based on the principle of `HOD conservation', which preserves  the moments of the standard  
HOD formalism: namely,  it is required that the marginalized moments of a new decorated
framework are equal to those of the standard HOD model,
in order to minimize the modifications needed for assembly bias.
In this regard, any model $P_{\rm dec} (N_{\rm g} | M_{\rm h}, x)$ with marginalized moments that satisfy the
HOD conservation preserves the moments of $P_{\rm std} (N_{\rm g} | M_{\rm h})$, where
 $P_{\rm std}$ is the occupation statistics of a standard HOD,
$P_{\rm dec}$ is the occupation statistics of a decorated HOD model, and $x$
represents a secondary halo property such that the HOD depends 
both on $x$ and $M_{\rm h}$, and the clustering of halos depends upon $x$. 
Within this formalism, a standard HOD is recovered in 
the limiting condition that
the strength of the assembly bias is zero. 

For central galaxies, in order to construct decorated HOD models that preserve
the full $P_{\rm std} (N_{\rm cen} |M_{\rm h})$ one just needs to ensure that the first order decoration function 
$\delta N_{\rm cen}^{1}$ satisfies the integral relation \citep{Hearin2016}:
\begin{equation}
\int \delta N_{\rm cen}^{\rm 1} (M_{\rm h}, x) P(x|M_{\rm h}) {\rm d} x = 0 .
\end{equation}
For satellites the situation is more complex, as
it is not possible to conserve the HOD under the assumption
that both $P(N_{\rm sat} | M_{\rm h})$ and  $P(N_{\rm sat} | M_{\rm h}, x)$
obey Poisson statistics -- as typically done: intuitively,  
this is because there is an additional source of variance
associated with the allocation of satellites into sub-populations at a given halo mass.  
Moreover, 
under HOD conservation, the average number of central-satellite pairs in massive halos for a decorated model is identical 
to that of its standard baseline model, except for the narrow range in halo masses for which
$0 \le \langle N_{\rm cen}|M_{\rm h} \rangle \le 1$.   

For our purposes, we consider the Hearin model in its simplest formulation, by assuming two
discrete halo sub-populations with different occupation statistics at fixed mass; this is 
essentially a perturbation of the \citet{Leauthaud2011} formalism, with the 
addition of assembly bias both in centrals and satellites. 
We choose the halo NFW concentration as
the secondary halo property ($x$) used to modulate the assembly bias.
Specifically, the first halo sub-population (indicated as `type 1' halos)
contains a fraction $P_1$ of all halos at fixed mass,
for which $x > \bar{x} (M_{\rm h})$;  the
second sub-population (`type 2' halos) contains $P_2=1-P_1$ of all halos at fixed mass,
for which $x < \bar{x} (M_{\rm h})$.
The halo population is split into the $P_1$ percentile of highest-concentration halos, and assigned a satellite galaxy occupation
enhancement, while the remaining $P_2=1-P_1$ percentile of lowest-concentration halos receive a satellite galaxy occupation
decrement. 
Essentially, we require halos at fixed mass above- or below-average concentration to have above- or below-average
mean occupation. For simplicity, we assume a 50/50 split at 
each halo mass based on the conditional secondary percentiles: halos
within the top 50 per cent of concentration at fixed $M_{\rm h}$
are assigned to the first subpopulation, and the remaining to the second population 
(so $P_1=P_2 = 0.5$). 
The strength of assembly bias
in the occupation statistics of centrals and satellite galaxies is 
modulated with two free parameters $A_{\rm bias}^{\rm cen}$ and $A_{\rm bias}^{\rm sat}$, respectively, 
where   $-1 \le A_{\rm bias}^{\rm cen} \le 1$ and $-1 \le A_{\rm bias}^{\rm sat} \le 1$.
With this choice, a positive value for $A_{\rm bias}$ 
implies that halos with above-average  concentration have boosted galaxy occupations;
note also that more positive values of $A_{\rm bias}$ correspond to models in which 
more concentrated halos host more galaxies relative to less concentrated halos
of the same mass.
When both of these parameters are set to zero, the model is formally equivalent to the
baseline `no assembly bias model'  of \citet{Leauthaud2011}.
We consider a constant assembly bias strength at all masses for simplicity, and 
the sign convention is to choose type-1 halos in the upper percentile of the secondary property.
Moreover, we assume that both $P_{\rm std}(N_{\rm sat} | M_{\rm h})$ and 
$P_{\rm dec} (N_{\rm sat}|M_{\rm h}, x)$ are Poisson distributions, so that the decorated HOD is entirely specified by
$A_{\rm bias}^{\rm cen}$ and $A_{\rm bias}^{\rm sat}$. 

In our mock-making procedure, we consider two cases for the 
strength of assembly bias related to centrals and satellites: in the first case (more conservative), we simply set 
$A_{\rm bias}^{\rm cen} = A_{\rm bias}^{\rm sat}=0.5$, namely the strength of assembly bias is equal for both centrals and satellites, with the
boost to their mean occupation equal to 50\% of the maximum allowable strength at each mass; 
in the second case (less conservative), we set different assembly bias strengths for centrals and satellites, namely
$A_{\rm bias}^{\rm cen} =1.0$ and $A_{\rm bias}^{\rm sat}=0.2$.
This latter choice is shown in the lower panels of Figure \ref{fig_lth_hod_A},
where we display the shapes of the 
Hearin HODs for the upper- and lower-percentile split in halo concentration, respectively, as indicated in the plot.
In this case, the satellite HODs are also modulated by their corresponding central distributions.
The same 3 thresholds in mass described before for the Leauthaud framework are adopted,
at $z=0.695$, and as usual we display the central occupation statistics (dotted lines),
satellite occupation statistics (dashed lines), as well as the global HOD shapes (solid lines). 

As noted by \citet{Hearin2016,Hearin2017} and \citet{Tinker2019}, 
assembly bias can enhance or diminish the clustering on large scales, 
but in general it increases the clustering on scales below Mpc -- being qualitatively different al large and small scales. Also,
assembly bias in satellites versus centrals imprints 
a distinct signature on galaxy clustering as well as lensing, and
the degree to which assembly bias alters galaxy clustering statistics can be quite sensitive to the underlying baseline
mass-only HOD of the galaxy population under consideration.
In particular, the impact of assembly bias on galaxy clustering is quite sensitive to
the steepness of the transition from $\langle N_{\rm cen} | M_{\rm h} \rangle_{\rm std}=0$ at low host masses to
$\langle N_{\rm cen} | M_{\rm h} \rangle_{\rm std}=1$ at high host masses.
This steepness is controlled by the level of stochasticity in the central galaxy stellar mass at fixed halo mass,
parameterized in our baseline model by $\sigma_{\log \rm M_{*}}$. 
Note that changing the values of $A_{\rm bias}$ does not change
$\langle N_{\rm g}|M_{\rm h} \rangle$, the mean number of galaxies averaged over all halos of fixed mass: this is
the defining feature of the  decorated HOD, and the meaning of the principle of HOD conservation.



\section{Modeling the Galaxy-Halo Connection: Tools and Methodology} \label{sec_tools_methodology}

 
In this section, we briefly describe the tools and methodologies
behind our mock-making procedure, the main $N$-body simulation used, and
the pipeline to produce novel heterogeneous sets of Outer Rim-based galaxy catalogs. 

 
\subsection{Outer Rim Mocks} \label{subsec_OR}


\begin{figure*}
\centering
\includegraphics[angle=0,width=1.00\textwidth]{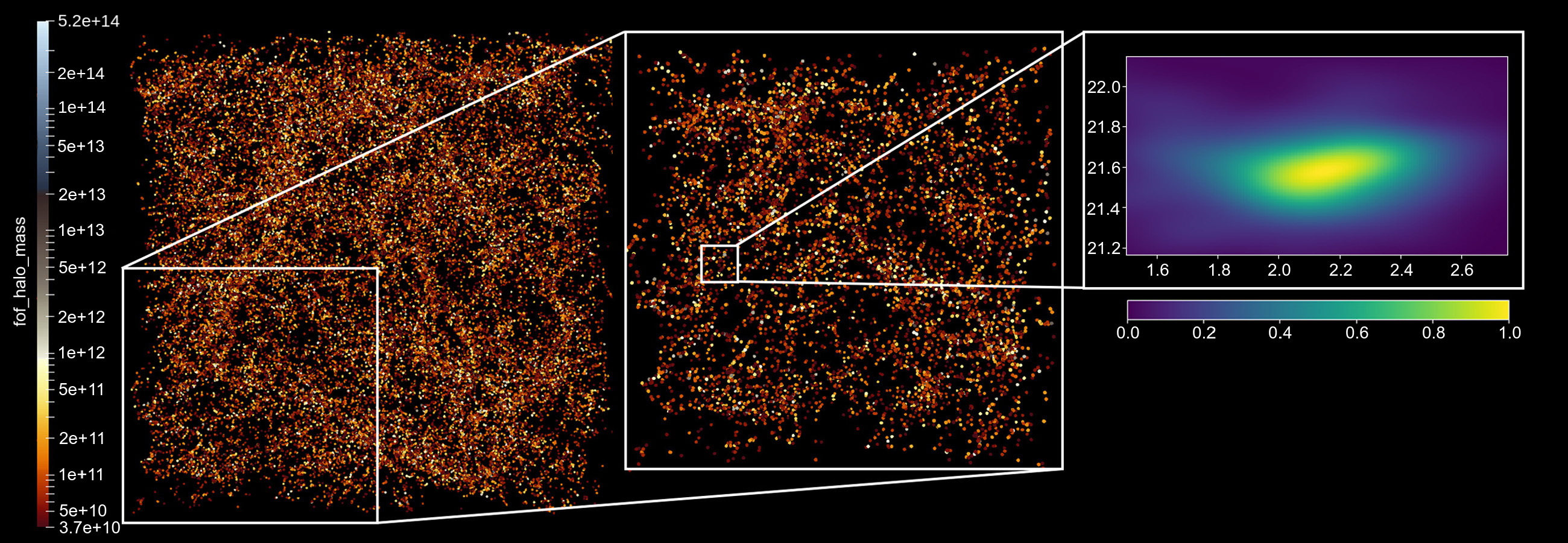}
\caption{Small portion of the Outer Rim halo catalog at $z=0.865$. The left panel is a $100 \times 100 ~[h^{-1} {\rm Mpc}]^2$  projection along $x$ and $y$ and across $z$, with thickness $\Delta z=50~h^{-1} {\rm Mpc}$, 
while the middle panel is a progressive zoom into a $50 \times 50 ~[h^{-1} {\rm Mpc}]^2$ block. Points the figure are FOF halos, color coded by their mass. 
Zooming into a smaller $7 \times 7~[h^{-1} {\rm Mpc}]^2$  inset of the halo catalog, the right panel displays the ellipsoidal shape of a halo of mass $4.938 \times 10^{13}~h^{-1}{\rm M_{\odot}}$ contained inside that area,
rendered with a 1\%  random particle subsample. Length units displayed in the left panel are in $h^{-1} {\rm Mpc}$. 
The high resolution of the simulation, down to $1.85 \times 10^9 h^{-1} {\rm M_{\odot}}$, 
allows one to resolve accurately also relatively low-mass halos.}
\label{fig_or_visualization}
\end{figure*}


The baseline simulation used for all our mock-making procedure is the Outer Rim (OR) run,
extensively described in \citet{Heitmann2019}. The simulation has been developed along the glorious tradition
of the   Millennium simulation \citep{Springel2005}, with similar mass resolution but a volume coverage increase by more than a factor of 200.
Currently, the Outer Rim is among the largest high-resolution gravity-only $N$-body  simulations
ever performed, spanning a ($3 h^{-1}{\rm Gpc}$)$^3$ volume, and characterized by an unprecedented combination of volume and mass resolution
(down to $1.85 \cdot 10^9 h^{-1} M_{\odot}$) 
evolving 1.07 trillion particles  -- i.e., $10,240^3$. The
actual size of the simulation was chosen to cover a volume large enough to enable synthetic sky catalogs for eBOSS, DESI and LSST, while maintaining
adequate mass resolution to capture halos reliably down to small masses. 
The entire run was carried out at the Argonne Leadership Computing Facility on Mira,
a Blue-Gene/Q many-core supercomputer. 
The simulation code adopted is an optimized version of the Hardware/Hybrid Accelerated Cosmology Code (HACC), designed to
overcome numerical challenges;  see \citet{Habib2016} for all the details. 
The cosmology of the simulation is close to the best-fit WMAP-7 fiducial model \citep{Komatsu2011}, 
namely $\omega_{\rm c} = 0.1109$, $\omega_{\rm b} = 0.02258$,
$n_{\rm s} =0.963$, $h=0.71$, $\sigma_8=0.8$, $w=-1$, with no massive neutrinos
($\Omega_{\nu}=0$) and assuming flatness.  
The dynamical range of the simulation is remarkable, spanning $10^6$ orders of magnitudes,
with a force resolution of $6h^{-1}{\rm kpc}$. 
Initial conditions are fixed at
$z_{\rm in} = 200$ with the Zel'dovich approximation \citep{Zeldovich1970}. 
Transfer functions are generated via CAMB \citep{Lewis2000}. 

A total of 101 redshifts in output were originally saved, from $z = 10$ to $z = 0$, evenly spaced in 
$\log a$, with $a$ the scale factor. 
In our study, we mainly focus on 2 redshift intervals, namely
$z=0.695$ and $z=0.865$, although we also consider a variety 
of other redshifts from some of the realizations to assess redshift evolution.   
Each snapshot encompasses globally about 40TB of data, and the entire data volume of the simulation is more than 5PB; 
all particle information for halos with more than 100,000 particles is stored 
for substructures and shape studies, as well as a random selection of 1\% of particles in each halo (with a minimum of 5 particles per halo).

For our mock-making purposes, we had access to the friends-of-friends (FOF) halo catalog at various redshifts,
generated using a linking length $b=0.168$.\footnote{The Outer Rim halo catalogs used here are publicly available at \url{https://cosmology.alcf.anl.gov}} 
Halos are defined by more than 20 particles, and found with 
a customized FOF finder \citep{Woodring2011,Heitmann2020}
that follows the standard
implementation, having the linking length
defined with respect to the mean inter-particle spacing.
All the centers of the halos are determined by the location of the FOF halo's minimum gravitational potential, and the center-of-mass and the halo velocities
are obtained by summing over all positions and velocities
and dividing by the number of particles. 
The FOF halo mass is simply determined by the number count of particles in each halo.

For a given redshift, our halo catalog (split into 110 subfiles, stored in a way such that they are not contiguous volumes) 
contains the number of particles in halo (Halo Count),
the halo ID  (Halo Tag), the halo FOF mass in $h^{-1}{\rm M}_{\odot}$ units, 
the comoving halo center positions from the potential minimum (in $h^{-1}{\rm Mpc}$),
and the comoving peculiar velocities of halo centers (in km/s). 
Note that the halo center is defined by its potential minimum  most bound particle, since 
accurate center-finding is important for measuring the halo concentration, 
for halo stacking, and for placing central galaxies from HOD modeling. 
 
Figure \ref{fig_or_visualization} visualizes a small portion of
the Outer Rim halo catalog at $z=0.865$. Specifically, the left panel shows a $100 \times 100 ~[h^{-1} {\rm Mpc}]^2$ 
projection along $x$ and $y$ and across $z$, having thickness $\Delta z=50~h^{-1} {\rm Mpc}$;
points in the figure are FOF halos, color coded by their mass.
The middle panel is a progressive zoom into a $50 \times 50 ~[h^{-1} {\rm Mpc}]^2$ block having the same depth as the left one, 
while the right panel shows an individual
halo of mass $4.938 \times 10^{13}~h^{-1}{\rm M_{\odot}}$ rendered with 1\% of random particles, located 
inside the smaller $7 \times 7~[h^{-1} {\rm Mpc}]^2$  white inset: it is possible to appreciate the neat ellipsoidal halo shape. 
Length units displayed in the left panel are in $h^{-1} {\rm Mpc}$. 
The {\bf high} resolution of the simulation, down to $1.85 \cdot 10^9 h^{-1} M_{\odot}$, 
allows one to resolve accurately also relatively low-mass halos.

 
\subsection{\textsc{Nseries} Mocks} \label{subsec_Nseries}

In addition to the heterogeneous sets of high-fidelity Outer Rim mocks developed in this work,
we also exploit
a small homogeneous set indicated as the \textsc{Nseries}, which has been
previously used in the SDSS DR12 galaxy clustering analysis \citep{Alam2017}.
The homogeneous set, comprised of 84 mocks in total, is particularly suitable to address cosmic variance and modeling systematics at the sub-percent level, 
since all mocks have the same underlying galaxy bias model built upon on
the same cosmology, but each mock is a quasi-independent realization -- thus not sharing the same LSS. 
Moreover, these mocks are cut sky: they have the same angular and radial selection function as the 
NGC DR12 CMASS sample within the redshift range $0.43 < z < 0.70$, and therefore they include observational artifacts closer to the eBOSS DR16 sample. 
The $N$-body simulations from which these cut-sky mocks were created
have been produced with \textsc{Gadget2} \citep{Springel2005}, 
with input parameters to ensure sufficient mass and spatial resolution to resolve the halos that BOSS galaxies occupy. 
Specifically, the \textsc{Nseries} cosmology is characterized by $\Omega_{\rm m} = 0.286$, $h = 0.7$, $\Omega_{\rm b} = 0.047$, $\sigma_8 = 0.820$, and $n_{\rm s} = 0.96$.
The main difference with respect to the Outer Rim mocks -- apart from being cut sky and not built on periodic cubes --  is that 
the \textsc{Nseries} derive from multiple realizations of the dark matter 
field on a larger volume with different random seeds (i.e., a series of $N$-body simulations identical in all but in the initial random seed),
which allows one to address the impact of cosmic variance: 
this is not achievable with only a halo catalog nor a single $N$-body simulation at hands.

 
\subsection{\textsc{EZmocks}}

For determining the rescaled covariance matrices functional to the subsequent analyses, 
we also make use of a new series of DR16 \textsc{EZmocks}, thoroughly described in \citet{Zhao2020}. 
These large number of galaxy catalogs (1000 per tracer), having accurate clustering properties,
are generated with a complex methodology built around the
Zel'dovich approximation \citep{Zeldovich1970},
 and effectively including stochastic scale-dependent, non-local, 
and non-linear biasing contributions; extensive details on the methodology can be found in the first release paper by \citet{Chuang2015}.
Non-local effects, such as tidal fields not included in linear Lagrangian Perturbation Theory (LPT) or other biasing contributions, are effectively included in both 
the scatter relation and the tilting of the initial power spectrum. The missing power towards small scales of perturbative approaches is included in the 
modulation of the initial power spectrum, when fitting for the resulting halo populations. 
These mocks have 
accurate clustering properties -- nearly indistinguishable from full $N$-body solutions -- 
in terms of the one-point, two-point, and three-point statistics. 
The underlying cosmology is based a flat $\Lambda$CDM model,
with $\Omega_{\rm m} =  0.307115$, $\Omega_{\rm b} = 0.048206$, $h = 0.6777$, $\sigma_8=0.8225$, and
$n_{\rm s} = 0.9611$. 
Specifically for LRGs, they contain the complexity of blending the CMASS plus eBOSS LRG samples,
as well as all the realistic effects of mask, cut sky, and observational systematics (i.e., fiber completeness, spectroscopic success rate, redshift failures, photometric systematics).
For our analysis, we adopt dedicated cubic \textsc{EZmocks} rather than the cut sky set for covariance estimations, 
to comply with the characteristics of the high-fidelity Outer Rim-based realizations.
The \textsc{EZmocks} are extensively used in all the supporting eBOSS DR16 papers and in the final eBOSS consensus analysis.
For additional technical details, we refer the reader to the companion paper by \citet{Zhao2020}. 

 
\subsection{Galaxy Mock-Making Procedure: Methods}  \label{sec_tools_methodology_lrgs}

Our synthetic high-fidelity galaxy mocks are primarily produced
exploiting the standard \textsc{Halotools}\footnote{See \url{https://github.com/astropy/halotools}}  
framework \citep{Hearin2017}, and by introducing a 
number of customizations depending on the desired challenge and model explored 
(see Section \ref{sec_unblind_challenge},
as well as the previous theoretical part). In particular, 
we interface \textsc{Halotools} capabilities with the   
Outer Rim halo catalog at different redshifts. 
\textsc{Halotools}
is an open-source, community-driven Python powerful package for studying the
galaxy-halo connection, which provides a highly modular, object-oriented
platform for building HOD models, so that individual modeling features can easily be swapped in and out.
This modularity facilitates rigorous study of all the components that makes up a 
halo occupation model, and has been designed from the ground-up with assembly bias applications in mind.
In this view, although our main products are
based on the Outer Rim simulation, following the \textsc{Halotools} philosophy the pipelines developed here are 
written in a general and flexible manner, so that any type of customization
is readily achievable with minimal efforts and modifications; hence, 
our modular-approach procedure is quite general, 
and readily applicable to any halo catalog and survey design in mind. 
The concept of generality and reusability of the code is in fact what has driven this
design from the start. In this view, although we limit here our modeling approach to HOD-based techniques 
(mainly due to limitations in our available halo catalog products), we plan to pursue
subhalo and abundance matching methods in follow-up studies, using the same code and modular structure. 

Adopting \textsc{Halotools} conventions, 
 three main  primary keyword arguments are used to
customize all the  instances retrieved by the mock factory, common to all the different HOD models developed here
(besides specific HOD parameters), namely: {\bf redshift}, {\bf threshold}, and {\bf modulate\_with\_cenocc} -- the latter being the
 modulation of the satellite distribution
with the central one.	
In our mock-making procedure, except for the Hearin framework, all other satellite HODs are not modulated by their corresponding central distributions.
Also, as previously mentioned, we treat the conventional Zheng model separately since
its HOD is effectively redshift-independent and  the meaning of `threshold' in the model is based on luminosity
rather than stellar mass, unlike for the other 3 frameworks considered (i.e., Leauthaud, Tinker, Hearin). 
Specifically,  \textsc{Halotools} is used to populate dark matter halos in the Outer Rim simulation with galaxies
having a stellar mass $M_{*} >10^{10.0}~h^{-1} M_{\odot}$ (`Threshold 1'),
$M_{*} >10^{10.5}~h^{-1} M_{\odot}$ (`Standard'), and
$M_{*} >10^{11.0}~h^{-1} M_{\odot}$ (`Threshold 2'). 
This roughly correspond to 
 `Threshold 1'  ($M_{\rm r}=-19$),
`Standard' ($M_{\rm r}=-20$), and `Threshold 2' ($M_{\rm r}=-21$), respectively, 
in the Zheng formalism.  
 
A composite HOD model is fully defined once one specifies the occupation statistics
and phase space prescription for centrals and satellites.
The theoretical formalism related to each individual model has been 
presented in Section \ref{sec_theory}.
The corresponding numerical implementation is briefly explained in what follows -- noting that    
dealing with the Outer Rim  simulation poses several non-trivial challenges in handling massive datasets. 
Specifically, at the highest level, we select the redshift, threshold, and number of desired mocks to produce.
Then, to populate  halos with central galaxies we first calculate the value of $\langle N_{\rm cen} \rangle$ for
every halo in the simulation according to the HOD formulas in our different prescriptions (Section \ref{sec_theory}).
For every halo in the simulation, we draw a random number $r$ from ${\cal{U}}[0,1]$, 
a uniform distribution between zero and unity.
For all halos with $r \le \langle N_{\rm cen} \rangle$, we place a central galaxy at the halo center, leaving all other halos devoid of centrals.
Populating satellites is more complicated, because the spatial distributions are nontrivial.
The first step is similar to that of the centrals, namely compute $\langle N_{\rm sat} \rangle$ for every halo using our specified formulas for a given HOD model 
(see again Section \ref{sec_theory}).
For each halo, the number of satellites that 
will be assigned to the halo is then determined by drawing an integer from the assumed satellite occupation distribution 
$p(N_{\rm sat}|M_{\rm h})$ or $p(N_{\rm sat}|M_{\rm h}, x)$. 
Satellites are modeled as being isotropically distributed within their halos according to a NFW profile with concentration
equal to the parent halo, using the Dutton-Macci{\`o} model \citep{Dutton2014}.
Monte Carlo realizations of both radial and angular positions are generated via the method of inverse transformation sampling. Briefly, first one
generates realizations of points uniformly distributed on the unit sphere.
 These halocentric (x,y,z) coordinates are then multiplied  by the corresponding realization of the radial position $r$, which is determined as follows: 
 first, calculate $P_{\rm NFW}(< \tilde{r}|c)$ where $c$ is the concentration, 
$\tilde{r} =r/R_{\rm vir}$ is the scale radius, $R_{\rm vir}$ the virial radius of the halo, 
and $P_{\rm NFW}(< \tilde{r}|c)$ is the cumulative probability distribution 
function of the mass profile of a NFW halo:
\begin{equation}
P_{\rm NFW} (< \tilde{r} | c) = {M_{\rm NFW}  (< \tilde{r} | c)  \over M_{\rm tot} } = {g(c \tilde{r}) \over g(c)}
\end{equation}
where
\begin{equation}
g(x) = \ln (1+x) - {x \over 1 + x}.
\end{equation}
Then, for a halo with concentration $c$ populated by $N_{\rm sat}$,  draw $N_{\rm sat}$ random numbers 
$p$ from   ${\cal{U}}[0,1]$. Each value of $p$ is 
interpreted as a probability where the corresponding value for the scaled radius $\tilde{r}$ 
comes from numerically inverting $p=P_{\rm NFW} (< \tilde{r} | c)$.
Scaling the (x,y,z) points on the unit sphere by the value $r$ gives the halocentric position of the satellites.
 
All these high-fidelity mocks have been produced
at the National Energy Research Scientific Computing Center (NERSC), a DOE Office of Science User Facility 
supported by the Office of Science of the U.S. Department of Energy, under Contract No. DE-AC02-05CH11231
using the Cori supercomputer, a Cray XC40 with a peak performance of about 30 petaflops. 
Cori is comprised of 2,388 Intel Xeon ``Haswell" processor nodes, and 9,688 Intel Xeon Phi ``Knight's Landing" (KNL) nodes. 
The system also has a large Lustre scratch file system and a first-of-its kind NVRAM ``burst buffer" storage device.
We devised new customized scripts and pipelines
to produce such mocks on Cori, exploiting especially the multi-thread architecture. Our mock-making
code/pipeline is  memory efficient and optimized to the machine.
Some additional supporting numerical work has also  been 
carried out using the Korea Institute of Science and Technology Information (KISTI) 
supercomputing infrastructure. 

In closing this part, we note that our main goal in the cubic $N$-body-based mock-making production and in the related mock challenge
is  to test  the  validity and robustness of 
different BAO and RSD fitting techniques on a common ground against a
series of different HOD prescriptions, and validate the clustering analysis pipelines and the various RSD models. 
Hence, we are not concerned with reproducing exactly all the features of the eBOSS DR16 LRG sample,
and this is why the various HOD parameters that enter in the models outlined in Section \ref{sec_theory}
have been maintained to their  corresponding literature values. 
Nevertheless, in Section \ref{sec_unblind_challenge} we show 
an  instructive comparison between eBOSS measurements and those obtained from Outer Rim TH2 mocks.
Note also that the HOD parameters adopted from the literature were chosen under cosmologies different from that of the Outer Rim, and therefore
we do not expect to find the same results as in the corresponding original publications: this is clearly not affecting the conclusions of our work.
Realistic observational artifacts related to the LRG sample, such as cut sky, matching number density, observational systematics, etc., 
are instead part of the EZmock release \citep{Zhao2020}.  
 


\section{Analysis: Methodology} \label{sec_analysis_methods}


In this section, we briefly describe the three 
configuration and Fourier space techniques 
used in the analysis of the challenge mocks,  
based on three different RSD analytical models --
exploiting  the FS information  in  the correlation function or power spectrum.  
The detailed BAO modeling is instead described in our LRG companion papers.
All these methods are 
adopted in the main analysis of the final eBOSS DR16 LRG sample. 

 
\subsection{CLPT-GS}

The CLPT-GS-based method is a combination of the
Convolutional Lagrangian Perturbation Theory (CLPT) and the RSD Gaussian Streaming (GS) formalism, 
originally developed by \citet{Reid2011}, \citet{Carlson2013}, and \citet{WangL2014}.
CLPT provides a non-perturbative resummation of Lagrangian perturbation to the two-point statistic in real space for biased tracers. 
In particular, the 
two-point correlation function is expanded in its Lagrangian coordinates considering the LRG tracer
to be locally biased with respect to the initial CDM overdensity, and the 
 expansion is performed over different orders of the Lagrangian bias function. 
The key equation for the two-point correlation
$\xi_{\rm LRG}({\bf r}) = \left< \delta_{\rm LRG} (\bf x) \delta_{\rm LRG} (\bf x + \bf r)\right>$, with
${\bf q}$ and ${\bf x}$ the  Lagrangian and Eulerian coordinates, respectively, $\delta$ the overdensity, and ${\bf r}$ the LRG separation,
is:   
\begin{equation}
1+\xi_{\rm LRG}({\bf r})=\int M({\bf r, \bf q}) {\rm d}{\bf q},
\label{xi_model}
\end{equation}
where $ M({\bf r, \bf q})$ is the convolution kernel taking into account the displacements and bias expansion up to its second derivative term. 
The bias derivative terms are computed using  a linear power spectrum, obtained with \textsc{CAMB} \citep{Lewis2000}
for a fixed cosmology -- namely, the fiducial cosmology of the analysis.
The peculiar velocity effect on  clustering statistic is also modeled, and
the pairwise velocity distribution ${\bf v}_{12}$  and velocity dispersion $\sigma_{12}$
are given by \citep{WangL2014}: 
\begin{equation}
{\bf v}_{12}(r)=[1+\xi_{\rm LRG}({\bf r})]^{-1}\int {M_1}({\bf r}, {\bf q}) {\rm d} {\bf q}
\end{equation}
and
\begin{equation}
\sigma_{12}(r)=[1+\xi_{\rm LRG}({\bf r})]^{-1}\int M_2({\bf r}, {\bf q}){\rm d}{\bf q}
\end{equation}
where  the kernels $M_{1}({\bf r},{\bf q})$ and $M_{2}({\bf r},{\bf q})$
also depend on the first two derivatives of the Lagrangian bias, which are free parameters in the model, in addition to the growth factor. 
CLPT generates more accurate multipoles than linear theory and even the Lagrangian Resummation Theory \citep[LRT;][]{Matsubara2008},
but a better performance is needed in order to study the smaller scales of quadrupoles.  
To achieve such precision, the real space CLPT models of the two-point statistics are mapped into redshift space following the 
Gaussian Streaming Model (GSM) formalism proposed by \citet{Reid2011}. 
In particular, the pairwise velocity distribution is assumed to 
have a Gaussian shape dependent on both the angle $\mu$ between the separation vector and the line-of-sight (LOS), and
the LRG separation $r$ in its parallel ($r_{||}$) and perpendicular ($r_{\perp}$) components with respect to the LOS.
The main equation for the correlation function is given by: 
\begin{equation}
\begin{split}
1+\xi_{\rm LRG}(r_\perp,r_\parallel)= & \int \frac{1}{\sqrt{2\pi (\sigma_{12}^2(r,\mu)+\sigma^2_{\rm FoG})}}[1+\xi_{\rm LRG}(r)]\\
& \times \exp \Big [{-\frac{[r_\parallel-y-\mu v_{12}(r,\mu)]^2}{2(\sigma_{12}^2(r,\mu)+\sigma^2_{\rm FoG})}} \Big ] {\rm d}y,
\end{split}
\label{gsrd_integral}
\end{equation}
where $\xi_{\rm LRG}(r)$, $v_{12}(r)$, and $\sigma_{12}(r)$ are computed from CLPT
as previously indicated, and $\sigma_{\rm FOG}$ is the 
Fingers of God (FoG) parameter to account for an additional contribution
to the velocity dispersion given by satellite galaxies. 
For the RSD model, the \citet{Alcock1979} effect implementation follows that of \citet{Xu2013}. 
The AP distortions are modeled through the $\alpha$ and $\epsilon$ parameters, which 
characterize respectively the isotropic and anisotropic distortion components.  

With this technique, the FS RSD analysis in configuration space is performed, and for a given cosmology
the model has 4 free parameters, namely ($f \sigma_8$, $F'$, $F''$, $\sigma_{\rm FOG}$), with
$f$ the linear growth factor and $F'$ and $F''$ the first and second derivatives of the Lagrangian bias function $F$. 
For extensive details on this method see  \citet{LRG_corr2020} and \citet{Icaza-Lizaola2020}. 

 
\subsection{TNS in Configuration Space}

The Modified TNS-based method  (also indicated in this paper as `CF-TNS', where `CF' stands for `correlation function')
is a combination of the Taruya, Nishimichi \& Saito \citep[TNS;][]{Taruya2010} 
technique and a galaxy non-linear bias prescription \citep{Beutler2017a,delaTorre2017}. 
This model is based on the conservation of the number density in real- and $z$-space \citep{Kaiser1987}. 
In this framework, the anisotropic power spectrum for unbiased matter tracers ($P^{\rm s}$) follows the general form of \citet{Scoccimarro1999},
which in the approximation proposed by  \citet{Taruya2010} reads: 
\begin{multline}
    P^{\rm s}(k,\mu)= D(k\mu\sigma_{\rm v})\big[ P_{\delta\delta}(k) +2\mu^2f P_{\delta\theta}(k) + \mu^4f^2P_{\theta\theta}(k)+ \\ C_{\rm A}(k,\mu,f)+C_{\rm B}(k,\mu,f) \big]\,.
\end{multline}
In the previous expression, $f$ is the linear growth factor; 
$\theta$ is the divergence of the velocity field defined as $\theta = -\nabla {\bf \cdot \mu}/(aHf)$; 
$\mu=k_\parallel/k$, with $k_\parallel$  the line-of-sight component of the wave vector $k$; $H$ is the Hubble constant at the considered redshift;
$a$ is the scale factor; $\delta$ is the matter density field;
$P_{\delta\delta}$,  $P_{\theta\theta}$, and $P_{\delta\theta}$  
are the non-linear matter density, velocity divergence, and density-velocity divergence power-spectra, respectively;
$C_A(k,\mu,f)$ and $C_B(k,\mu,f)$ 
are two correction terms expressed as integrals of the matter power spectrum -- see \citet{Taruya2010} for their detailed expressions; 
and $D(k\mu\sigma_{\rm v})$ is a phenomenological damping function
modeled as a Lorentzian so that 
$D(k,\mu,\sigma_{\rm v}) = [1+(k\mu\sigma_{\rm v})^2]^{-1}$, with 
$\sigma_{\rm v}$ an effective pairwise velocity dispersion that is later treated as a nuisance parameter in the cosmological inference.

This model can be generalized for biased tracers via the inclusion of a galaxy biasing model, so
that  the anisotropic galaxy power spectrum becomes
\citep{Beutler2014,Gil-Marin2017}:
\begin{multline}
P^{\rm s}_{\rm g}(k,\mu) = D(k\mu\sigma_{\rm v}) \big[ P_{\rm gg}(k) + 2\mu^2fP_{\rm{g} \theta} + \mu^4f^2 P_{\theta\theta}(k) + \\
C_{\rm A}(k,\mu,f,b) + C_{\rm B}(k,\mu,f,b) \big]
\label{eq_psg}
\end{multline}
with $b$ the galaxy linear bias. 
Specifically, here we assume a non-linear, non-local, galaxy biasing
prescription that follows the work of \citet{McDonald2009} and \citet{Assassi2017}.
Explicit expressions for  $C_{\rm A}(k,\mu,f,b)$ and $C_{\rm B}(k,\mu,f,b)$ that enter in Equation (\ref{eq_psg})
can be found in \citet{delaTorre2012}, while detailed expressions for the galaxy-galaxy and galaxy-velocity divergence power spectra
-- $P_{\rm gg}(k)$ and $P_{\rm g\theta}(k)$, respectively -- for a 1-loop perturbative expansion of the biasing function
are given in \citet{LRG_corr2020}.  

The linear and nonlinear matter power spectra entering in the model are computed with \textsc{Camb} and the 
\textsc{Halofit} semi-analytical prescription, respectively. To obtain $P_{\theta\theta}$ and 
$P_{\delta\theta}$, we use the universal fitting functions provided by \citet{Bel2019}. In particular,
the overall degree of nonlinear evolution is encoded via the amplitude of the matter fluctuation at the 
 effective redshift considered.   
Finally, the multipole moments of the anisotropic correlation function are obtained by performing the Hankel transform of the model,
and regarding the RSD part, the implementation of the AP effect follows the formalism of \citet{Xu2013}:
the AP distortions are modeled through the $\alpha$ and $\epsilon$ parameters, which 
characterize the isotropic and anisotropic distortion components, respectively.  

With this technique, the FS RSD analysis in configuration space is performed, and for a given cosmology
the model has 5 free parameters, namely ($f$, $\sigma_8$, $b_1$, $b_2$, $\sigma_{\rm v}$) --
 although since $f$ and $\sigma_8$ are degenerate they are thus combined at the level of the likelihood into the single parameter $f\sigma_8$. 
For extensive details on this method see \citet{delaTorre2017}, \citet{Mohammad2018},
and \citet{LRG_corr2020}. 

 
\subsection{TNS in Fourier Space}

While the previous techniques are used to carry out the analysis of the LRG sample in configuration space, 
the method described here -- also based on the TNS model and indicated as  `$P_{\rm k}$-TNS'  --
is performed in Fourier space. 
To this end, the modeling of the BAO signal within this framework -- along with the BAO fitting procedure in Fourier space --
are described in \citet{Gil-Marin2020}. 
Here, we briefly illustrate only the strategy adopted for the 
RSD and AP analysis,  exploiting  the  
FS information  in  the power spectrum.  

Specifically,  the FS formalism employed to describe power spectrum multipoles is the same as the one previously
used in BOSS and eBOSS studies for galaxies \citep{Gil-Marin2016}  and quasars \citep{Gil-Marin2018}. 
We adopt the Eulerian non-linear bias model of \citet{McDonald2009}, consisting of
4 bias parameters: namely, 
the linear galaxy bias $b_1$, the non-linear galaxy bias $b_2$, and two non-local galaxy bias parameters, 
$b_{\rm s^2} =-4/7\,(b_1-1)$ \citep{Baldauf2012}  and $b_{3{\rm nl}}=32/315\,(b_1-1)$ \citep{Saito2014} -- with $b_1$ and $b_2$ considered as 
free nuisance parameters in the fitting. 
The density-density ($\delta\delta$), density-velocity ($\delta\theta$), and velocity-velocity ($\theta\theta$)  real space DM auto- and cross-power spectra
are obtained via 2-loop resummation perturbation theory, as in \citet{Gil-Marin2012}: 
these moments accurately 
describe the DM clustering up to $k\simeq 0.15$ at $z=0.5$, $k\simeq 0.20$ at $z=1.0$, and $k\simeq 0.30$ at $z=1.5$, respectively. 
Expressions for those galaxy power spectra -- with no velocity bias -- are given by \citep{Beutler2014}: 
\begin{eqnarray}
\nonumber P_{\rm g,\,\delta\delta}(k)&=&b_1^2P_{\delta\delta}(k)+2b_2b_1P_{\rm b2,\,\delta}(k)+2b_{\rm s2}b_1P_{\rm bs2,\,\delta}(k)+\\
\nonumber&\quad&b_2^2P_{\rm b22}+2b_2b_{\rm s2}P_{\rm b2s2}(k)+b^2_{\rm s2}P_{\rm bs22}(k)+\\
&\quad& 2b_1b_{3\rm nl}\sigma_3^2(k)P_{\rm lin}(k) \\
\nonumber P_{\rm g,\,\delta\theta}(k)&=&b_1P_{\delta\theta}(k)+b_2P_{\rm b2,\,\theta}(k)+b_{\rm s2}P_{\rm bs2,\,\theta}(k)+\\
&\quad&b_{3\rm nl}\sigma_3^2(k)P_{\rm lin}(k)\\
P_{\rm g,\,\theta\theta}(k)&=&P_{\theta\theta}(k).
\label{eq:Pthetatheta}
\end{eqnarray}

RSD effects are incorporated following \citet{Taruya2010}, who extended the 
original methodology of \citet{Scoccimarro2004}, so that the redshift space galaxy power spectrum reads:
\begin{eqnarray}
\label{eq_2loop_rsd}
\nonumber P^{(\rm s)}_{\rm g}(k,\,\mu)&=&D_{\rm FoG}(k,\,\mu)\left[P_{\rm g\,\delta\delta}(k)+2f\mu^2P_{\rm g,\,\delta\theta}(k) +\right. \\
\nonumber&\quad&f^2\mu^4P_{\theta\theta}(k)+b_1^3A^{\rm TNS}(k,\mu,f/b_1)+\\
&\quad&\left. b_1^4B^{\rm TNS}(k,\,\mu,f/b_1) \right].
\end{eqnarray}
In particular, galaxy real space quantities are computed as previously described, assuming a fixed linear power spectrum template (obtained with \textsc{Camb}) at the 
fiducial cosmology. The various 
power spectrum multipoles encode the coherent velocity 
field through the redshift space displacement and the logarithmic growth of structure parameter, which 
boosts the amplitude of the isotropic power spectrum and generates
an anisotropic component. 
In Equation (\ref{eq_2loop_rsd}), the term  $D_{\rm FoG}$
accounts for  FoG effects along the LOS direction, and it is 
 modeled as a Lorentzian, while 
$A^{\rm TNS}$ and $B^{\rm TNS}$ are second-order corrections. 
Finally, the AP effect is added when computing the multipoles as:
\begin{equation}
\label{eq:rsdaniso}P_{\rm g}^{(\ell)}(k)=\frac{2\ell+1}{2\alpha_\parallel\alpha_\perp^2}\int_{-1}^{1}d\mu \mathcal{L}_{\ell}(\mu)P_{\rm g}^{(s)}[k'(k,\mu),\mu'(\mu)],
\end{equation}
where explicit expressions for $k'(k,\mu)$ and $\mu'(\mu)$ are given 
in \citet{Gil-Marin2020}.
We also consider that the shot noise contribution in the power spectrum monopole may differ from a Poisson sampling prediction, and
parameterize this potential deviation with a free parameter ($A_{\rm noise}$), which modifies the shot noise amplitude without
introducing any scale dependence. 
By default, our measured power spectrum monopole has a fixed Poissonian shot noise contribution subtracted, whereas 
this is not the case for higher other multipoles.

With this technique, the FS RSD analysis in Fourier space is performed, and for a given cosmology
the model has 7 free parameters, namely  ($\alpha_{\parallel}$, $\alpha_{\perp}$, $f \sigma_8$) and
($b_1$, $b_2$, $A_{\rm noise}$, $\sigma_{\rm FoG}$).
Note that while  the BAO analysis consists of using a fixed and arbitrary template  to  compare  the  relative  BAO-peak  positions  in  the  
power spectrum multipoles, the FS analysis allows for a full modeling of the shape and amplitude of the power spectrum multipoles, taking into account 
DM non-linear effects, galaxy bias and RSDs.
For extensive details on this method see \citet{Gil-Marin2020}. 



\section{The Galaxy Mock Challenge} \label{sec_unblind_challenge}


In this section, we present the main outcomes of the galaxy mock challenge. 
After a brief description of the 
mock products directly useful in the actual fits considered in the HOD systematic error budget, we show 
selected results in configuration and Fourier space. We eventually 
compare the complementary BAO/RSD models 
adopted for the analysis of the complete DR16 eBOSS LRG sample, assessing the theoretical systematic budget.
Our findings demonstrate that 
all the methods are mutually consistent, with comparable systematic errors on the AP parameters 
and the growth of structures, and robust to 
different HOD prescriptions -- thus validating the clustering analysis pipelines.

 
\subsection{Mock Products Used in the Analysis}

 
\begin{figure*}
\begin{center}
\includegraphics[angle=0,width=0.75\textwidth]{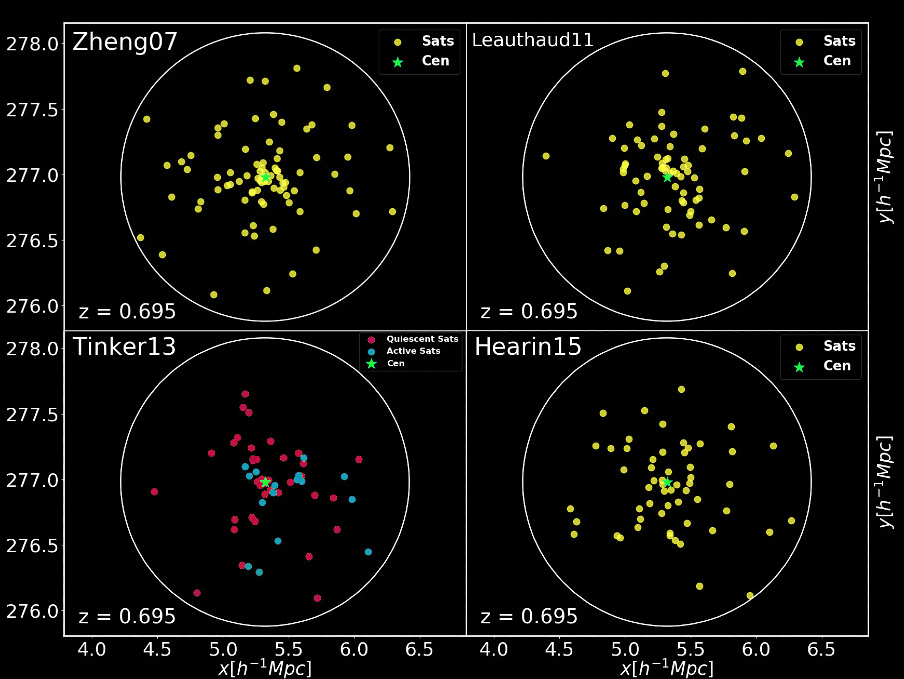}
\caption{Example of the distribution of satellite galaxies within the same and randomly chosen Outer Rim halo, at $z=0.695$,
according to the different HOD prescriptions presented in Section \ref{sec_theory}. 
The plot represents a spatial projection ($x-y$) of the halo, and its spherical shape as determined by its virial radius is also indicated in the various panels. 
Clockwise, starting from the upper left corner, the 
 Zheng, Leauthaud, Hearin, and Tinker models are shown, respectively.}
\label{fig_or_halo_hods}
\end{center}
\end{figure*}


For the galaxy mock challenge, we devised 
three sets of heterogeneous Outer Rim-based galaxy mocks (indicated as `Challenge Set 1', 
`Challenge Set 2', `Challenge Set 3', respectively).\footnote{We have also devised an additional set which includes a variety of customizations, 
beyond the scope of the current analysis, exploring extreme variations in HOD parameters for all of the models, quenching, 
assembly bias, and modulations with the central distribution.} These are cubic mocks, 
in the Outer Rim cosmology, obtained by  populating Outer Rim halo catalogs with galaxies 
using the Zheng, Leauthaud, Tinker, and Hearin HOD prescriptions -- as explained in Section \ref{subsec_OR}.
Extensive details regarding each set are provided in Appendix \ref{sec_appendix_mock_products}.
For the main analysis presented here, focused on 
  testing the BAO templates 
and the RSD models adopted for the characterization of LRG clustering systematics,
we only use a 
 subset of those mocks drawn from `Challenge Set 1' at $z=0.695$ assuming the `Threshold 2' (Th2) flavor.
 As explained in Sections \ref{sec_theory} and \ref{sec_tools_methodology_lrgs},
the meaning of `flavor' is related to the 
\textsc{Halotools}  key parameter  `threshold', which 
globally sets all the individual HOD parameters as best fit realizations from the corresponding literature dictionary of each HOD model
(unless customizations are introduced).
Specifically, we select Th2 mocks with the Leauthaud, Tinker, and Hearin prescriptions, respectively,  
since their number density is closer to the eBOSS LRG sample (see Table \ref{table_number_density_unblind_set}). 
Moreover, since we require fully independent mocks (i.e., not sharing the same DM field), we only 
consider 27 realizations per HOD per flavor: such realizations are 
obtained by 
populating once the full $3 h^{-1} {\rm Gpc}$ Outer Rim periodic halo catalog box with galaxies, and
then by cutting the full box into 27 non-overlapping subcubes of $1h^{-1}{\rm Gpc}$ side and rescaling the various spatial positions accordingly. 
In fact, by construction, 
at a fixed $z$ and for a fixed set of HOD parameters, the central galaxies will always reside at the center of their hosted halos, 
inheriting the same halo velocity; hence, additional subcubes  
would be fully or highly correlated in the central galaxy population, depending on how the box is cut. 
We then add RSDs to each individual mock in two different ways: radially, or with the usual plane-parallel approximation.

In Figure \ref{fig_or_halo_hods}, we show an example 
on how satellite galaxies are distributed within the same Outer Rim halo, 
according to the different HOD prescriptions presented in Section \ref{sec_theory}, 
to convey some intuition on  the galaxy-halo connection modeling and the spatial location of satellites. 
The plot displays the x-y spatial projection at $z=0.695$ of 
a randomly chosen halo, with its spherical shape determined by its
virial radius. 
The standard Zheng model is shown in the upper left panel, and clockwise
the Leauthaud, Hearin, and Tinker models are displayed, respectively. While in all the HOD schemes the satellite 
phase space statistics follow an unbiased NFW profile with a
phase space distribution  
in isotropic Jeans equilibrium and galaxy concentration identical to that of the parent halo, more
sophisticated frameworks such as the Tinker model (lower left corner) are able to
distinguish between active and quiescent populations (indicated with different colors in the panel), thus providing additional
useful physical insights. 

Moreover, in addition to the heterogeneous Outer Rim mocks, as detailed in Section \ref{subsec_Nseries}
we also  exploit 84 homogeneous  cut-sky \textsc{Nseries} mocks, which have been
previously used in the SDSS DR12 galaxy clustering analysis \citep{Alam2017},
to address cosmic variance in the various methods -- since 
the \textsc{Nseries} derive from multiple realizations of the dark matter 
field with different random seeds. Here we show only one global \textsc{Nseries} application, 
while in \citet{Gil-Marin2020} and \citet{LRG_corr2020} those mocks are 
extensively used to assess systematics related to each individual fitting 
method in configuration or Fourier space, respectively. 


\subsection{Galaxy Mock Challenge: BAO Analysis and HOD Systematics} \label{sec_bao_team_analysis}


 
Approximate catalogs such as the EZmocks \citep{Zhao2020} 
are in principle sufficient for covariance estimates and 
for quantifying systematic biases in BAO studies, while 
the analysis of the FS of the correlation function and power spectrum 
requires high-fidelity ($N$-body-based) mocks to precisely test the modeling.
Nevertheless, using high-resolution mocks, we are able to characterize the impact of
systematics in HOD modeling both on BAO and RSD constraints with high-accuracy. 
Specifically, the
main goals are to quantify possible effects induced by different galaxy HOD schemes on
the cosmic growth rate and obtain useful information
on parameter inference based on HOD variations,
to assess the impact of an arbitrary choice 
of the BAO reference template on the inferred cosmological parameters,
and more generally to determine the theoretical systematic budget and validate the
clustering analysis pipelines.  
 
The standard procedure common to all BAO fitting methods is to
assume a  fixed and arbitrary template, and compare the relative
BAO peak positions in the correlation function or power spectrum multipoles.   
Reconstruction techniques such as those presented in  \citet{Burden2014,Burden2015}
are then applied to the density field,
in order to remove a fraction of the RSDs and the nonlinear motions of galaxies.
The BAO feature in the 2-point statistics (both in configuration and Fourier space) is sharpened, 
increasing the precision of the measurement of the acoustic scale.

The BAO scale measurement in configuration space adopted here 
is the same as the one described in previous SDSS publications \citep[i.e.,][]{Alam2017,Ata2018,Bautista2018}, 
and thoroughly illustrated in our companion paper \citet{LRG_corr2020}, while the 
modeling of the BAO signal along with the BAO fitting procedure in Fourier space
are explained in detail in \citet{Gil-Marin2020}.  In particular, for the latter case, 
the power spectrum anisotropic signal is modeled in order to
measure the BAO peak position and marginalize over the broadband information --
taking into account the BAO signal both in the radial and transverse LOS directions.
Generally, BAO results are obtained from pre- and post-reconstructed data, while 
RSD results use only the non-reconstructed sample. In the following analyses, we 
assume standard dependency of the growth rate $f$, and adopt a smoothing scale of 15 $h^{-1}$Mpc.
Whenever required, galaxy redshifts  are converted into
radial comoving distances  for clustering measurements, using the cosmological parameters
of the OR simulation. As shown in \citet{LRG_corr2020}, the analysis methodology is 
insensitive to the choice of a fiducial cosmology. 
  


\begin{figure*}
\centering
\includegraphics[angle=0,width=0.7\textwidth]{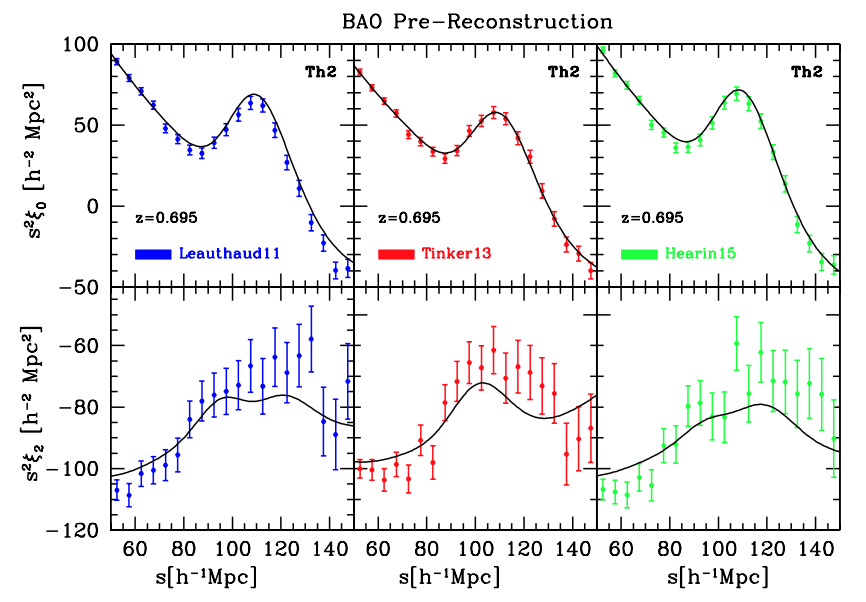} 
\includegraphics[angle=0,width=0.7\textwidth]{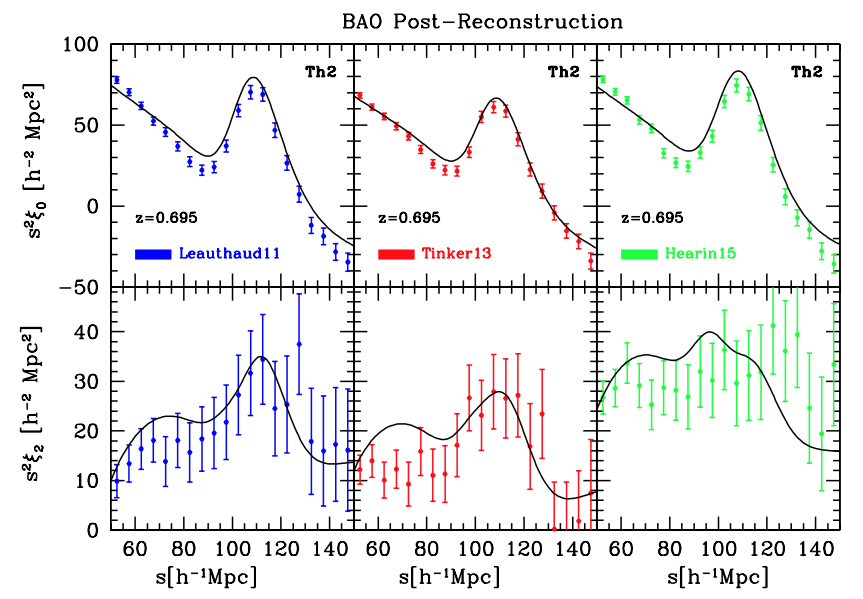}  
\caption{Monopole and quadrupole of the average 2PCFs as computed from a subset of 27 OR-based mocks per HOD type, and
fits of the BAO feature as seen in the correlation function multipoles. Top panels show results for the pre-reconstruction case, bottom
panels refer to the reconstructed density field. The HOD models of  Leauthaud, Tinker, and Hearin are shown -- from left to right, respectively --
for the `Th2' flavor at $z=0.695$. Note that the BAO feature (around $100 h^{-1}{\rm Mpc}$) appears much sharper after application of the reconstruction procedure, as expected.
Results of these fits are reported in Table \ref{table_lrg_team_1}.}
\label{fig_lrg_team_1}
\end{figure*}


\begin{figure}
\centering 
\includegraphics[angle=0,width=0.43\textwidth]{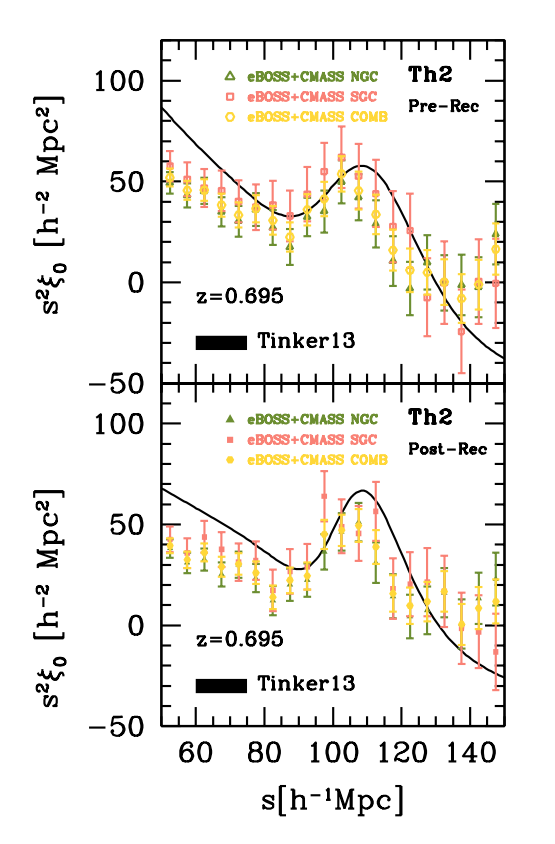} 
\caption{Illustrative example showing
the comparison between eBOSS measurements and those made from Th2 OR mocks at $z=0.695$.
The top panel displays pre-reconstruction calculations (reported with  open symbols), while the
bottom panel refers to post-reconstruction results (indicated with filled symbols).
Green triangles in the figure show eBOSS+CMASS measurements for the NGC, 
pink squares are used for eBOSS+CMASS SGC measurements, and yellow circles
represent the combined samples. 
Errorbars are estimated from 1000 EZmocks. 
The solid black line in both panels is the redshift-space 2PCF monopole expectation assuming the 
Tinker HOD recipe, derived from 27 Th2 OR mocks, without including 
 the complications of masks, cut sky, and observational artifacts.}
\label{fig_lrg_team_1_bis}
\end{figure}


\begin{table}
\centering
\caption{BAO fits to the average pre- and post-reconstructed 2PCFs for different HOD prescriptions over 27 corresponding Outer Rim mock 
realizations, at $z=0.695$, with the `Th2' flavor -- as displayed in Figure \ref{fig_lrg_team_1}.}
\doublerulesep2.0pt
\renewcommand\arraystretch{1.5}
\begin{tabular}{ccccccc} 
\hline \hline  
{\bf CF [Th2]} & Leauthaud &   Tinker &  Hearin\\
\hline
&  {\it BAO} & {\it Pre-Rec} &   \\
\hline
$\alpha_{\perp}$ &  $0.9990 \pm 0.0080$  &  $0.9922 \pm 0.0073$ &  $1.0074 \pm 0.0083$  \\
$\alpha_{\parallel}$ &  $1.0084 \pm 0.0164$ &  $1.0234 \pm 0.0147$ &   $1.0040 \pm 0.0157$  \\
$\Sigma_{\perp}$ & $6.7663 \pm 1.1320$ & $7.6741 \pm 1.2932$ &   $7.4335 \pm 1.0760$ \\
$\Sigma_{\parallel}$ & $10.5333 \pm 1.5795$  &  $8.4652 \pm 1.5436$ & $9.5687 \pm 1.6354$ \\
$\beta$ & $0.2685 \pm 0.0965$ &  $0.2404 \pm 0.1017$ &  $0.2137 \pm 0.0910$ \\
$b$ & $2.6748 \pm 0.1315$ & $2.4769 \pm 0.1393$ & $2.7525 \pm 0.1339$ \\
$\chi^2$ & $129.4 $& $115.4$ &  $126.0$ \\
\hline
& {\it BAO} & {\it Post-Rec} &   \\
\hline
$\alpha_{\perp}$ &    $1.0045 \pm 0.0056$ &  $1.0014 \pm 0.0060$ &  $1.0066 \pm 0.0057$ \\
$\alpha_{\parallel}$ & $0.9937 \pm 0.0084$ &  $0.9976 \pm 0.0090$ &  $1.0115 \pm 0.0094$ \\
$\Sigma_{\rm rec}$ & $15.0 \pm 1.5$ & $15.0 \pm 1.5$ & $15.0 \pm 1.5$  \\
$\Sigma_{\perp}$ &   $2.0002 \pm 11.9307$ & $2.0000 \pm 0.9844$ & $2.9483 \pm 1.3018$  \\
$\Sigma_{\parallel}$ & $4.0169  \pm 1.6667$  & $4.9272 \pm 1.3161$ & $6.8683  \pm 1.1810$ \\
$\beta$ & $0.4018 \pm 0.0911$ & $0.5127 \pm 0.0863$ & $0.4157 \pm 0.0849$  \\
$b$ & $2.3873 \pm 0.0744$ & $2.1669 \pm 0.0710$ & $2.5137 \pm 0.0879$ \\
$\chi^2$ & $131.3$ & $130.0$ &  $164.5$ \\
\hline
\hline
\label{table_lrg_team_1}
\end{tabular}
\end{table} 


\begin{figure*}
\includegraphics[angle=0,width=0.7\textwidth]{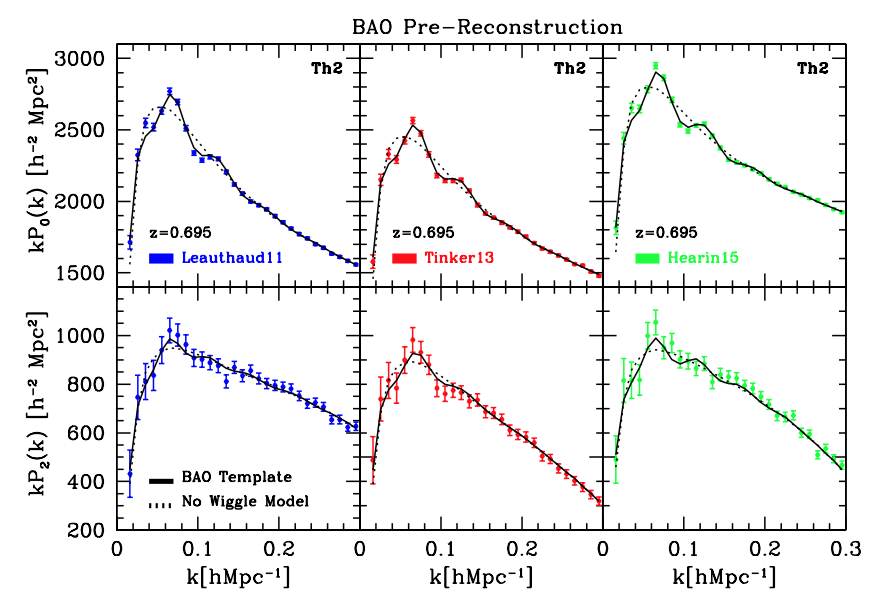} 
\includegraphics[angle=0,width=0.7\textwidth]{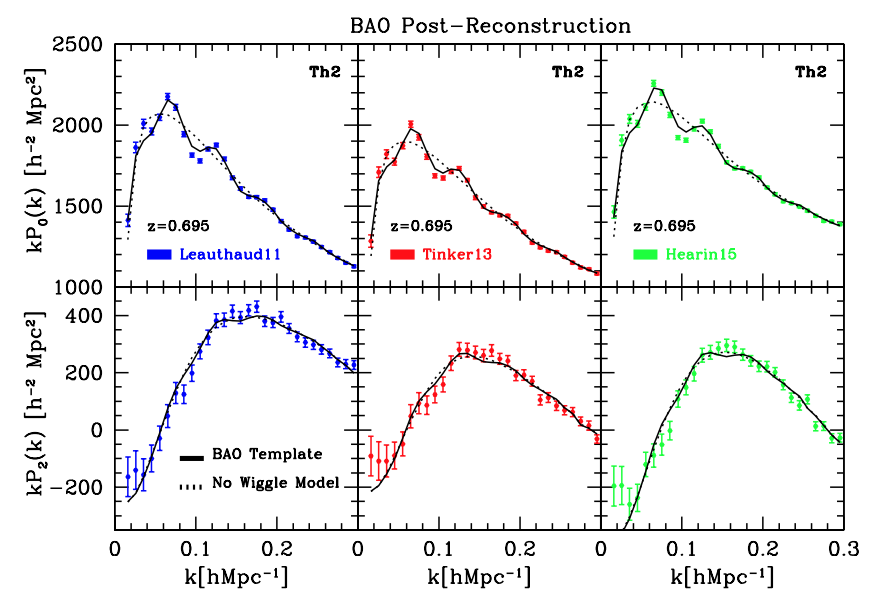} 
\caption{Monopole and quadrupole of the average power spectra as computed from a subset of 27 OR-based mocks per HOD type, and
fits of the BAO feature as seen in the power spectrum multipoles (solid lines). The corresponding no-wiggle model is also reported in the figure, with dotted lines.
Top panels show results for the pre-reconstruction case, bottom
panels refer to the reconstructed density field. The HOD models of Leauthaud, Tinker, and Hearin are shown -- from left to right, respectively --
for the `Th2' flavor at $z=0.695$. Results of these fits are reported in Table \ref{table_lrg_team_2}.}
\label{fig_lrg_team_2}
\end{figure*}


Figure \ref{fig_lrg_team_1} is an example of the redshift-space
galaxy clustering (monopole and quadrupole), along with corresponding BAO fits, 
as inferred from the average of 27 Th2 OR-based mocks at $z=0.695$ having different HOD schemes.
From left to right, the Leauthaud, Tinker, and Hearin models are displayed, respectively. Top panels
are for the pre-reconstructed fields, while bottom panels refer to post-reconstructed fields.
Specifically, 
for cosmological analyses the information 
contained in the anisotropic 2-point correlation function $\xi(s, \mu)$ -- decomposed into 
polar coordinates $(s, \mu)$ aligned with the LOS direction, with $\mu$ the cosine of the angle 
between the LOS and separation vector directions, and $s$ the norm of the galaxy separation vector ${\bf s}$ --
is compressed 
into the correlation function multipole moments ${\xi}_{\ell}$,  obtained by 
decomposing $\xi(s, \mu)$ on the basis of Legendre polynomials $P_{\ell}$ as:
\begin{equation}
{\xi}_{\ell} (s) = {(2 \ell + 1)} \sum_{\rm i} \xi(s, \mu_{\rm i}) P_{\ell} (\mu_{\rm i}) {\rm \Delta}\mu. 
\label{eq_cf_legendre}
\end{equation}
In the previous expression, only 
even multipoles do not vanish,
and the correlation function is binned according to the 
absolute value of $\mu$. 
In our analyses, we only consider the $\ell=0,2, 4$ moments, namely
monopole, quadrupole, and hexadecapole (whenever specified), and  $\xi(s, \mu)$ is 
quantified with the classical \citet{Landy1993} estimator.
The pair counts are binned into $5h^{-1}{\rm Mpc}$ bins in separation and $0.01$ in $\mu$.
In the panels of Figure \ref{fig_lrg_team_1}, the BAO feature is clearly seen at $s \simeq 100 h^{-1} {\rm Mpc}$, as well as the impact  
of the reconstruction procedure: the BAO feature appears in fact much sharper in the bottom panels.
The various fits shown in the figure are obtained with the BAO technique adopted in \citet{LRG_corr2020}
for the analysis of the correlation function in configuration space, and 
 correlation function multipoles $\xi_{\ell}(s)$ are rendered
as a function of separations $s$ relevant for BAO
($30 \le s \le 180 h^{-1}{\rm Mpc}$), starting from the modeling of the
redshift-space anisotropic power spectrum.
In particular, 
as explained in \citet{LRG_corr2020}, the  
non-linear broadening of the BAO peak is modeled by multiplying the ``peak-only'' power spectrum $P_{\rm
peak}$  by a Gaussian term with $\Sigma_{\rm nl}^2(\mu) =
\Sigma_\parallel^2 \mu^2 + \Sigma^2_\perp(1-\mu^2)$, with 
$\Sigma_\parallel$ and $\Sigma_\perp$ the BAO damping terms, and
the non-linear
random motions on small scales are rendered with a Lorentzian term
parametrized by $\Sigma_s$.  When performing fits to the multipoles of a  single realization of the survey, the values of 
$(\Sigma_\parallel, \Sigma_\perp, \Sigma_s)$ are maintained fixed
to improve convergence.  
Moreover, 
the BAO peak position is parameterized via two dilation
parameters that scale separations into transverse ($\alpha_{\perp}$)
and radial ($\alpha_{\parallel}$) directions.
These quantities are related 
 to the comoving angular diameter distance $D_{\rm M}(z) = (1+z)D_{\rm A}(z)$
and to the Hubble distance $D_{\rm H}(z)=c/H(z)$ as:
\begin{equation} 
\alpha_{\perp} = {D_{\rm M}(z_{\rm eff})/r_{\rm drag}  \over D_{\rm M}^{\rm fid}(z_{\rm eff})/r_{\rm drag}^{\rm fid} },
\label{eq_bao_1}
\end{equation}
\begin{equation} 
\alpha_{\parallel} = {D_{\rm H}(z_{\rm eff})/r_{\rm drag}  \over D_{\rm H}^{\rm fid}(z_{\rm eff})/r_{\rm drag}^{\rm fid} },
\label{eq_bao_2}
\end{equation}
with $r_{\rm drag}$ the comoving horizon scale at the drag epoch. 
Fits on mock multipoles are performed -- including hexadecapole, which however does not add extra information.
In the procedure, BAO broadband parameters are let free while 
both dilation parameters are allowed to vary between 0.5 and 1.5. A
total of 9 parameters are fitted simultaneously. 
Table \ref{table_lrg_team_1} contains the results of such BAO fits in configuration space, where in particular $b$ is the linear bias
and  $\beta=f/b$ is the RSD parameter. 
The covariance matrix used for the fit is obtained from 1000 \textsc{EZmocks}, properly rescaled by 
the difference in particle number to match the characteristics of the OR-based mocks. 
Note that expected statistical errors in the eBOSS LRG data sample 
are of the order of $\sim1.9\%$ for $\alpha_{\perp}$ and $\sim2.6\%$ for $\alpha_{\parallel}$,
and that reconstruction improves constraints on $\alpha_{\perp}$ and $\alpha_{\parallel}$, as expected.  


As mentioned in Section \ref{sec_tools_methodology_lrgs}, the OR-based mocks have not been designed to
reproduce all of the features of the eBOSS DR16 LRG sample, since that level of complexity is not
necessary for our subsequent analysis. Hence, we clearly do not expect to exactly match the clustering properties of the eBOSS LRG sample.
Nevertheless,  as an  instructive  example, we show in Figure \ref{fig_lrg_team_1_bis}
a comparison between eBOSS measurements and those made from Th2 OR mocks at $z=0.695$.
Specifically,  green triangles in the figure display eBOSS plus CMASS measurements for the NGC, 
pink squares are used for eBOSS plus CMASS SGC measurements, and yellow circles
show results obtained from the combined samples. 
Open symbols refer to pre-reconstruction calculations (top panel), while
filled symbols indicate post-reconstruction results (bottom panel). 
Errorbars are estimated from analogous measurements performed on 1000 EZmocks, and via averaging.
The solid black line in both panels represents the redshift-space 2PCF monopole expectation assuming the 
Tinker HOD recipe, derived from 27 Th2 OR mocks.  Note that, besides 
not tuning the various HOD parameters to reproduce the exact clustering properties of the eBOSS DR16 LRG sample,
our OR-based realizations are cubic mocks, and do not contain 
the complications of masks, cut sky, and observational artifacts (as opposed to the  EZmocks and \textsc{Nseries}) -- which would be instead 
more appropriate for a fair data-mock comparison.

 

\begin{table}
\centering
\caption{
BAO fits to the average pre- and post-reconstructed power spectra for different HOD prescriptions over 27 corresponding Outer Rim mock realizations, at $z = 0.695$, with the `Th2' 
flavor -- as displayed in Figure \ref{fig_lrg_team_2}.}
\doublerulesep2.0pt
\renewcommand\arraystretch{1.5}
\begin{tabular}{ccccccc} 
\hline \hline  
{\bf PS [Th2]} & Leauthaud &   Tinker &  Hearin\\
\hline
&  {\it BAO} & {\it Pre-Rec} &   \\
\hline
$\alpha_{\perp}$ &  $1.0028 \pm 0.0107$ & $0.9953 \pm 0.0124$ & $1.0108 \pm 0.0099$ \\
$\alpha_{\parallel}$  & $0.9885 \pm 0.0178$ & $1.0023 \pm 0.0177$ & $0.9779 \pm 0.0143$ \\
$\Sigma_{\perp}$ &  $4.2992 \pm 1.8875$ & $5.9568 \pm 2.0364$ & $4.1865 \pm 1.8043$ \\
$\Sigma_{\parallel}$ & $9.6867 \pm 1.8250$ & $7.5044 \pm 2.2301$ & $7.0045 \pm 2.0438$ \\
$\chi^2$ &$44.3$ & $34.5$&  $57.8$\\
\hline
&  {\it BAO} & {\it Post-Rec} &   \\
\hline
$\alpha_{\perp}$ &  $0.9976 \pm 0.0075$ & $0.9975 \pm 0.0088$ & $1.0122 \pm 0.0075$ \\
$\alpha_{\parallel}$  & $0.9938 \pm 0.0104$ & $0.9976 \pm 0.0121$ & $1.0002 \pm 0.0113$ \\
$\Sigma_{\rm rec}$ & $15.0 \pm 1.5$ & $15.0 \pm 1.5$ & $15.0 \pm 1.5$  \\
$\Sigma_{\perp}$ &  $1.0585 \pm 0.7909$ & $1.3377 \pm 0.9904$ &  $1.1071 \pm 0.8221$ \\
$\Sigma_{\parallel}$ & $1.5451 \pm 1.1110$  & $1.9511 \pm 1.3408$ & $2.5084 \pm 1.5258$ \\
$\chi^2$ & $94.4$& $69.0$& $120.1$ \\
\hline
\hline
\label{table_lrg_team_2}
\end{tabular}
\end{table} 


Figure \ref{fig_lrg_team_2} shows examples of
redshift-space galaxy power spectra as computed  from the average of 27 OR-based mocks, each set being characterized by a different HOD scheme,
at $z=0.695$ for the `Th2' flavor. The plot represents the analogous, in Fourier space, of the previous correlation function estimates in configuration space. 
The pre- (top panels) and post-reconstructed (bottom panels) monopoles and quadrupoles of the power spectra
are shown, for the Leauthaud, Tinker, and Hearin models -- from left to right, respectively. 
Fits are obtained with the BAO theoretical model of \citet{Gil-Marin2020}, considering wave numbers between $0.02 \le k [h {\rm Mpc}^{-1}] \le 0.30$,
and the corresponding results are reported in Table \ref{table_lrg_team_2}.
Unlike correlation function calculations, Discrete Fourier Transform (DFT) methods 
used to compute the power spectrum multipoles are quite sensitive to the assumption of 
periodic boundary conditions, and therefore a 
procedure denoted as `padding' is applied in this process, to mitigate non-periodicity effects.  
The detailed effects of non-periodicity 
on BAO measurements are discussed in \citet{Gil-Marin2020}. In particular, 
results of such analyses show that no significant changes are
observed in terms of $\alpha_\perp$, while shifts at the level of $2-3\%$
can be systematically seen in $\alpha_\parallel$ if padding is not applied. Hence, non-periodic effects are relevant
in determining  $\alpha_\parallel$, but they do not impact significantly  $\alpha_\perp$. 
Moreover, no relative shifts in any of the $\alpha$ parameters are seen when the HOD model or flavor is varied -- as we show next.

 


\begin{figure*}
\includegraphics[angle=0,width=0.4973\textwidth]{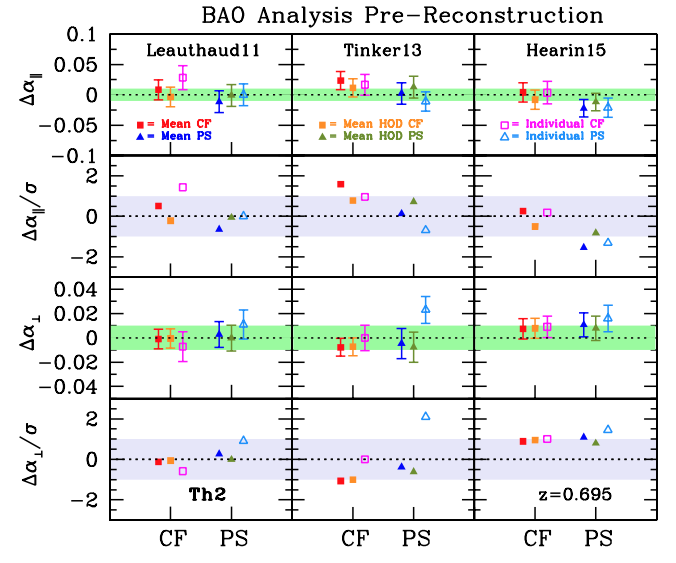}
\includegraphics[angle=0,width=0.4973\textwidth]{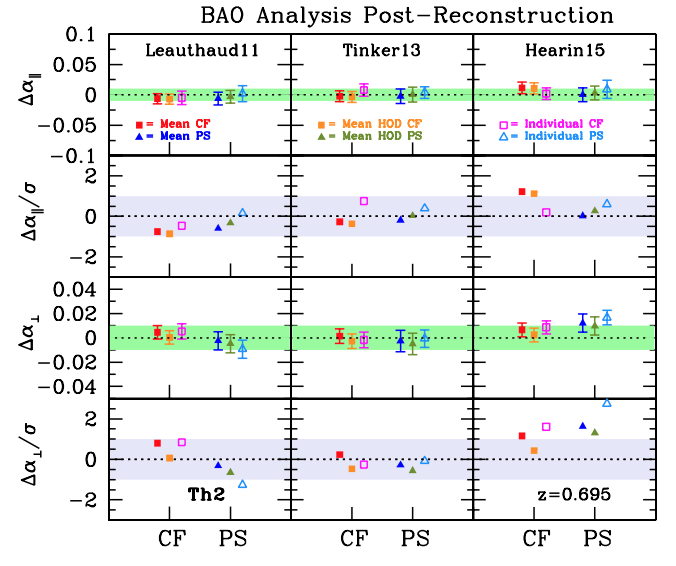} 
\caption{Performance of the BAO fitting methods in configuration and Fourier space, with respect of variations in the underlying HOD model.
Fits to individual mock realizations as well as to the mean 
of a set of 27 independent realizations 
of the OR mocks  (from the `Challenge Set 1') are carried out (see Table \ref{table_lrg_team_3}). 
The difference between the measured 
$\alpha_\parallel$ and $\alpha_\perp$ from 
pre- (left panels) and post-reconstructed (right panels) catalogs are displayed with different symbols and colors, as indicated in the figure. 
This BAO fitting methodology is adopted in the analysis of the final eBOSS LRG sample.
Shaded dark-green areas in the figure highlight the $1\%$ error level, 
while horizontal grey bands display $1\sigma$ error levels.
Overall, we do not detect any significant systematics due to different HOD prescriptions:
after application of the BAO reconstruction
procedure, all of the mean measurements are within $0.5-1.2\%$ of their expected values,
thus below the statistical precision of the eBOSS LRG sample.
See the main text for more details.}
\label{fig_lrg_team_3a}
\end{figure*}


Figure \ref{fig_lrg_team_3a} summarizes and confronts
the performance of the BAO fitting technique 
adopted in the analysis of the final eBOSS LRG sample, in configuration and Fourier space, with respect of variations in the underlying HOD model.
For each mock realization, at a fixed HOD scheme and threshold flavor, 
the correlation function and power spectrum are computed along with their multipoles, respectively.
Subsequently, fits for the BAO peak position are performed -- both to the pre- and post-reconstructed synthetic catalogs -- 
to determine the dilation parameters $\alpha_{\parallel}$
and $\alpha_{\perp}$ and their corresponding errors.
The expected values for the dilation parameters are computed in the OR cosmology, at the
effective redshift $\bar{z} =0.695$.
In addition, fits to the  average multipoles of a given set 
of mocks (27 mocks per set) are also carried out, to probe biases in a very high precision configuration.
In particular, the BAO pipeline on the power spectrum monopole and quadrupole is run
in the interval $0.02 \le k [h {\rm Mpc}^{-1}] \le 0.30$. 
The various BAO  fittings are performed by fixing the BAO damping
parameters ($\Sigma_\parallel, \Sigma_\perp$) at their best-fitting
values on the mean of the pre- and post-reconstructed mocks, and
the analysis is done in terms of the scaling parameters $\alpha_\perp$ and $\alpha_\parallel$.
The various covariances used in the analysis of the OR-based mocks
are derived from the \textsc{Ezmocks}, properly rescaled to account for differences in number density
(see Table \ref{table_number_density_unblind_set}).
As explained in detail in \citet{LRG_corr2020}, the final BAO model is
a combination of the cosmological multipoles 
$\xi_{\ell}$ and a smooth function of separation, which accounts for 
unknown systematic effects in the survey that can potentially contaminate the results.

Table \ref{table_lrg_team_3} contains the results when fitting the mean correlation functions and power spectra 
(rows labeled `Mean') of the 27 independent realizations 
of the OR-based mocks with different HOD prescriptions, as well as   
the mean of the fits of individual realizations (rows labeled `Individual').
Those data are shown in  
Figure \ref{fig_lrg_team_3a}, where
each sub-panel displays 
the difference between the measured 
$\alpha_\parallel$ and $\alpha_\perp$: 
their expected value are inferred for the pre- (left panels) and post-reconstructed (right panels) catalogs. 
Mean estimates are displayed  in red with filled rectangles  as derived from configuration space techniques,
and in blue with filled triangles as determined with Fourier space techniques; in the figure, these points are indicated as `MEAN CF' or `MEAN PS', respectively. 
The associated errors are consistently the errors of the mean, 
obtained by rescaling the related \textsc{EZmocks} covariance by the number of realizations, $N_{\rm OR}=27$. 
Therefore, these errors are a factor of $\sqrt{N_{\rm OR}}$ smaller than the error one would obtain for a single realization of these mocks.  
In the figure, the dark-orange filled rectangles (related to configuration space measurements, and denoted `MEAN HOD CF') and light-green filled triangles 
(related to Fourier space measurements, and termed `MEAN HOD PS') 
show the same mean estimates but now normalized by the corresponding mean expectation value averaged over all the 3 HODs: 
hence, by construction, the sum of those points for a given method (CF or PS) will be zero. In this way, one can  better disentangle
the systematics 
introduced by the HOD modeling  versus the theoretical systematics related to BAO fitting methodologies.
Analogous empty symbols are used to display the corresponding individual 
measurements for pre-and post-reconstruction catalogs, respectively.
In this case the error associated is the root mean square ({\it rms}) of all the 
individual fits, 
scaled by the square root of the number of realizations ($\sqrt{N_{\rm OR}}$).
The shaded dark-green areas represent the $1\%$ error level on the $\alpha$ parameters. 
The second and fourth sub-panels in Figure \ref{fig_lrg_team_3a}  show the difference between the measured and the expected 
values of $\alpha_\parallel$ and $\alpha_\perp$ in terms of number of statistical $\sigma$ 
of the error of the mean, and the ${ rms}/\sqrt{N_{\rm OR}}$. The
horizontal grey bands highlight the $1\sigma$ error level.
In general, considering fits to the mean over 27 realizations, 
the reported dilation parameters for all the different HODs are
consistent with their expected values within $ 0.8\%$
for $\alpha_{\perp}$
and  $1.2\%$ for $\alpha_{\parallel}$. 
Recall that the expected statistical errors in the eBOSS LRG data sample 
are of the order of $\sim 1.9\%$ for $\alpha_{\perp}$ and $\sim 2.6\%$ for $\alpha_{\parallel}$.
From the $N$-body mocks we do not observe any significant BAO peak position shift with respect to their corresponding 
expected value in any of the post-reconstructed catalogs analyzed. The BAO pipeline performs well with different HOD models;
some fluctuations are present, but their values lie always below the $\pm2\sigma$ limit, hence the shifts are not significant.  
Overall, we do not detect any relative systematics due to different HOD modeling, although 
the statistical precision of the Outer Rim-based mocks is comparable to the statistical precision of the LRG sample (see Section \ref{sec_global_sys}). 
Interestingly, from Figure \ref{fig_lrg_team_3a} it is evident that the reconstruction procedure (right panels) generally
ameliorates the agreements of the $\alpha$-parameters with their expected values. Moreover,
it is also worth noticing that for the average values most of the detected discrepancy (after reconstruction) arises from the Hearin HOD model;
this is not unexpected,  since 
we have considered a quite extreme case of assembly bias both in the central and 
satellite galaxy population -- as explained in Section \ref{sec_theory}.
Finally, after application of the BAO reconstruction
procedure, we find all of the mean measurements to be within $0.5-1.2\%$ of their expected values, thus
below the statistical precision of the eBOSS LRG sample. Hence, for BAO-only fitting methods, 
both modeling and HOD systematics
are subdominant to the global systematic error budget and the BAO analysis is unbiased.
However, sub-percent level corrections may become relevant for
future surveys like DESI, that are expected to achieve sub-percent statistical precision on the galaxy sample. 


\subsection{Galaxy Mock Challenge: RSD Analysis and HOD Systematics} \label{sec_rsd_team_analysis}
 

\begin{figure*}
\includegraphics[angle=0,width=1.0\textwidth]{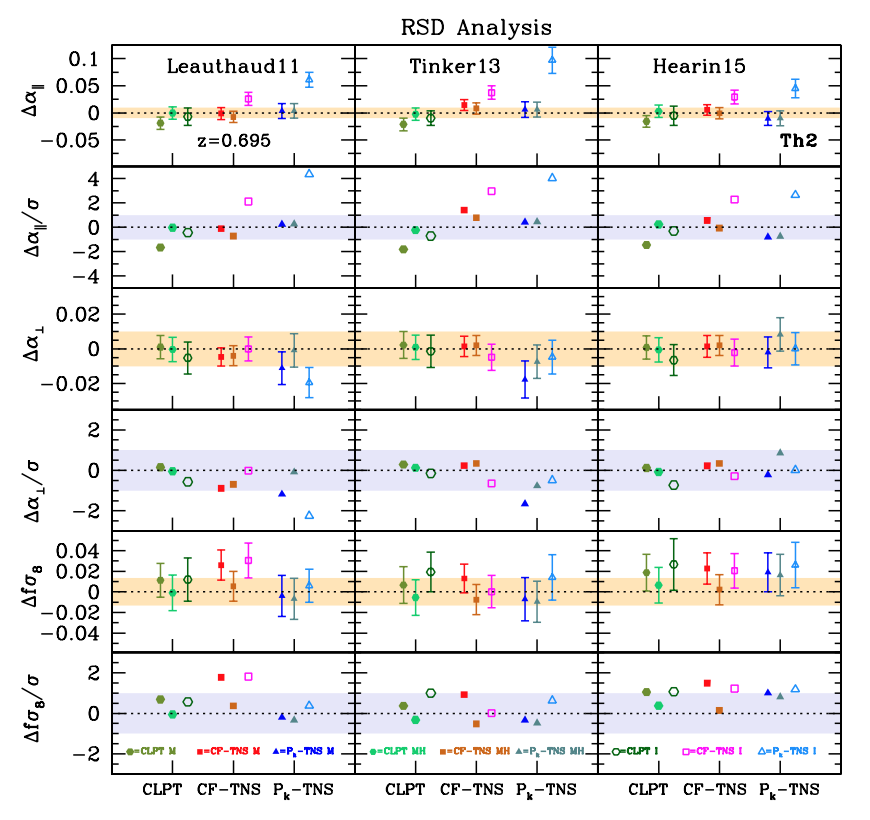}
\caption{Full shape RSD analysis: main results on the AP parameters 
and the growth of structure. 
The three techniques adopted for the
analysis of the final eBOSS LRG sample, two in configuration space (CLPT-GS and CF-TNS) and one in Fourier space (P$_{k}$-TNS), 
are confronted on a series of Outer Rim mocks having different HOD prescriptions.
From left to right, the Leauthaud, Tinker, and Hearin models corresponding to
`Th2' are analyzed, as they are closer to the characteristics of the eBOSS LRG sample.
Individual fits on each of the 27 realizations per model are performed (open points), as well as fits on the mean of the mocks (filled points),
allowing one to obtain  accurate estimates of $\alpha_{\parallel}, \alpha_{\perp}$, and $f \sigma_8$. 
Scatter plots in terms of $\sigma$-deviations are also shown. 
Shaded orange areas represent the $1\%$ error levels on the $\alpha$ parameters and the 
$3\%$ error on $f \sigma_8$, while shaded grey areas  highlight the $1\sigma$ error level.
The corresponding numerical results are reported in Table \ref{table_lrg_team_4}. Errorbars follow the same conventions as in Figure \ref{fig_lrg_team_3a}.
See the main text for more details.}
\label{fig_team_4}
\end{figure*}



In Section \ref{sec_analysis_methods}, we have briefly described the three RSD theoretical models
adopted for the analysis of the final eBOSS DR16 LRG sample. Here, we
confront those models and show that they are 
mutually consistent, with comparable systematic errors on the AP parameters 
and the growth of structure -- as well as robust to 
different HOD prescriptions. While
for the previous BAO-only analysis simply the BAO peak position has been
taken into account, here we consider a 
full modeling of
the shape and amplitude of the correlation function and power spectrum multipoles, including nonlinear DM effects,
galaxy bias, and RSDs. We generically refer to this methodology as the `full shape' (FS) analysis.
Quantitative investigations involving the correlation function or power spectrum FS
require high fidelity $N$-body-based mocks  
to test and validate the underlying RSD models, and typically such
analyses are only performed over pre-reconstructed synthetic catalogs.
The overall aim is to quantify the impact of the different HOD prescriptions 
used to populate simulated halos with galaxies on RSD constraints.
Our primary focus here is thus on modeling and HOD systematics, while
 \citet{LRG_corr2020}, \citet{Gil-Marin2020}, and \citet{Icaza-Lizaola2020}
also examined the impact of the choice of scales in the fits and the choice of a fiducial cosmology. 
In particular, their conclusions (directly relevant for the current work)
suggest that the most robust results and optimal configuration for the FS analysis of the correlation function
are obtained with monopole and quadrupole in the range $20 \le s {\rm [} h^{-1}{\rm Mpc} {\rm ]} \le 130$,
and hexadecapole in the interval $25 \le s {\rm [} h^{-1}{\rm Mpc} {\rm ]} \le 130$ for the TNS model in configuration space, 
while using the interval $25 \le s {\rm [} h^{-1}{\rm Mpc} {\rm ]} \le 130$ for all the moments when considering CLPT-GS.
Regarding power spectrum computations, the optimal range of scales 
are $0.02 \le k [h/{\rm Mpc}] \le 0.15$, and results include the hexadecapole.  
In what follows, the analysis is carried out in the Outer Rim fiducial cosmology,  and
the monopole, quadrupole, and hexadecapole ranges are those previously specified -- always set for optimal 
performance.  Moreover, for the analysis of the challenge mocks in Fourier space, a procedure called 
padding is applied, in order to prevent the impact of non-periodicity 
to affect results when applying the discrete Fourier transform. 
Also, the mock covariances adopted in these investigations are 
derived from a set of 1000 \textsc{EZmocks}, and properly rescaled by the difference
in particle number. 

 

Figure \ref{fig_team_4} summarizes the main results of the RSD FS analysis,
confronting the three different modeling techniques: 
2 in configuration space (CLPT-GS and CF-TNS), and one
in Fourier space (P$_{\rm k}$-TNS).
Specifically,  we analyzed    
the Leauthaud, Tinker, and Hearin HODs (from left to right in the plots) corresponding to
`Th2', both in configuration and Fourier space -- closer to the characteristics of the eBOSS LRG sample. 
Individual fits on each of the 27 realizations per model are performed (i.e., open symbols in the panels,
denoted with the letter `I' following the specific model), 
as well as fits on the mean of the mocks
(i.e., filled symbols indicated with the letter `M' in the panels, adopting similar conventions).
Results are also reported in Table \ref{table_lrg_team_4}. 
In detail,  circles refer to CLPT-GS measurements, squares are used for the CF-TNS model, and
triangles indicate results from the P$_{\rm k}$-TNS technique. Moreover, open symbols are related to individual fits, 
while filled symbols display fits on the mean. In the latter case, similarly to what has been done in Figure \ref{fig_lrg_team_3a}, we
also show  fits on the mean
 normalized by the corresponding mean expectation value averaged over all the 3 HODs:
such points are denoted as `MH' (or `Mean HOD'), and are useful to
separate the 
systematics  introduced by the HOD modeling versus the theoretical systematics related to individual RSD FS fitting methodologies.
In the figure, shaded orange areas represent the $1\%$ error levels on the $\alpha$ parameters and the 
$3\%$ error on $f \sigma_8$, while shaded grey areas  highlight the $1\sigma$ error level.

Overall, from the fit of the mean, biases observed for CLPT-GS and CF-TNS are mostly within $1.5\sigma$ away
from the expected values for all the parameters across all models (and always less than $2\sigma$ away),
and errors estimated from the different RSD models
are mutually compatible.
In general, CF-TNS seems to imply slightly larger errors than CLPT-GS.  
Comparing fits on the mean with the mean of individual fits, for $\alpha_\perp$ and $f \sigma_8$
there is good agreement in values and errors. Also, for $\alpha_\parallel$, the fit of the mean is a more robust estimate of potential biases.
The most significant differences are found for the  best fits of $\alpha_\parallel$, but this comes with no surprise:
it is in fact expected that fits on individual mocks
are dominated by the low signal to noise, as the effective volume of a single OR mock is $1.10~{\rm Gpc}^3$ -- thus relatively small.
Namely, given the low volume spanned by OR mocks, the fit to the mean and the mean of the fits differ, primarily because
 individual subcubes are dominated by noise and there are not enough subcubes for this effect to effectively cancel out.
A good proof of this fact is provided by Figure \ref{fig_lrg_team_3a}: when reconstruction is applied to the mocks
 (i.e., signal-to-noise increased), the differences between fits to the mean and mean of the fits shrink -- 
 as can be inferred by comparing the left and right panels of Figure \ref{fig_lrg_team_3a}. Moreover, 
 the $\sim 4\sigma$ discrepancy shown in 
 Figure \ref{fig_team_4} from individual fits to the Tinker model 
 happens only for $P(k)$ measurements, and affects primarily $\alpha_\parallel$ --
 the variable with the lowest signal (hence providing noisy spectra), as it depends only on the radial direction. 
 In such case,  very likely the fitting procedure tends to 
(incorrectly) identify the BAO signal in those noisier regions of the spectra; however, 
 this does not happen for configuration space analyses, 
 since the BAO signal is localized.  In other terms, 
 this is a clear example of inadequate BAO detections in individual fits.
Finally,  another source of discrepancy could be due to the rescaling for the EZmock covariance adopted in the analysis, that 
may not provide a full accurate description and causes a much larger variation on individual mocks.
 
In summary, from the results of fitting the mean, since we do not observe biases larger than 2$\sigma$, we conclude that different HODs do not have
a significant impact in the fits even from a FS RSD analysis. This type of systematics is always below the statistical error of the LRG sample.
It is also quite interesting to notice the
remarkable consistency and (nontrivial) 
agreement between RSD FS techniques in configuration and Fourier space.
In general, as demonstrated by Figure \ref{fig_team_4}, the HOD systematics derived from fits on the mean
are within the $\sim 1\%$ level, even  
smaller than the modeling systematics, and always below the statistical precision of the eBOSS LRG sample.
The modeling systematics instead could reach the percentage level particularly in $\alpha_\parallel$ and $f\sigma_8$,
and could represent a dominant source of systematics.  
From a more extensive FS analysis, we thus 
conclude that 
while HOD systematics are 
subdominant to the global systematic error budget,
the modeling systematics should be taken into account -- although both are
 below the statistical precision of the eBOSS LRG sample.
Moreover, from the FS study, we conclude that the
different methodologies  adopted for the analysis 
of the final eBOSS LRG sample are mutually consistent and robust,
thus validating the clustering analysis pipelines.


\subsection{Galaxy Mock Challenge: Modeling Systematics} \label{sec_systematics_analysis}

\begin{figure}
\includegraphics[angle=0,width=0.48\textwidth]{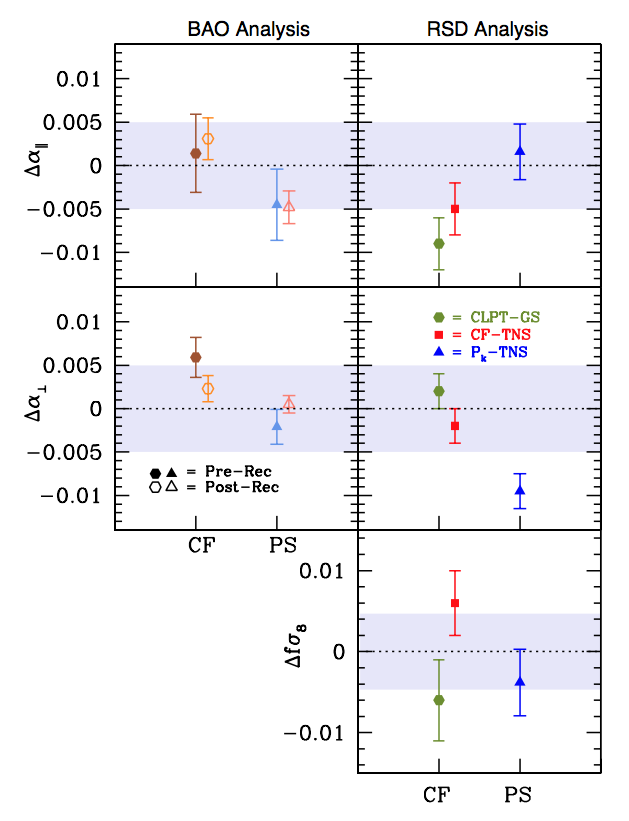}
\caption{Comparing modeling systematics in BAO and RSD methods, 
estimated from 84 \textsc{Nseries} mocks.
Left panels  show the 
AP parameters 
derived from BAO-only fits
in configuration and Fourier space, respectively.
Right panels refer to RSD full shape analyses. 
Filled symbols are used for
pre-reconstructed catalogs, while 
open symbols refer to
post-reconstructed catalogs.
Gray areas in the figure highlight the $0.5\%$ error
level on $\alpha_\parallel$ and $\alpha_\perp$, and the
$1.0\%$ error on $f\sigma_8$. These numerical results are 
reported in Table \ref{table_lrg_team_5}. 
All the different methods adopted for the clustering analysis of the eBOSS LRG sample are 
mutually consistent, showing a remarkable accuracy in recovering the
expected cosmological parameters at an exquisite level of precision.}
\label{fig_team_5}
\end{figure}

The heterogeneous set of 
Outer Rim mocks previously adopted for assessing 
possible systematic effects in the galaxy-halo connection 
and imprecisions related to the galaxy clustering
modeling (i.e., impact of HODs on BAO and RSD methods, and RSD modeling systematics)
 is sub-optimal in accuracy at the sub-percent level, although this type of accuracy is well-below the statistical sensitivity of the eBOSS LRG sample.
 This is  mainly because of the relatively 
small effective volume spanned by each individual independent mock, 
due to the limitations posed by having only a single Outer Rim halo catalog at $z=0.695$ combined with the constraints intrinsic to the LRG modeling (see Section \ref{sec_theory}).
For this reason, only 27 mocks were used in the previous analyses, as fully independent realizations (i.e. not sharing the same DM field) are required to properly assess cosmic variance.
In terms of errorbars, the resolution limit of the OR mocks is in fact around the $\sim 1-2\%$ level. 
In order to  evaluate the performance of the BAO and RSD modeling at a sub-percent level, a more
suitable choice is to abandon a single simulation -- although of exquisite mass-resolution such as the Outer Rim --  
and opt instead for multiple realizations of the same box (i.e., identical initial conditions in all but the random seeds) at
a lower mass-resolution and with a larger effective volume.\footnote{Another alternative would be to pursue instead  a subhalo-type modeling
approach, rather than the more traditional HOD framework, but we do not have access to full merger trees from the Outer Rim simulation.}
This is the logic beyond the \textsc{Nseries}, 
a small homogeneous  set of 84 pseudo-independent  mocks 
constructed from 7 independent  periodic boxes of $2.6h^{-1}{\rm Gpc}$ side, projected through 12 
different orientations and cuts per box; the mass resolution of these periodic boxes is 
 much lower than that of the Outer Rim run, but it is still sufficient 
for resolving LRG-type halos ($1.5 \times 10^{11} h^{-1}M_{\odot}$, with $2048^3$ particles per box). The global
effective volume spanned is $84 \times 3.67 ~{\rm [Gpc]}^3$. 
The \textsc{Nseries} are characterized by 
the same underlying galaxy bias model built upon 
the same cosmology,
but each mock is a quasi-independent realization -- thus not sharing exactly the same LSS --
and including observational artifacts closer to the eBOSS DR16 sample, with similar angular and radial selection function
of the observed sample. The HOD used is targeted to  BOSS CMASS galaxies, at an effective redshift of $\bar{z}=0.56$.
Although this set was originally devised for BOSS galaxies, it is still useful for
evaluating modeling systematics at the sub-percentage level also for eBOSS tracers. 
This is why the \textsc{Nseries} is extensively used in \citet{LRG_corr2020} and in \citet{Gil-Marin2020}
for addressing the modeling systematics related to each complementary analysis in configuration and Fourier space, respectively.
Here, we show only an interesting combined example, confronting the
fitting methodologies in configuration and Fourier space and the performance of the RSD
models previously introduced in Section \ref{sec_analysis_methods}.

Figure \ref{fig_team_5} provides a summary result
obtained by running the various BAO and RSD FS analysis pipelines on
the average of 84 \textsc{Nseries} mocks. 
Specifically, the left panels display the 
AP parameters 
derived from a BAO-only fit 
in configuration and Fourier space, respectively.
Filled symbols are used for
pre-reconstructed catalogs, while 
open symbols refer to
post-reconstructed catalogs.
The left panels show analogous quantities, as well as
the growth of structure in terms of $f \sigma_8$, 
derived from  RSD FS fits. 
In this case the analysis is performed only over pre-reconstructed catalogs, and 
carried out in the \textsc{Nseries} cosmology (see Section \ref{subsec_Nseries}).
Results from the three different RSD models -- two in configuration space (CLPT-GS and CF-TNS)
and one in Fourier space (P$_{k}$-TNS), all set for optimal performance as explained before, including the hexadecapole -- are
displayed with filled symbols. 
The corresponding numerical values are reported in Table \ref{table_lrg_team_5}. 
The gray areas in the figure highlight the $0.5\%$ error
level for the AP parameters,
and the $1.0\%$ error
level for $f\sigma_8$.
As clearly seen, all the different methods are 
mutually consistent, showing a remarkable accuracy in recovering the
expected cosmological parameters ($\alpha_{\parallel}, \alpha_{\perp}, f\sigma_8$)
at an exquisite level of precision, within at worst $0.9\%$ of their expected values for the 
$\alpha$'s and within $1.5\%$ for $f\sigma_8$ -- as a conservative estimate.
These results can be compared with 
analogous measurements performed on the Outer Rim mocks
displayed in Figures  \ref{fig_lrg_team_3a} and \ref{fig_team_4}.
Although here at a sub-percentage level precision, results
from the two different sets of mocks are 
consistent: we observe a similar trend at a fixed HOD recipe, indicating an
impressive level of consistency between techniques in 
configuration and Fourier space.
Clearly, the modeling  systematics is addressed here with higher accuracy, showing deviations at the sub-percent level 
for the AP parameters and $f\sigma_8$. 
While this type of systematics may be a dominant source of error in the global 
systematic budget (despite sub-percent deviations),
the   LRG sample is 
primarily dominated by the statistical error of the data.   




\section{Systematic Error Budget} \label{sec_global_sys}


\begin{table}
\caption{Global error budget for the final eBOSS DR16 LRG sample, as derived from configuration and Fourier space analyses.}
\begin{center}
\begin{tabular}{|c|c|c|c|c|}
\hline
\hline
RSD-FS Analysis & Global Syst. \\
\hline
Error Type & Model &  $\sigma_{\alpha_\parallel}$ &  $\sigma_{\alpha_\perp}$  &  $\sigma_{\rm f \sigma_8}$   \\
\hline
\hline
RSD Modeling        						&  CLPT-GS  & 0.0090 & 0.0040 & 0.0100  \\
$[\sigma_{\rm syst}^{\rm model, \textsc{NS}}]$	&  CF-TNS    & 0.0060 & 0.0040 & 0.0080  \\
									&  P$_{\rm k}$-TNS & 0.0064 & 0.0095 & 0.0082  \\
\hline
RSD Additional         			   &  CLPT-GS  &  0.0156 & 0.0127 & 0.0220  \\
$[\sigma_{\rm syst}^{\rm other}]$ &  CF-TNS   &  0.0153 &  0.0112 & 0.0216  \\
						   &  P$_{\rm k}$-TNS &  0.0117 & 0.0068 & 0.0155  \\
\hline
RSD Systematics      &  CLPT-GS  &  0.0180 &   0.0133&   0.0242  \\
$[\sigma_{\rm syst}]$ &  CF-TNS   & 0.0164 &   0.0119 &   0.0230   \\
				&  P$_{\rm k}$-TNS &  0.0133 &   0.0117 &   0.0175   \\

\hline
RSD Statistical          &  CLPT-GS  &  0.0280 &  0.0200 & 0.0450  \\
$[\sigma_{\rm stat}]$ 	&  CF-TNS   &  0.0310 &  0.0180 & 0.0400  \\
				&  P$_{\rm k}$-TNS &  0.0360 & 0.0270 & 0.0420  \\
				
\hline
RSD Total          	&  CLPT-GS  &  0.0333 &  0.0240 &   0.0511   \\
$[\sigma_{\rm tot}]$ &  CF-TNS   &  0.0351 &    0.0216 &    0.0462  \\
				&  P$_{\rm k}$-TNS &  0.0384 &    0.0294&   0.0455  \\
\hline
\hline
 		            											&  CLPT-GS  &  0.3214 &   0.2000 &   0.2222   \\
$\sigma_{\rm syst}^{\rm model, \textsc{NS}}$/$\sigma_{\rm stat}$ &  CF-TNS   &  0.1935 &   0.2222 &   0.2000  \\
														&  P$_{\rm k}$-TNS &  0.1778  &  0.3518 &   0.1952   \\	
\hline
 		            											&  CLPT-GS  & 0.5571 &   0.6350 &   0.4889   \\
$\sigma_{\rm syst}^{\rm other}$/$\sigma_{\rm stat}$ &  CF-TNS   &  0.4935 &   0.6222 &   0.5400 \\
														&  P$_{\rm k}$-TNS &  0.3250 &   0.2518 &   0.3690  \\
														\hline
 		            				      &  CLPT-GS  &  0.6432 &   0.6658 &    0.5370   \\
$\sigma_{\rm syst}$/$\sigma_{\rm stat}$ &  CF-TNS   &  0.5301 &   0.6607 &  0.5758  \\
							     &  P$_{\rm k}$-TNS &  0.3704 &   0.4327 &   0.4175   \\
\hline
 		            											&  CLPT-GS  &  0.2703  &   0.1665 &   0.1958   \\
$\sigma_{\rm syst}^{\rm model, \textsc{NS}}$/$\sigma_{\rm tot}$ &  CF-TNS   &  0.1710  & 0.1854  &  0.1733  \\
														&  P$_{\rm k}$-TNS &  0.1667 &   0.3229 &   0.1802   \\	
\hline
 		            											&  CLPT-GS  &  0.4686 &    0.5286 &   0.4307   \\
$\sigma_{\rm syst}^{\rm other}$/$\sigma_{\rm tot}$ &  CF-TNS   &  0.4361 &  0.5191 &   0.4679   \\
														&  P$_{\rm k}$-TNS &  0.3048 &   0.2311 &   0.3406   \\
\hline
 		            				      &  CLPT-GS  &  0.5410  &   0.5542 &   0.4731   \\
$\sigma_{\rm syst}$/$\sigma_{\rm tot}$ &  CF-TNS   & 0.4684 &   0.5513 &   0.4990   \\
							     &  P$_{\rm k}$-TNS &  0.3474 &    0.3971 &   0.3853  \\	
\hline
 		            				      &  CLPT-GS  &  0.8410 &   0.8324 &   0.8810  \\
$\sigma_{\rm stat}$/$\sigma_{\rm tot}$ &  CF-TNS   &  0.8835 &   0.8343 &   0.8666    \\
							     &  P$_{\rm k}$-TNS &  0.9378  & 0.9178 &  0.9228   \\
\hline
\hline		
\end{tabular}
\end{center}
\label{table_6_systematics}
\end{table}


Finally, we address here the LRG global error budget with a major focus on theoretical systematics, and
also summarize the previous mock challenge results in term of 
biases in the estimation of $\alpha_\parallel$, $\alpha_\perp$, and $f\sigma_8$.


\subsection{Global Error Budget}

In our companion papers \citet{LRG_corr2020} and \citet{Gil-Marin2020}, besides
modeling and HOD imperfections, detailed investigations regarding
the impact of a fiducial cosmology, the optimal fitting range of scales, effects on non-periodicity, and
observational artifacts such as redshift failures, completeness, close-pair collisions, and radial integral constraint \citep{DeMattia2020}
are carried out in configuration and Fourier space, respectively, and the associated errors are carefully quantified using all the available types of mocks. 
In the following, we indicate the contribution of
all these additional systematics as $\sigma_{\rm syst}^{\rm other}$,   
while we use $\sigma_{\rm syst}^{\rm model}$  for denoting
the theoretical systematics ascribed to imperfections in the RSD modeling.
Adopting similar conventions as in the companion papers, for a given cosmological parameter $x_{\rm p}$ measured with error $\sigma_{\rm p}$ whose reference
value is $x_{\rm p}^{\rm ref}$, the systematic error assigned is:
\begin{equation}
\sigma_{\rm p, syst} = 2 \sigma_{\rm p} ~~ {\rm if} ~ |x_{\rm p} - x_{\rm p}^{\rm ref}| < 2 \sigma_{\rm p};   
\label{eq_systematics_A}
\end{equation}
\begin{equation}
\sigma_{\rm p, syst} =  |x_{\rm p} - x_{\rm p}^{\rm ref}| ~~ {\rm if} ~ |x_{\rm p} - x_{\rm p}^{\rm ref}| \ge 2 \sigma_{\rm p}.   
\label{eq_systematics_A}
\end{equation}
In essence, anything above the $2\sigma$ level is considered as a detected systematics (corresponding to a $95\%$ confidence level on the mean of the mocks), 
and the maximal value is always used as a conservative choice.
The statistical properties of the LRG sample are also characterized in  \citet{LRG_corr2020} and in  \citet{Gil-Marin2020}, and the
consensus statistical error related to each individual method is denoted here as $\sigma_{\rm stat}$.
Recall again that from a joint BAO and RSD FS analysis, both in configuration and Fourier space, the
statistical consensus errors are $1.9\%$ on $\alpha_\perp$ and $2.6\%$ on $\alpha_\parallel$, respectively.
In the subsequent analysis, we always consider fits to the mean,
as they are less sensitive to noise effects compared to individual fits, and 
only focus on RSD FS results.

Table \ref{table_6_systematics} summarizes
the global error budget for the eBOSS DR16 LRG sample.
Here, the modeling systematics is inferred from the \textsc{Nseries} 
(see Table \ref{table_lrg_team_5}) and indicated as $\sigma_{\rm syst}^{\rm model, \textsc{NS}}$: we
explain later on the reason behind this choice, and why the Outer Rim contribution is not included here. 
The comprehensive systematic error budget intrinsic to each RSD
method ($\sigma_{\rm syst}$) is simply obtained by 
summing in quadrature the modeling and additional systematics,
and the total error budget $\sigma_{\rm tot}$ is also derived
in quadrature from  the contributions of $\sigma_{\rm syst}$
and  $\sigma_{\rm stat}$. In the table, we provide some useful ratios as well,
that allow one to directly compare the contribution of systematics
or statistics to the total error estimate.  
While the BAO-only pipeline is essentially unbiased, 
from the RSD FS analyses we conclude that  systematic errors account for a
significant fraction of the total error budget, contributing up to $50\%$ (or more) to 
the uncertainties associated with the AP parameters and the growth of structures (see the various ratios).
The impact in the determination of  $\alpha_\parallel$ and $\alpha_\perp$ is at the $\sim1.0\%$ level,
and it can reach even $\sim 1.5-2.0\%$ for $f\sigma_8$.
From configuration space analyses, the most relevant contribution to systematics is caused by
observational artifacts. In Fourier space, the most 
dominant systematic is arising from the
assumption of a reference cosmology, that can bias in particular the estimation of $f\sigma_8$ up to $2.0\%$.
All of the other effects, including modeling systematics, are within the $1.0\%$ range or below. 
Eventually, systematic errors are added only to the diagonal of the
covariance of each measurement, assuming that all the contributions to systematics are independent.


\begin{table}
\caption{Modeling systematics derived from RSD FS analyses of the `Th2' Outer Rim challenge mocks (see Tables \ref{table_OR_unblind_set} and \ref{table_number_density_unblind_set}).}
\begin{center}
\begin{tabular}{|c|c|c|c|c|}
\hline
\hline
RSD-FS Analysis & Model. Syst. \\
\hline
Systematic Type & Model &  $\sigma_{\alpha_\parallel}$ &  $\sigma_{\alpha_\perp}$  &  $\sigma_{\rm f \sigma_8}$   \\
\hline
\hline
RSD Modeling 	         						&  CLPT-GS  &  0.0228 & 0.0134   &    0.0330  \\
$[\sigma_{\rm syst, LH}^{\rm model, \textsc{OR}}]$ 	&  CF-TNS   &   0.0220 &   0.0104 &  0.0292   \\
										&  P$_{\rm k}$-TNS &  0.0276 &  0.0188  & 0.0400  \\
\hline
RSD Modeling 	         						&  CLPT-GS  &   0.0234 & 0.0154 &   0.0356 \\
$[\sigma_{\rm syst, TK}^{\rm model, \textsc{OR}}]$ 	&  CF-TNS   &  0.0204 & 0.0118 & 0.0280  \\
										&  P$_{\rm k}$-TNS &  0.0288 &  0.0214 & 0.0420 \\
\hline
RSD Modeling 	         						&  CLPT-GS  &  0.0218 & 0.0134 & 0.0356   \\
$[\sigma_{\rm syst, HE}^{\rm model, \textsc{OR}}]$ 	&  CF-TNS   &  0.0196 &  0.0124 &  0.0304  \\
										&  P$_{\rm k}$-TNS &  0.0258 & 0.0178 &  0.0380 \\
\hline
RSD Modeling 		    						&  CLPT-GS  &   0.0227 &  0.0141 &   0.0347    \\
$[\sigma_{\rm syst}^{\rm model, \textsc{OR}}]$ 		&  CF-TNS   &    0.0207 &  0.0115 &   0.0292   \\
										&  P$_{\rm k}$-TNS &   0.0274 &   0.0193 &   0.0400   \\
\hline
\hline
 		            				      					      		&  CLPT-GS  &  0.8143  & 0.6700  & 0.7334   \\
$\sigma_{\rm syst, LH}^{\rm model, \textsc{OR}}$/$\sigma_{\rm stat}$ 	&  CF-TNS   &  0.7097 &  0.5778  & 0.7300   \\
							  					     		&  P$_{\rm k}$-TNS &  0.7667  & 0.6963  & 0.9524  \\
\hline
 		            				      					      &  CLPT-GS  &  0.8357  & 0.7700    & 0.7911   \\
$\sigma_{\rm syst, TK}^{\rm model, \textsc{OR}}$/$\sigma_{\rm stat}$ &  CF-TNS   &  0.6581  & 0.6556  & 0.7000   \\
							  					     &  P$_{\rm k}$-TNS & 0.8000  & 0.7926  & 1.0000  \\
\hline
 		            				      					      &  CLPT-GS  & 0.7786  & 0.6700 &  0.7911   \\
$\sigma_{\rm syst, HE}^{\rm model, \textsc{OR}}$/$\sigma_{\rm stat}$ &  CF-TNS   &  0.6323  & 0.6889  & 0.7600   \\
							  					     &  P$_{\rm k}$-TNS &  0.7167&  0.6593 &  0.9048   \\
\hline
 		            				      					      &  CLPT-GS  & 0.8095 &   0.7033 &   0.7718    \\
$\sigma_{\rm syst}^{\rm model, \textsc{OR}}$/$\sigma_{\rm stat}$ &  CF-TNS   &   0.6667 &   0.6407 &   0.7300    \\
							  					     &  P$_{\rm k}$-TNS &  0.7611 &   0.7160 &   0.9524   \\
\hline
 		            				      										  &  CLPT-GS  &  2.5334 &  3.3500  & 3.3000  \\
$\sigma_{\rm syst, LH}^{\rm model, \textsc{OR}}$/$\sigma_{\rm syst}^{\rm model, \textsc{NS}}$ &  CF-TNS   &  3.6667 &  2.6000 &   3.6500   \\
							  										  &  P$_{\rm k}$-TNS &  4.3125 &  1.9789 &  4.8781   \\
\hline
 		            				      										  &  CLPT-GS  &  2.6000  & 3.8000  & 3.5600   \\
$\sigma_{\rm syst, TK}^{\rm model, \textsc{OR}}$/$\sigma_{\rm syst}^{\rm model, \textsc{NS}}$ &  CF-TNS   &  3.4000  & 2.9500   & 3.5000   \\
							  										  &  P$_{\rm k}$-TNS &  4.5000 &  2.2526   &5.1219  \\
\hline
 		            				      										  &  CLPT-GS  &  2.4222  & 3.3500   & 3.5600  \\
$\sigma_{\rm syst, HE}^{\rm model, \textsc{OR}}$/$\sigma_{\rm syst}^{\rm model, \textsc{NS}}$ &  CF-TNS   & 3.2667   & 3.1000  &  3.8000  \\
							  										  &  P$_{\rm k}$-TNS & 4.0312  & 1.8737   & 4.6341   \\
\hline
 		            				      										  &  CLPT-GS  &  2.5185 &   3.5167 &   3.4733  \\
$\sigma_{\rm syst}^{\rm model, \textsc{OR}}$/$\sigma_{\rm syst}^{\rm model, \textsc{NS}}$ &  CF-TNS   &   3.4445  &  2.8834  &  3.6500  \\
							  										  &  P$_{\rm k}$-TNS &  4.2812  & 2.0351 &  4.8780 \\
\hline
\hline		
\end{tabular}
\end{center}
\label{table_7_systematics}
\end{table}


\subsection{Impact of Modeling Systematics}

The modeling systematics estimated from `Th2' Outer Rim mocks analyzed in Section \ref{sec_unblind_challenge}
are detailed in  Table \ref{table_7_systematics}, where we list all
the contributions inferred from the individual HODs of Leauthaud (LH),
Tinker (TK), and Hearin (HE), respectively, as well as the combined  
theoretical systematics derived by 
simply averaging those contributions ($\sigma_{\rm syst}^{\rm model, OR}$).
We also report some useful ratios, for the ease of comparison. 
Not surprisingly, the
modeling systematics 
obtained from Outer Rim mocks
are
much larger than those derived from the \textsc{Nseries}.
The reason is related to the difference in effective volume, 
combined with the limited number of fully independent realizations available (27 synthetic catalogs per flavor).
In fact, the global effective volume of `Th2' Outer Rim mocks is 
$29.7$ ${\rm Gpc}^3$, about 11 times bigger than the combined CMASS plus eBOSS LRG sample,
but $\sim10.27$ times smaller than the global effective volume spanned by the \textsc{Nseries} --
which is $308.28$ ${\rm Gpc}^3$, and thus $113$ times larger than the combined DR16 LRG sample.
In this respect, the statistical threshold of the  \textsc{Nseries}  is at the $0.1-0.5\%$ level, while
the resolution of Outer Rim mocks is around $1.0-1.5\%$. 
As evident from Table \ref{table_7_systematics}, the modeling
systematics inferred from Outer Rim mocks is closer to the 
statistical error of the LRG sample (see the various ratios $\sigma_{\rm syst, LH}^{\rm model, OR}/\sigma_{\rm stat}$,
$\sigma_{\rm syst, TK}^{\rm model, OR}/\sigma_{\rm stat}$, $\sigma_{\rm syst, HE}^{\rm model, OR}/\sigma_{\rm stat}$),
and a factor $\sim 2-5$ times bigger than uncertainties 
derived from the \textsc{Nseries}. The size of the errorbars simply scales with the 
global effective volume and the number of available realizations, as clearly 
highlighted by the various ratios $\sigma_{\rm syst, LH}^{\rm model,OR}/\sigma_{\rm syst}^{\rm model,NS}$,
 $\sigma_{\rm syst, TK}^{\rm model,OR}/\sigma_{\rm syst}^{\rm model,NS}$, and  $\sigma_{\rm syst, HE}^{\rm model,OR}/\sigma_{\rm syst}^{\rm model,NS}$.   
This is the main reason why we omit to
include these theoretical systematics in the 
previous global error budget, as the larger errorbars 
are primarily due to limited volume and statistics. Note, however, that the Outer Rim mocks are fully independent, and not pseudo-independent realizations, 
hence the errorbars are completely uncorrelated.   
After all, although closer to the statistical limit of the sample,
the Outer Rim mocks allowed us to test and validate the 
robustness of the LRG analysis pipelines, the sensitivity to a number of HOD prescriptions, and to confirm 
the remarkable consistency across different methods in configuration and Fourier space. 



\section{Conclusions and Outlook} \label{sec_conclusions}


 
In support of the final analysis of the  eBOSS DR16 galaxy sample,
we have carried out an extensive $N$-body data challenge
with the aim of testing and validating the robustness of the LRG clustering pipelines of  \citet{LRG_corr2020}
in configuration space, and of  \citet{Gil-Marin2020} in Fourier space. We have also quantified
the theoretical systematics related to 
BAO  and RSD fitting methodologies, and the bias intrinsic to the modeling of the galaxy-halo connnection.

To this end, we have constructed new heterogeneous galaxy mocks 
from the Outer Rim simulation spanning different redshift intervals, using a  variety of HOD schemes of increasing complexity and characterized by analogous clustering properties.
The theoretical  foundation for modeling the galaxy-halo connection is laid out in Section \ref{sec_theory}, and 
the mock-making procedure is explained in detail in Section \ref{sec_tools_methodology}.   
Moving from the most conventional HOD approach, we have considered more sophisticated scenarios able to 
distinguish between quiescent or star-forming galaxies, and with the inclusion of assembly bias that generalize further the standard HOD framework.
Our Outer Rim-based mocks cover a range of number densities and effective volumes, and are well-suited for a variety of studies. 
In this work, we  have mainly focused on a subset at $z=0.695$, with characteristics closer to the eBOSS LRG sample 
(i.e., `Th2' flavor with the Leauthaud, Tinker, and Hearin prescriptions). 
We have also briefly exploited 
a small homogeneous synthetic set (the \textsc{Nseries}), which has been previously used in the SDSS DR12 galaxy clustering analysis
and  is more suitable to assess theoretical systematics at the sub-percent level, thanks to a larger effective volume.  


In our challenge, detailed in  Section  \ref{sec_unblind_challenge}, 
we have tested the performance of BAO and RSD fitting techniques against different galaxy population schemes and bias 
models having analogous clustering properties, with the main objective of validation and calibration of such methods and the quantification of theoretical systematics. 
The mock products have allowed us to confront on a common ground and 
assess the performance of the BAO fitting methodologies for the LRG sample, and 
of three complementary RSD models in configuration and Fourier space -- denoted as CLPT-GS, CF-TNS, and P$_{\rm k}$-TNS, respectively. 
Overall, we have found a remarkable agreement at the sub-percent level
between different techniques in configuration and Fourier space
(see in particular Figures \ref{fig_lrg_team_3a}, \ref{fig_team_4} and \ref{fig_team_5}), along with an impressive 
level of consistency among
 BAO fitting and reconstruction procedures and from all the RSD 
models used in FS analyses. All of the methods
performed equally well, with comparable errors on the 
AP parameters and the growth of structure. Moreover, 
reconstruction significantly improved  the constraints on both $\alpha_\parallel$ and $\alpha_\perp$. 
We have thus validated the robustness of the LRG clustering analysis pipelines. 

Including in our analysis the complexity of  more sophisticated  HODs schemes 
that go beyond the traditional mass-only ansatz, such as models with  
central and satellite velocity bias \citep{Tinker2007,Guo2015}, 
generalizations of the standard five parameter HOD model with various halo-scale physics 
\citep{Yuan2018, Duan2019, XuZheng2020,Xu+2020}, and
forward-modelling approaches using satellite kinematics \citep{Lange2019a,Lange2019b,vdb2019}
is left to future studies. However, as pointed out by \citet{Duan2019}, 
only a rather unrealistic extreme level of velocity bias of the central galaxies would produce a shift of 0.7\% in the 
LOS acoustic scale, and basically other bias models including satellite velocity bias are consistent with zero shift at the 0.2\% level after reconstruction. 
Since current surveys such as eBOSS measure the acoustic scale to $\sim 1\%$ precision, they are insensitive to the galaxy bias effect --
and we expect that similar conclusions will hold for the related AP parameters.

Regarding systematics and the global error budget (Section \ref{sec_global_sys}), we have
found that the impact of different HOD prescriptions is always sub-dominant 
to the total systematics, and that 
modeling systematics in the estimation of  $\alpha_\parallel$ and $f \sigma_8$, although at worst around $\sim 1.5\%$, may be  
a dominant source of error in the comprehensive quantification of systematics. 
In particular, from the analysis in configuration space of 
pre-reconstructed mocks (considering only fits to the mean), biases in the recovered $\alpha$ values reach up to $0.5\%$
in $\alpha_\perp$ and $1.0\%$ in $\alpha_\parallel$.
After reconstruction, there is a reduction of the biases to less than $0.2\%$, hence the  BAO analysis is unbiased.
For RSD analyses in configuration space,  the most significant contribution to systematic errors arises from observational effects.
From the Fourier space methodology, 
for the post-reconstruction BAO analysis we detected a $0.5\%$ systematic shift induced by modeling systematic on $\alpha_\parallel$, 
and none for $\alpha_\perp$, with a resolution limit of $0.2\%$ for the \textsc{Nseries} mocks.
The systematic shift is of order $1.5\%$ for the FS analysis from Outer Rim mocks instead. Moreover, we did not detect 
any significant relative shift on the cosmological parameters when either the HOD model or the flavor is
varied. Such results put constrains in the upper limit of systematic errors in the modeling, as a result of different HODs,  
with upper limits of order $0.5-1.1\%$ systematic shifts.
In any case, both HOD and modeling systematics are below the statistical error of the eBOSS LRG data.  
The expected statistical errors in the eBOSS LRG data sample are in fact of the order of $\sim 1.9\%$ 
for $\alpha_\perp$, and $\sim 2.6\%$ for $\alpha_\parallel$.
Eventually, these systematic corrections in the  AP parameters and the growth of structure
are combined with additional sources of systematics (Table \ref{table_6_systematics}), and such
errors are accounted for in the final consensus results \citep{eBOSS_Cosmology2020}
from the analysis of the LRG DR16 galaxy sample.
Finally, our analysis  provides a global and complementary perspective of the
systematic studies carried out in \citet{LRG_corr2020} in configuration space, and
in  \citet{Gil-Marin2020} in Fourier space: their overall agreement at such level of precision is remarkable. 


Quantifying the modeling systematics in BAO clustering estimators and in RSD methods for all the eBOSS tracers, as well as characterizing the robustness 
of the analysis pipelines, are essential tasks in order to obtain unbiased cosmological parameters, accurate $f \sigma_8$ constraints, and reliable consensus likelihoods.
In this respect, besides being relevant for the 
final eBOSS DR16 `consensus cosmology' -- as the systematic error budget is informed by testing the results of analyses against these high-resolution mocks --
our study represents also a testbed for future large-volume surveys. 
In particular,  similar mock-making techniques and systematic corrections can be readily extended to model for instance the DESI galaxy sample,
and we expect that more extensive mock challenges along these lines  will be necessary and progressively relevant
in the next few years. In fact, mock challenges designed to validate data analysis pipelines and assess the impact of 
systematics in massive datasets
are becoming increasingly important for large-volume surveys -- see for example the recent 
works by \citet{MacCrann2018} for the
Dark Energy Survey (DES; \citealt{DES2005}), and by
\citet{Sanchez2020} for LSST \citep{Ivezic2019}.
In this view, while the sub-percent level theoretical systematic corrections quantified in this study  may not be relevant for the current state-of-the-art (as they are always inferior to the statistical precision of the data),
soon they will become relevant for DESI and LSST, that are expected to achieve sub-percent statistical precision on the galaxy sample; for such surveys, it will be crucial to control the systematics at an extremely low level. 
To this end, our flexible and highly modular pipeline for building complex HODs offers 
several  directions of extension, as well as applications that go beyond the modeling of LRGs --
toward more elaborated galaxy-halo connection physics, particularly in relation to ELGs.



\section*{Data Availability}

All of the SDSS mock products developed in this study, listed in Table \ref{table_OR_unblind_set}, are stored at the National Energy Research 
Scientific Computing Center (NERSC) and are available upon request. The Outer Rim halo catalogs used to produce
the Outer Rim-based mocks are publicly available at \url{https://cosmology.alcf.anl.gov}.



\section*{Acknowledgements}

GR, PDC, and JM acknowledge support from the National Research Foundation of Korea (NRF) through Grants No. 2017R1E1A1A01077508 
and No. 2020R1A2C1005655 funded by 
the Korean Ministry of Education, Science and Technology (MoEST), and from the faculty research fund of Sejong University. 
GR was on sabbatical leave in 2020, when this research was completed.  
Some numerical work performed in this paper was carried out using the Korea Institute of Science and Technology Information (KISTI) 
supercomputing infrastructure under allocations KSC-2018-G3-0008 and KSC-2018-T1-0061, and with
computing resources at Sejong University (Xeon Silver 4114 master node and Xeon Gold 6126 computing node architecture).

HGM acknowledges support from la Caixa Foundation (ID 100010434) which code LCF/BQ/PI18/11630024. 

RP, SdlT, and SE acknowledge support from the eBOSS ANR grant (under contract ANR-16-CE31-0021) and the OCEVU LABEX (Grant No. ANR-11-LABX-0060) of the French National Research Agency.

Funding for the Sloan Digital Sky Survey IV has been provided by the Alfred P. Sloan Foundation, the U.S. Department of Energy Office of Science, and the Participating Institutions. 
SDSS-IV acknowledges
support and resources from the Center for High-Performance Computing at
the University of Utah. The SDSS web site is www.sdss.org.

SDSS-IV is managed by the Astrophysical Research Consortium for the 
Participating Institutions of the SDSS Collaboration including the 
Brazilian Participation Group, the Carnegie Institution for Science, 
Carnegie Mellon University, the Chilean Participation Group, the French Participation Group, Harvard-Smithsonian Center for Astrophysics, 
Instituto de Astrof\'isica de Canarias, The Johns Hopkins University, Kavli Institute for the Physics and Mathematics of the Universe (IPMU) / 
University of Tokyo, the Korean Participation Group, Lawrence Berkeley National Laboratory, 
Leibniz Institut f\"ur Astrophysik Potsdam (AIP),  
Max-Planck-Institut f\"ur Astronomie (MPIA Heidelberg), 
Max-Planck-Institut f\"ur Astrophysik (MPA Garching), 
Max-Planck-Institut f\"ur Extraterrestrische Physik (MPE), 
National Astronomical Observatories of China, New Mexico State University, 
New York University, University of Notre Dame, 
Observat\'ario Nacional / MCTI, The Ohio State University, 
Pennsylvania State University, Shanghai Astronomical Observatory, 
United Kingdom Participation Group,
Universidad Nacional Aut\'onoma de M\'exico, University of Arizona, 
University of Colorado Boulder, University of Oxford, University of Portsmouth, 
University of Utah, University of Virginia, University of Washington, University of Wisconsin, 
Vanderbilt University, and Yale University.

In addition, this research relied on resources provided to the eBOSS 
Collaboration by the National Energy Research Scientific Computing 
Center (NERSC). NERSC is a U.S. Department of Energy Office of Science 
User Facility operated under Contract No. DE-AC02-05CH11231. 



\bibliographystyle{mnras}
\bibliography{references} 



\appendix



\section{Mock Products} \label{sec_appendix_mock_products}

 
\begin{table*}
\centering
\caption{List of Outer Rim synthetic products developed for the galaxy mock challenge.}
\doublerulesep2.0pt
\renewcommand\arraystretch{1.5}
\begin{tabular}{cccccc} 
\hline \hline
Set & HOD Style & HOD Flavor & Redshift  & Box [$h^{-1}{\rm Gpc}$] & Total Mocks \\
\hline \hline
                       & Zheng07 & Th1, Std, Th2 & 0.695, 0.865 & 1.0 & 810 \\
{\bf Challenge Set 1} & Leauthaud11 & Th1, Std, Th2 & 0.695, 0.865 & 1.0 & 810 \\
                      & Tinker13 & Th1, Std, Th2 & 0.695, 0.865 & 1.0 & 810 \\
                      & Hearin15 & Th1, Std, Th2 & 0.695, 0.865 & 1.0 & 810 \\
\hline 
                      &   &  &   &  &  {\bf 3240} \\
\hline \hline 
                              &  Leauthaud11 & Th1 & 0.402, 0.502, 0.618, 0.695, 0.779, 0.865, 1.006 & 1.0 & 945   \\
{\bf Challenge Set 2} &  Tinker13 & Th1 & 0.402, 0.502, 0.618, 0.695, 0.779, 0.865, 1.006 & 1.0 & 945   \\
                              & Hearin15 & Th1 & 0.402, 0.502, 0.618, 0.695, 0.779, 0.865, 1.006 & 1.0 & 945   \\
\hline 
                      &   &  &   &  &  {\bf 2835} \\
\hline \hline 
                       & Zheng07 & Th1, Std, Th2 & 0.695, 0.865 & 3.0 & 600 \\
{\bf Challenge Set 3} & Leauthaud11 & Th1, Std, Th2 & 0.695, 0.865 & 3.0 & 600 \\
                      & Tinker13 & Th1, Std, Th2 & 0.695, 0.865 & 3.0 & 600 \\
                      & Hearin15 & Th1, Std, Th2 & 0.695, 0.865 & 3.0 & 600 \\
\hline 
                      &   &  &   &  &  {\bf 2400} \\
\hline \hline
\label{table_OR_unblind_set}
\end{tabular}
\end{table*}


We have devised 
three sets of heterogeneous Outer Rim-based galaxy mocks (indicated as `Challenge Set 1', 
`Challenge Set 2', `Challenge Set 3', respectively) for the galaxy mock challenge. These are cubic mocks, 
in the Outer Rim cosmology, obtained by  populating Outer Rim halo catalogs with galaxies as explained in Section \ref{subsec_OR}.
Details regarding each set are provided next. 

Specifically, \textit{`Challenge Set 1}' (HOD VARIATIONS)
contains a total of 3240 mocks (1620 at $z=0.695$, and 1620 at $z=0.865$), 
grouped into 4 model categories according to the underlying HOD 
scheme (i.e., Zheng, Leauthaud, Tinker, Hearin); each model category 
consists of 3 `flavors', denoted as `Standard' (Std), `Threshold 1' (Th1), and `Threshold 2' (Th2).
As explained in Sections \ref{sec_theory} and \ref{sec_tools_methodology_lrgs},
the meaning of `flavor' is related to the 
key parameter  `threshold', which 
globally sets all the individual HOD parameters as best fit realizations from the corresponding literature dictionary of each HOD model
  (unless specific customizations are introduced).
At a given redshift, we produced 135 mocks per model flavor, by populating the full $3 h^{-1} {\rm Gpc}$ Outer Rim periodic halo catalog box 5 times, and
then cutting each full box into 27 subcubes of $1h^{-1}{\rm Gpc}$ side and rescaling the various spatial positions accordingly. 
This means that effectively we have  27 fully independent (i.e., not sharing the same DM field) mocks per realization, and each of these 27 mocks
will have 5 different replicas. According to the modeling explained in Section \ref{sec_theory}, 
central galaxies are always located at the center of their parent halos with identical velocities, 
while the satellite population is statistically different in all the realizations -- assuming a NFW profile. 
We then add RSDs to each individual mock in two different ways: radially, or with the usual plane-parallel approximation. 
This is the primary set considered in our main analysis presented in Section \ref{sec_unblind_challenge}.

\textit{`Challenge Set 2}' (REDSHIFT EVOLUTION) 
is similar to the previous one, but now the redshift evolution is taken into account for one threshold flavor
and 3 different HOD prescriptions. In detail, 
we consider 7 redshift intervals, namely $z= 0.402, 0.502, 0.618, 0.695, 0.779, 0.865, 1.006$,
and produced a set of 2835 mocks ($135 \times 7 \times 3$) with 3 HOD  schemes (Leauthaud, Tinker, Hearin), for the 
`Th1'  flavor. 
Even in this case, we consider subcubes of $1h^{-1}{\rm Gpc}$ side. 

Finally, \textit{`Challenge Set 3}'  (HOD VARIATIONS / LARGE BOX)
is similar to the first one, but in this case we exploit the full Outer Rim box ($3h^{-1}{\rm Gpc}$)
with periodic boundary conditions rather than subcubes, and produced 100 realizations 
per flavor for all the HODs and thresholds considered in the first set -- for a total of
2400 mocks. 
If desirable, these large-box realizations can be casted into smaller 
pseudo-independent mocks by performing cuts along different directions of the boxes,
and also via the inclusion of partial overlaps in order to maximize the effective volume.

Regardless of the specific set, each mock contains the following information:
galaxy spatial positions (in $h^{-1}{\rm Mpc}$), galaxy velocities (in comoving $km/s$), the galaxy type (central, satellite),
the number of centrals that a halo hosts (either 0 or 1), 
the number of satellites per halo, the 
global ID of the halo a galaxy belongs to,
the halo mass and virial radius, the  
central star formation designation for some models (active, quiescent),
the number of active or quiescent satellites, and
the percentile spit in concentration for models with assembly bias.

A summary of all the synthetic products available, categorized by HOD and redshift, is
provided in Table \ref{table_OR_unblind_set}. 
While only a subset of these mocks is used for testing the BAO templates and the RSD models adopted for the characterization of LRG clustering systematics,
with this work we release the entire suite of products -- that are suitable to several interesting applications.


Table \ref{table_number_density_unblind_set} reports the 
number densities of the challenge mocks, expressed in units of
$10^{-4} [h^3 {\rm Mpc}^{-3}]$, and ordered by HOD type and flavor.
The `Th2' models of Leauthaud, Tinker, and Hearin are those
characterized by a number density closer to the eBOSS LRG sample (see e.g. Figure 1 in
\citealt{Gil-Marin2020} and Figure 1 in \citealt{LRG_corr2020} for details), and are
extensively used in Section \ref{sec_unblind_challenge} to test the LRG BAO and RSD analysis pipelines. 
The other threshold levels are more suitable for example  for assessing the details of the satellite distributions and for studying the 
galaxy-halo connection, particularly in relation to ELGs; 
they are only marginally explored in this work, although of high interest, and left to future applications.  


\begin{table}
\centering
\caption{Number densities of the challenge mocks listed in Table \ref{table_OR_unblind_set}, ordered per HOD type and flavor.}
\doublerulesep2.0pt
\renewcommand\arraystretch{1.5}
\begin{tabular}{cccc} 
\hline \hline  
Number Density [$10^{-4} (h^3 {\rm Mpc}^{-3})$]  \\ 
\hline
HOD Model & Th1 & Std & Th2 \\
\hline
Zheng07 & 67.05 & 34.26 & 5.64  \\
Leauthaud11 & 50.55 & 11.85 & 0.69 \\
Tinker13 & 26.37 & 8.12 & 0.79 \\
Hearin15  & 44.93 & 10.82 & 0.69 \\
\hline
\hline
\label{table_number_density_unblind_set}
\end{tabular}
\end{table}


\begin{figure*}
\centering
\includegraphics[angle=0,width=0.80\textwidth]{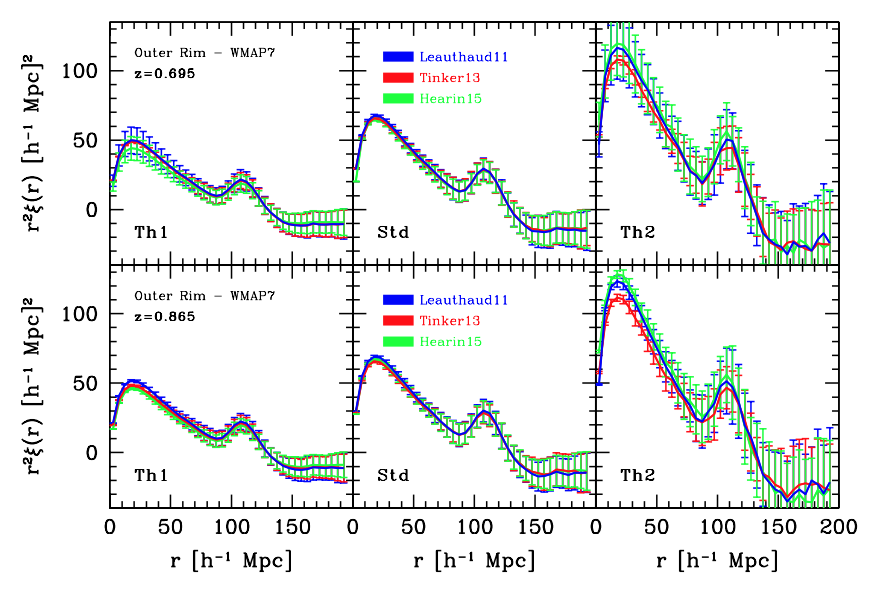}
\caption{Clustering properties of the \textit{`Challenge Set 1}'. 
Examples of 2-point spatial correlation functions computed at $z=0.695$ (top panels) and at $z=0.865$ (bottom panels), for the
three threshold levels denoted as `Th1', `Std', and `Th2'  -- in decreasing number density order, from left to right. 
The Leauthaud (blue), Tinker (red), and Hearin (green) HOD models are
displayed with different colors. They are characterized by approximately
similar clustering properties, at a fixed threshold level.}
\label{fig_or_clustering_global}
\end{figure*}

 
\begin{figure}
\centering
\includegraphics[angle=0,width=0.41\textwidth]{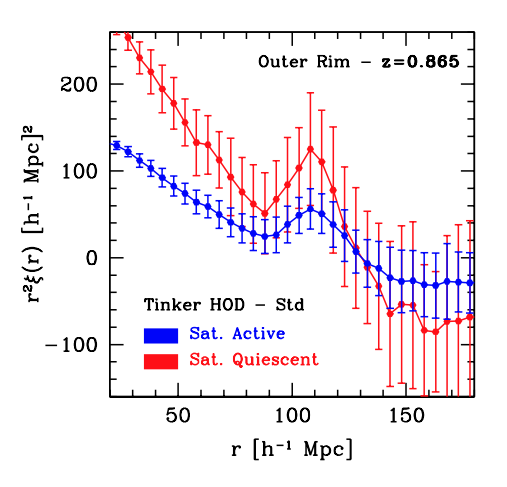}
\caption{Spatial clustering of satellite galaxies at $z=0.865$, split by active/star-forming (blue) and quiescent (red) -- as described by the Tinker model for the `Std' threshold level. 
This more complex HOD framework provides interesting insights on the galaxy-halo connection physics.}
\label{fig_or_clustering_elgs}
\end{figure}


Figure \ref{fig_or_clustering_global} shows examples of the three-dimensional galaxy clustering  quantified by
 the 2-point spatial correlation $\xi(r)$, as a function of separation $r$. The various measurements are performed
at $z=0.695$ (top panels)
and at $z=0.865$ (bottom panels), for the mocks belonging to \textit{`Challenge Set 1}'. 
From left to right, the `Th1', `Std', and `Th2' flavors are displayed, respectively. 
The HOD models of Leauthaud (blue), Tinker (red), and Hearin (green) are
displayed with different colors: they are characterized by approximately
similar clustering properties, at a fixed threshold level. 
Each measurement represents an average over 135 mocks, according to the specific HOD style and flavor, and errorbars are $1\sigma$
variations.  The effect of a decreasing number density (from left to right) is of an overall increase in the clustering and BAO peak 
amplitude, with a relatively  small redshift dependence.  

Finally, Figure \ref{fig_or_clustering_elgs} displays an 
example of the 2-point spatial clustering of satellites, split by 
active/star-forming (blue) and quiescent (red) galaxies, at $z=0.865$ for the
`Std' threshold level (see Table \ref{table_number_density_unblind_set}). 
The Tinker HOD formalism used in this work -- primarily for LRG studies --  could also potentially provide interesting applications to ELGs that go beyond the scope of this paper, 
and that highlight the flexibility of our mock-making procedure and mock products in exploring the physics of the galaxy-halo connection.



\section{Useful Tables} \label{sec_appendix_tables}


We provide here some useful tables 
with numerical results that appear in the various plots displayed in the manuscript, 
for the ease of direct comparisons,
or that include additional information not
reported in the main text. 

Specifically,  Table \ref{table_lrg_team_zheng} contains
results from the BAO and RSD analyses in Fourier space 
related to the Zheng model -- see also our companion paper \citet{Gil-Marin2020} for additional extensive details. 

Table \ref{table_lrg_team_3} reports the results when fitting the mean correlation functions and power spectra 
(rows labeled `Mean') of a set of 27 independent realizations 
of the OR-based mocks with different HOD prescriptions (`Challenge Set 1'), as well as   
the mean of the fits of individual realizations (rows labeled `Individual'), in relation to the 
BAO analyses presented in Section \ref{sec_bao_team_analysis}.

Table \ref{table_lrg_team_4} reports the 
main results of the RSD FS analyses presented in Section \ref{sec_rsd_team_analysis},
confronting the three different modeling techniques: 
2 in configuration space (CLPT-GS and CF-TNS), and one
in Fourier space (P$_{\rm k}$-TNS). 
Individual fits on each of the 27 realizations per model are performed, as well as fits on the mean of the mocks.

Finally, Table  \ref{table_lrg_team_5} 
provides the numerical results 
obtained by running the various BAO and RSD FS analysis pipelines on
the average of 84 \textsc{Nseries} mocks, as
detailed in Section \ref{sec_systematics_analysis}.
The analysis is performed only over pre-reconstructed catalogs, and 
carried out in the \textsc{Nseries} cosmology (Section \ref{subsec_Nseries}).


\begin{table*}
\caption{Impact of the Zheng HOD model, presented in three different flavors (see Table \ref{table_zheng07}), 
on pre-reconstructed BAO and FS fits on OR cubic boxes (1$h^{-1}{\rm Gpc}$; `Challenge Set 1'), from a Fourier space analysis. 
For simplicity, only the fit to the mean is provided -- see \citet{Gil-Marin2020} for extensive details.}
\begin{center}
\begin{tabular}{|c|c|c|c|c|c|}
\hline
\hline
Analysis Type & HOD Type & HOD Flavor & $\alpha_\parallel -\alpha_\parallel^{\rm exp}$ & $\alpha_\perp -\alpha_\perp^{\rm exp}$  & $f\sigma_8 - f\sigma_8^{\rm exp}$\\
\hline
\hline
BAO Fourier Space & Zheng & Th1 & $0.0270 \pm 0.0110$ & $-0.0006 \pm 0.0062$ &  -- \\
$[$Pre-Rec$]$ &Zheng& Std & $0.0240 \pm 0.0100$ &  $-0.0005 \pm 0.0060$ &  -- \\
&Zheng & Th2 & $0.0310 \pm 0.0130$ & $-0.0014 \pm 0.0075$&  -- \\
\hline
RSD Fourier Space & Zheng & Th1 &   $0.0107 \pm 0.0069$ &    $-0.0035 \pm 0.0052$   &    $-0.0085 \pm 0.0073$   \\
$[$Pre-Rec$]$ & Zheng & Std &   $0.0108 \pm 0.0067$    &  $-0.0025 \pm 0.0051$     &   $-0.0083 \pm 0.0065$  \\
&Zheng & Th2 &  $0.0138 \pm 0.0095$ & $-0.0049 \pm 0.0072$      &   $0.0080 \pm 0.0130$      \\
\hline
\hline
\end{tabular}
\end{center}
\label{table_lrg_team_zheng}
\end{table*}


\begin{table*}
\caption{Performance of the BAO templates  on OR-based mocks, for the `Th2' flavor and different HOD models. 
Mocks are analyzed in their own Outer Rim cosmology, so the expected values for both $\alpha_\parallel$ and  $\alpha_\perp$ are $1$.
For each set of mocks, the results from pre- and post-reconstruction catalogs are presented. 
We report both the results of fitting the mean of all the mocks, indicated with `Mean', and the mean of individual fits on the mocks, indicated as `Individual'. 
For the fit to the mean, the error quoted is the $1\sigma$ of the error on this fit, where  the covariances are scaled by the 27 \textsc{Outer Rim}  realizations per HOD
used to compute the mean.  
For the mean of individual best-fits, the error quoted is the {\it rms} divided by $\sqrt{N_{\rm OR}}$, where $N_{\rm OR}=27$.  
The average of the best-fits is then performed over $N_{\rm OR}$. 
Consequently, the errors of `Mean' and `Individual' are comparable. Results of these fits are displayed in Figure \ref{fig_lrg_team_3a}.}
\begin{center}
\begin{tabular}{|c|c|c|c|c|c|c|}
\hline
\hline
BAO Analysis Type & HOD Type & HOD Flavor &  Analysis Details & $\alpha_\parallel -\alpha_\parallel^{\rm exp}$ & $\alpha_\perp -\alpha_\perp^{\rm exp}$  & $N_{\rm det}/ N_{\rm tot}$\\
\hline
\hline
Configuration Space & Leauthaud & Th2 & Mean & $0.0084 \pm 0.0164$ & $-0.0010 \pm 0.0080$&  $1/1$\\
$[$Pre-Rec$]$ &Leauthaud & Th2 & Individual &  $0.0283  \pm 0.0197$ & $-0.0072 \pm  0.0123 $ & $27/27$\\
&Tinker & Th2 & Mean & $0.0234 \pm 0.0147$& $-0.0078 \pm 0.0073$&   $1/1$\\
&Tinker & Th2 & Individual & $0.0166  \pm 0.0172$  & $0.0000 \pm  0.0104$ &  $27/27$\\
& Hearin & Th2 & Mean &   $0.0040 \pm 0.0157$ &   $0.0074 \pm 0.0083$& $1/1$\\
& Hearin & Th2 & Individual  &   $0.0035 \pm  0.0185$  & $0.0090  \pm 0.0089$  &   $27/27$\\
\hline
Configuration Space & Leauthaud & Th2 & Mean &$-0.0063 \pm 0.0084$ & $0.0045 \pm 0.0056$&  $1/1$\\
$[$Post-Rec$]$ & Leauthaud & Th2 & Individual &  $-0.0050 \pm  0.0108$  & $0.0053 \pm  0.0063$ & $27/27$\\
&Tinker & Th2 & Mean & $-0.0024 \pm 0.0090$& $0.0014 \pm 0.0060$&   $1/1$\\
&Tinker & Th2 & Individual &   $0.0078  \pm 0.0103$ & $-0.0017 \pm  0.0065 $ & $27/27$\\
&Hearin & Th2 & Mean &   $0.0115 \pm 0.0094$ &   $0.0066 \pm 0.0057$& $1/1$\\
& Hearin & Th2 & Individual  &   $0.0019  \pm 0.0097$ &  $0.0087 \pm  0.0054$  &   $27/27$\\
\hline
Fourier Space & Leauthaud & Th2 & Mean & $-0.0114 \pm 0.0178$  & $0.0028 \pm 0.0107$   &$1/1$\\
$[$Pre-Rec$]$ & Leauthaud & Th2 & Individual & $0.0000 \pm0.0180$  & $0.0110 \pm0.0120$   &$27/27$\\
& Tinker & Th2 & Mean & $0.0023 \pm 0.0177$ & $-0.0047 \pm 0.0124$ &   $1/1$\\
& Tinker & Th2 & Individual & $-0.0110 \pm0.0160$  & $0.0230 \pm 0.0110$ &  $27/27$\\
& Hearin & Th2 & Mean & $-0.0221 \pm 0.0143$ & $0.0108 \pm 0.0099$ &   $1/1$\\
& Hearin & Th2 & Individual & $-0.0210\pm 0.0160$ & $0.0160 \pm0.0110$ &   $27/27$\\
\hline
Fourier Space & Leauthaud & Th2 & Mean & $-0.0062 \pm 0.0104$ &  $-0.0024 \pm 0.0075$ &  $1/1$\\
$[$Post-Rec$]$ & Leauthaud & Th2 & Individual & $0.0020 \pm0.0130$  & $-0.0093 \pm0.0074$  & $27/27$\\
& Tinker & Th2 & Mean & $-0.0024 \pm 0.0121$ & $-0.0025 \pm 0.0088$ &  $1/1$\\
& Tinker & Th2 & Individual & $0.0038 \pm0.0097$  & $-0.0006 \pm0.0072$ & $27/27$\\
& Hearin & Th2 & Mean & $0.0002 \pm 0.0113$ & $0.0122 \pm 0.0075$ &   $1/1$\\
& Hearin & Th2 & Individual & $0.0090\pm 0.0150$ & $0.0167\pm 0.0061$  &  $27/27$\\
\hline
\hline
\end{tabular}
\end{center}
\label{table_lrg_team_3}
\end{table*}


\begin{table*}
\caption{Performance of the RSD FS methods  evaluated on the OR-based mocks, for the `Th2' flavor with different HOD models. 
Mocks are analyzed in their own Outer Rim cosmology, so the expected values are $1$ for the $\alpha$ parameters and $f\sigma_8^{\rm exp} = 0.447$. 
For each set of mocks, the results from pre-reconstructed catalogs are presented. 
We report both the results of fitting the mean of all the 27 realization per HOD, indicated with `Mean', and the mean of individual fits on the mocks, indicated as `Individual'. 
For the fit to the mean, the error quoted is the $1\sigma$ of the error on the fit, where the covariances are scaled by the 27 realizations.  
For the mean of individual best-fits, the error quoted is the {\it rms} divided by the number of realizations.  These results are visualized in Figure \ref{fig_team_4}. 
}
\begin{center}
\begin{tabular}{|c|c|c|c|c|c|c|c|}
\hline
\hline
RSD Analysis Type & HOD Type & HOD Flavor &  Analysis Details & $\alpha_\parallel -\alpha_\parallel^{\rm exp}$ & $\alpha_\perp -\alpha_\perp^{\rm exp}$  &  $f\sigma_8 - f\sigma_8^{\rm exp}$ & $N_{\rm det}/ N_{\rm tot}$\\
\hline
\hline
Configuration Space & Leauthaud & Th2 & Mean &  $-0.0189 \pm 0.0114$ &  $0.0010 \pm 0.0067$ &  $0.0113 \pm 0.0165$  & $1/1$\\
$[$Pre-Rec$]$  & Leauthaud & Th2 & Individual & $-0.0069  \pm 0.0159$   & $-0.0052 \pm  0.0092$  & $0.0119  \pm 0.0209$ & $27/27$\\
CLPT-GS &Tinker & Th2 & Mean &  $ -0.0211 \pm 0.0117$ &  $0.0022 \pm 0.0077$ & $0.0066 \pm 0.0178$  & $1/1$\\
&Tinker & Th2 & Individual & $-0.0096 \pm 0.0135$   & $-0.0014 \pm  0.0093$  & $0.0193 \pm   0.0193$   & $27/27$\\
& Hearin & Th2 & Mean & $-0.0158 \pm 0.0109$ & $ 0.0008 \pm 0.0067$  & $ 0.0187 \pm  0.0178$   & $1/1$\\ 
& Hearin & Th2 & Individual & $-0.0053 \pm   0.0176$  & $-0.0065 \pm 0.0089$  & $0.0267 \pm  0.0250$ & $27/27$\\
\hline
Configuration Space & Leauthaud & Th2 & Mean &$-0.0012 \pm 0.0110$ & $-0.0046 \pm 0.0052$ &  $0.0260 \pm 0.0146$ & $1/1$\\
$[$Pre-Rec$]$ &Leauthaud & Th2 & Individual &  $0.0259 \pm 0.0122$  &   $-0.0001 \pm 0.0069$   & $0.0305 \pm  0.0168$  & $27/27$\\
CF-TNS &Tinker & Th2 & Mean & $0.0144 \pm 0.0102$ & $0.0014 \pm 0.0059 $   & $0.0130 \pm 0.0140$ & $1/1$\\
&Tinker & Th2 & Individual & $0.0375 \pm 0.0126$  & $-0.0049 \pm 0.0076$  & $0.0002 \pm  0.0157$  & $27/27$\\
&Hearin & Th2 & Mean &  $0.0056 \pm 0.0098$  & $0.0014 \pm 0.0062$  & $0.0227 \pm 0.0152 $& $1/1$\\
& Hearin & Th2 & Individual  &  $0.0291 \pm  0.0127$  &  $-0.0022 \pm 0.0077$  & $0.0205 \pm 0.0168$ &  $27/27$\\
\hline
Fourier Space  & Leauthaud & Th2 & Mean & $0.0034 \pm 0.0138$   &  $-0.0111 \pm 0.0094$   & $-0.0040 \pm 0.0200$& $1/1$\\
$[$Pre-Rec$]$ & Leauthaud & Th2 & Individual &  $0.0610 \pm 0.0140$  & $-0.0195 \pm 0.0087$   & $0.0060 \pm 0.0160$ & $27/27$\\
P$_{\rm k}$-TNS & Tinker & Th2 & Mean &  $0.0060 \pm 0.0144$ & $-0.0177 \pm 0.0107$   & $-0.0070 \pm 0.0210$  & $1/1$\\
& Tinker & Th2 & Individual &    $0.0970 \pm 0.0240$ & $-0.0047 \pm 0.0097$   & $0.0140 \pm 0.0220$  & $27/27$\\
& Hearin & Th2 & Mean & $-0.0104 \pm 0.0129$  &  $-0.0020 \pm 0.0089$ &  $0.0190 \pm 0.0190$ & $1/1$\\
& Hearin & Th2 & Individual & $0.0450 \pm 0.0170$  & $0.0001 \pm 0.0093$  & $0.0260 \pm 0.0220$ &  $27/27$\\
\hline
\hline
\end{tabular}
\end{center}
\label{table_lrg_team_4}
\end{table*}


\begin{table*}
\caption{Modeling systematics related to BAO and RSD methodologies, addressed with the  \textsc{Nseries}. These numerical results are shown in Figure \ref{fig_team_5}.}
\begin{center}
\begin{tabular}{|c|c|c|c|c|c|c|}
\hline
\hline
Analysis Type & Analysis Space &  Analysis Method & Fitting Model & $\alpha_\parallel -\alpha_\parallel^{\rm exp}$ & $\alpha_\perp -\alpha_\perp^{\rm exp}$  &  $f\sigma_8 - f\sigma_8^{\rm exp}$ \\
\hline
\hline
BAO $[$Pre-Rec$]$ & Configuration Space & CF  -- BAO Peak & BAO Template & $0.0014 \pm 0.0045$  &  $0.0059 \pm 0.0023$ & -- \\
BAO $[$Pre-Rec$]$ & Fourier Space  & PS -- BAO Peaks  & BAO Template & $-0.0045 \pm 0.0041$ &   $-0.0021 \pm 0.0020$ & -- \\
\hline
BAO $[$Post-Rec$]$ & Configuration Space & CF -- BAO Peak  & BAO Template & $0.0031 \pm 0.0024$  &    $0.0023 \pm 0.0015$ & -- \\
BAO $[$Post-Rec$]$ & Fourier Space & PS -- BAO Peaks   & BAO Template & $-0.0048 \pm 0.0019$ &   $0.0005 \pm 0.0010$ & -- \\
\hline
RSD $[$Pre-Rec$]$ & Configuration Space & CF -- Full Shape & CLPT-GS  & $-0.0090 \pm 0.0030$ & $0.0020 \pm 0.0020$  & $-0.0060 \pm 0.0050$  \\
RSD $[$Pre-Rec$]$ & Configuration Space & CF -- Full Shape & CF-TNS  & $-0.0050 \pm 0.0030$ & $ -0.0020 \pm 0.0020$ & $0.0060 \pm 0.0040$ \\
RSD $[$Pre-Rec$]$ & Fourier Space & PS -- Full Shape & P$_{\rm k}$-TNS & $0.0016 \pm 0.0032$ & $-0.0095 \pm 0.0020$ & $-0.0038 \pm 0.0041$  \\
\hline
\hline
\end{tabular}
\end{center}
\label{table_lrg_team_5}
\end{table*}


\label{lastpage}


\end{document}